\documentclass[aps,amsmath,amssymb,prd,floatfix,preprint,superscriptaddress,nofootinbib,12pt]{JHEP3}
\usepackage{youngtab}
\usepackage{epsfig}
\usepackage{amsmath}
\usepackage{amssymb,amsfonts}
\usepackage{comment}
\usepackage{graphicx}
\usepackage{latexsym}
\usepackage{slashed}
\usepackage{xcolor}
\let\normalcolor\relax

\usepackage{subfigure}
\usepackage{subfig}
\usepackage{simpler-wick}
\usepackage{epstopdf}

\relax
\renewcommand{\theequation}{\arabic{section}.\arabic{equation}}
\def\be{\begin{equation}}
\def\ee{\end{equation}}
\def\bs{\begin{subequations}}
\def\es{\end{subequations}}
\newcommand{\een}{\end{subequations}}
\newcommand{\ben}{\begin{subequations}}
\newcommand{\beq}{\begin{eqalignno}}
\newcommand{\eeq}{\end{eqalignno}}

\DeclareMathOperator\Tr{Tr}

\def\square{\large\hbox{{$\sqcup$}\llap{$\sqcap$}}}

\def\l{\lambda}
\def\m{\mu}
\def\n{\nu}
\def\a{\alpha}
\def\b{\beta}
\def\t{\tau}

\def\sp{\;\;\;,\;\;\;}
\def\r{\rho}

\def\e{\epsilon}

\newcommand\fverb{\setbox\pippobox=\hbox\bgroup\verb}
\newcommand\fverbdo{\egroup\medskip\noindent%
                        \fbox{\unhbox\pippobox}\ }
\newcommand\fverbit{\egroup\item[\fbox{\unhbox\pippobox}]}
\newbox\pippobox

\def\hri#1#2{\href{http://arxiv.org/abs/#1}{[ArXiv:#1]#2}}
\def\hre#1#2{\href{http://arxiv.org/abs/#2/#1}{[ArXiv:#1/#2]}}

\def\hrj#1#2{\href{https://doi.org/#1}{#2}}

\def\beq{\begin{equation}}
\def\eeq{\end{equation}}

\def\4R{{{}^{(4)}R}}

\def\K5{{\kappa}}
\def\K52{{\kappa^2}}

\newcommand{\half}{\frac{1}{2}}

\newcommand{\B}{B}

\def\bea{\begin{eqnarray}}
\def\eea{\end{eqnarray}}
\def\nn{\nonumber}
\newcommand{\ba}{\begin{aligned}}
\newcommand{\ea}{\end{aligned}}

\makeatletter
\def\@fpheader{\relax}
\makeatother

\title{Interacting systems and wormholes}
\author{P. Betzios$^{\diamondsuit}$, E. Kiritsis$^{\flat,\natural}$ and O. Papadoulaki$^*$\\
~\\
$\diamondsuit$ \href{https://phas.ubc.ca/}{Department of Physics and Astronomy} , University of British Columbia, \\
6224 Agricultural Road, Vancouver, B.C. V6T 1Z1, Canada
~\\
~\\
$^\flat$ \href{http://hep.physics.uoc.gr}{Crete Center for Theoretical Physics}, Institute for Theoretical and Computational Physics,
Department of Physics,  P.O. Box 2208,\\
University of Crete, 70013, Heraklion, Greece
~\\
~\\
$^\natural$ \href{http://www.apc.univ-paris7.fr}{APC, AstroParticule et Cosmologie}, Universit\'e Paris Diderot, CNRS/IN2P3, CEA/IRFU,
Observatoire de Paris, Sorbonne Paris Cit\'e,\\
 10, rue Alice Domon et L\'eonie Duquet, 75205 Paris
Cedex 13, France\\
~\\
$^*$ \href{https://perimeterinstitute.ca/}{Perimeter Institute for Theoretical Physics} , Waterloo, \\
 Ontario N2L 2Y5, Canada
}

\abstract{We consider a class of tripartite systems for which two $d$-dimensional QFTs are cross-coupled via a third $d+1$-dimensional ``messenger" QFT. We analyse in detail the example of a pair of one-dimensional matrix quantum mechanics, coupled via a two-dimensional theory of the BF-type and compute its partition function and simple correlators. This construction is extendible in higher dimensions, using a Chern-Simons ``messenger" theory.  In all such examples, the exact partition function acquires a form, speculated to correspond to systems dual to Euclidean wormholes and the cross correlators are sufficiently soft and consistent with analogous gravitational calculations. Another variant of the tripartite system is studied, where the messenger theory is described by a non-self-interacting (matrix)-field, reaching similar conclusions. While the Euclidean theories we consider are perfectly consistent, the two possible analytic continuations into Lorentzian signature (messenger vs. boundary QFT directions) of the tripartite models, reveal physical features and ``pathologies" resembling those of the expected Lorentzian gravitational backgrounds.}


\vspace{10cm}

\dedicated{Dedicated to the memory of our colleague, Theodore Tomaras}

\begin{document}

\section{Introduction}\label{introduction}

Euclidean wormholes comprise exotic types of gravitational solutions, that still challenge our physical
intuition and understanding of gravitational theories. The main unsettled issue is determining their precise role in the gravitational path integral  and therefore the physical consequences of their presence both from a theoretical as well as a phenomenological perspective\footnote{Although the issue of stability is an important one raised long time ago, and recently revisited in~\cite{Hertog:2018kbz,Marolf:2021kjc,Mahajan:2021maz}, we do expect a subset of solutions to be perturbatively stable as shown in~\cite{Marolf:2021kjc}, so one cannot dismiss their role altogether.}.

In the context of holography, their existence gives rise to further paradoxes the most notable being the \emph{factorisation} paradox~\cite{Maldacena:2004rf}. Simply put, while the presence of distinct asymptotic $AdS$ boundaries
on the one hand, indicates a collection of decoupled QFT's\footnote{It should be stressed that there is a distinction between \emph{Euclidean} and \emph{Lorentzian} wormholes, such as the Einstein-Rosen bridge, the latter known to arise when a pair of QFT's is entangled~\cite{Maldacena:2001kr}.} with factorised generating source functionals $Z_1(J_1) Z_2(J_2)$, on the other hand computations of correlation functions and other observables using the connected bulk geometry indicates a form of cross-coupling between the two QFT's.

A further conceptual issue stems from the existence of (naively) different asymptotic global symmetries and conserved currents on the two boundaries, that are nevertheless related by the presence of common bulk gauge fields and associated constraints (such as the Gauss' law)~\cite{worm}. Finally it has been argued that wormholes are at clash with the principle of cluster decomposition~\cite{ArkaniHamed:2007js}, but this problem is related to the $\a$-parameter and baby universe interpretation of microscopic wormholes \'a la Coleman~\cite{Coleman:1988cy,Lavrelashvili:1987jg}\footnote{An important earlier paper pinpointing various inconsistencies of the $\a$-parameter description of Euclidean asymptotically AdS wormholes is~\cite{Rey:1998yx}. We analyse further these issues in a forthcoming work~\cite{BKP}.}, and not with the type of macroscopic Euclidean multi-boundary saddles\footnote{Or approximate saddles such as constrained wormholes~\cite{Cotler:2020lxj}.}, whose physics we would like to reproduce and
understand from a dual field theory point of view.

There have been various proposals in the literature so far, regarding the correct holographic interpretation of such
geometries and the resolution of these puzzles. They roughly fall into the following categories:
\begin{itemize}

\item A possibility that has recently attracted a lot of attention~\cite{Saad:2019lba,Marolf:2020xie,Betzios:2020nry,Johnson:2020heh,Mertens:2020hbs,Blommaert:2020seb} is that the quantum gravity path integral is dual to an appropriate \emph{average over theories} (or over states~\cite{Freivogel:2021ivu}). This can be shown to be possible in some simple two-dimensional models such as JT gravity or minimal models (and some rather exotic examples in $3d$, see~\cite{Maloney:2020nni} and refs.). It has a merit in that it also conforms with expectations based on the eigenstate thermalization hypothesis (ETH), quantum ergodicity and the complicated/chaotic nature of gravity~\cite{Pollack:2020gfa,Belin:2020hea,Belin:2020jxr,Altland:2020ccq}. It brings however a certain tension with the extremely well-studied paradigms of AdS/CFT, involving single partner duals (such as $\mathcal{N}=4$ SYM), as well as with the principle of unitarity, and therefore seems to make sense in an approximate statistical sense (it is perhaps better to call such wormholes as \emph{statistical wormholes}).

    One can also formulate additional arguments in support of an inherent difference between higher-dimensional theories and these lower-dimensional examples, based on the Weak Gravity and Swampland conjectures~\cite{McNamara:2020uza} and the fact that the two-dimensional theories can also be interpreted as models of 2d quantum gravity on the world-sheet of a string - the matrices triangulate the 2d geometry (fat-graphs) and one  cannot describe strings propagating on \emph{target space} wormhole geometries with these models~\cite{Betzios:2020nry}.

\item Another possibility is that once all possible topologies and geometries are included in the quantum gravity path integral (perhaps with the inclusion of semi-wormholes~\cite{Saad:2021rcu,Garcia-Garcia:2021squ} or other non-perturbative states), the partition functions do factorise~\cite{Maldacena:2004rf} and the paradox is hence an \emph{artifact of an incomplete effective semi-classical} description in the bulk. This is very hard to actually check, and in simple models of two-dimensional quantum gravity where this computation can be performed exactly (such as in the $c=1$ model~\cite{Betzios:2020nry}) it does not seem to be the case, unless picking special members of the ensemble as argued in~\cite{Saad:2021uzi}. Although Berry's ``diagonal approximation" is used as an argument in favor of this idea~\cite{Saad:2019lba,Saad:2021uzi}, there is no comprehensive theory of periodic orbits for QFT's and,  in addition,  it is not even clear whether the diagonal terms have a geometric interpretation as wormhole saddles of gravitational equations in higher dimensions.

\item It has also been argued that an inherently \emph{string theoretic description automatically includes such geometries} (large stringy corrections around a factorised saddle might have an interpretation as a non factorised background). In a sense, the geometric connection could be effectively arising from a condensate of strings~\cite{Eberhardt:2021jvj}. Again,  it is not clear if the specialised discussion about the symmetric product orbifold CFT's generalises in more realistic examples,  having a simple semi-classical low-energy gravitational description,  with a small $\alpha'/L_{AdS}^2$ ratio\footnote{Moreover,  in the example of the symmetric product orbifold, it is natural not to fix $N$ but a quantity similar to a dual chemical potential $\mu$ for the number of strings. Nevertheless, something analogous could make sense in M-theory or in the free fermion grand-canonical description of ABJM~\cite{Marino:2011eh}, which also incorporates non-perturbative effects.}. As discussed in the second bullet point above, in more general setups one expects the presence of various kinds of non-perturbative effects and objects being important in the resolution of the factorisation paradox.

\item  A conceptually straightforward resolution is that  Euclidean wormhole saddles, can be found in models of appropriately \emph{interacting QFTs} as first suggested in~\cite{worm}. A generic feature of wormhole geometries, uncovered in \cite{worm}, is that they generically confine, via the usual Wilson loop criterion. This suggests that (cross) - confinement (and the possible formation of an IR gap) must be an integral part of the dual quantum system, fusing together the colored states of the two boundary QFTs (see also~\cite{VanRaamsdonk:2020tlr}). This leads to very tight constraints and special properties that correlators of such an interacting pair should satisfy.

A first difficulty with this proposal, is that the constraints for the correlators found in
~\cite{worm} seem to be impossible to realise, if one couples the two $d$-dimensional QFT's with local interactions. These constraints thus indicate an inherent difference with analogous constructions of Lorentzian traversable wormholes~\cite{Gao:2016bin,Maldacena:2018lmt,Bzowski:2020umc}, that rely on such direct local cross couplings. Another difficulty, is that one would wish to be able to describe both the factorised geometries and the non-factorised geometries with a single field theoretic construction, and not change the definition of the model depending on the background geometry (if the different backgrounds are solutions of the same bulk theory).

\end{itemize}

In this work we shall improve this last proposal~\cite{worm} and address these issues. We shall base our analysis on an observation by Raamsdonk~\cite{VanRaamsdonk:2020tlr}, that coupled pairs of $d$-dimensional QFT's via an intermediate $d+1$ ``messenger" theory can potentially exhibit the desired features of cross correlators for the pairs of the $d$-dimensional QFT's~\cite{worm}, once the messenger theory is integrated out in the path integral\footnote{Wormholes can be expected to appear once $c_{messenger} \ll c_{QFT}$, so that the bulk geometry effectively remains $d+1$ dimensional. We should also mention that this system also belongs to the general category of systems having coupled ``sectors" via messenger fields, analysed in~\cite{Betzios:2020sro}.}. We shall henceforth define the system of the pair of QFT's and the messenger theory  as the \emph{tripartite system}. As for the second difficulty raised above, our proposed resolution is that the dual QFT tripartite system should exhibit different leading saddles at large-$N$, depending on the parameters and sources turned on, some of which correspond to dual wormhole backgrounds and others to factorised geometries. In fact this is something that has already been observed in the literature (from a gravitational perspective), for example in~\cite{Marolf:2021kjc}\footnote{From the examples studied in that work, there even exist cases where the connected solutions are perturbatively stable and can dominate the disconnected ones. One the other hand, when such solutions could be embeddable in UV complete settings, it was found that they do suffer from non-perturbative brane nucleation instabilities, implying the existence of an additional branch of solutions with lower action (that has not been constructed yet).}.

In addition, we shall prove that in cases where the intermediate theory has a topological nature without any local $(d+1)$-dimensional propagating degrees of freedom, the partition function of the combined system acquires a special form, whose relevance for a holographic description of two-boundary Euclidean wormhole geometries first appeared in the work by Maldacena and Maoz~\cite{Maldacena:2004rf}. More precisely we find that the general form for the Schwinger source functional of such systems is expressed as a weighted average over individual source functionals
\be\label{maineqn}
Z(J_1, J_2) = \sum_S e^{w(S)} Z^1_S (J_1) Z^2_S(J_2) \, ,
\ee
where $S$ labels some appropriate ``sector" of the system\footnote{Notice that this property is reminiscent to the holomorphic factorisation property of WZW models~\cite{Witten:1991mm}.}, and $w(S)$ introduces a weighting in the space of ``sectors" (the authors of~\cite{Maldacena:2004rf} had not introduced a weighting factor). 

It is interesting to note that if the messenger theory were classical, then unique boundary conditions for the messenger fields would lead to a unique saddle for the system. We are precisely in that case as our messenger theory is topological.
In this case the sum given by eqn.~\eqref{maineqn} resembles the sum over instanton/superselection sectors. 
However, even if the messenger theory has more than one saddle points, one may of may not obtain a sum like the one we obtained. Whether this happens or not, depends on whether the saddle points correlate with quantum aspects of the two boundary theories. For example a possible generalisation of eqn.\eqref{maineqn},  would involve an additional summation or integration for each fixed individual sector $S$.

In (\ref{maineqn}), such a decomposition of the source functional also guarantees that there are no short distance divergences for the two-point cross correlators (as expected from the gravitational computations~\cite{worm}). In particular one finds that
\be
\langle \mathcal{O}_1(x_1) \, \mathcal{O}_2(x_2) \rangle_c = \frac{1}{Z^2} \sum_S e^{w(S)} \langle \mathcal{O}_1 (x_1) \rangle^1_S \langle  \mathcal{O}_2 (x_2) \rangle^2_S \, .
\ee
It is therefore impossible for short distance singularities as $x_1 \rightarrow x_2$ to develop, if the individual one-point functions are well defined. In addition, depending on whether the states of the ``sector" $S$ preserve translational invariance, the one-point functions and the two-point cross correlator might simply be constants.

This structure persists to higher-point functions, since singularities can only develop when the operators whose points collide belong to the same $d$-dimensional theory/part of the tripartite model, as expected from the holographically dual computation on a two-sided Euclidean wormhole geometry. Some other appealing features of our construction in terms of a $(d+1)$-dimensional (quasi)-topological theory coupled to $d$-dimensional QFT's, is that it can also be naturally generalised in the case of multiple asymptotic boundaries. It is also possible to interpret the ``gluing" of the two boundary partition functions as arising from a form of topological entanglement~\cite{Balasubramanian:2016sro,Balasubramanian:2018por}\footnote{To avoid confusion, we should again emphasize that in our case,  the ($d+1$)-dimensional topological theory is part of the \emph{boundary} and not the gravitational bulk description.}. We briefly touch upon the resulting form of the multi-boundary partition function, that replaces eqn. \eqref{maineqn} in the conclusions~\ref{conclusions}.

The example we analyse in great detail in section~\ref{D0D1}, is a model of two matrix quantum mechanics (MQM's) coupled via a 2d theory of the BF-type. For this model, we shall explicitly describe what the sector $S$ in eqn. \eqref{maineqn} means. In particular, one has to sum over different $U(N)$ representations for the two MQM models, that are found to be ``tied" by common selection rules (and not by a direct interaction term in their Hamiltonian that would inevitably lead to short distance singularities in the cross correlators). This provides the softest possible  ``cross communication"  between the two MQM models.
The Hilbert space of the system is found to take the form of a direct sum of tensor products $\mathcal{H}_s = \sum_R \mathcal{H}^1_{R} \otimes  \mathcal{H}^2_{R} $, with $R$ a $U(N)$ representation. We believe that this is an important aspect that distinguishing the duals of Euclidean wormholes with those of the well understood Lorentzian black holes (Einstein-Rosen bridge), which can be described by the simple tensor product Hilbert space of the two boundary CFT's (and whose Euclidean continuation factorises into the product of two cigar geometries).

In addition, in a large representation limit that we describe in sections~\ref{HallLittlewood} and~\ref{LargeNn}, one can show the presence of competing saddles, some of which could correspond to connected and others to disconnected bulk geometries (each individual singlet MQM with an inverted oscillator potential is known to describe $c=1$-Liouville string theory on a linear dilaton background)\footnote{It might also be interesting to analyse the case of the \emph{usual oscillator potential} describing $1/2$-BPS states in $\mathcal{N}=4$ SYM and the corresponding LLM geometries~\cite{Berenstein:2004kk,Lin:2004nb,Berenstein:2017abm}.}. Unfortunately,  not much is known about the geometric interpretation of the non-singlet sector of MQM (see though~\cite{Maldacena:2005hi,Gaiotto:2005gd,Aniceto:2006rr,Betzios:2017yms} for some preliminary steps in this direction) and hence we cannot completely settle this question in the affirmative at the moment. In section~\ref{MQMcorrelators} we analyse some simple two- and four-point cross-correlators and demonstrate their expected properties.

In section~\ref{higherdim} we describe how such tripartite models can be naturally generalised in higher dimensions and describe in some detail the specific case where two BCFT's are cross coupled through a higher-dimensional Chern-Simons ``messenger" theory. The sectors in this case correspond to labels of the associated Chern-Simons wavefunctions, that are also related to group characters and representations.

In section~\ref{crosscoupledsimple} we analyse an additional class of weakly cross-coupled models, whose correlators again exhibit the desired features, in terms of a free ($d+1$)-dimensional messenger theory coupled with two $d$-dimensional theories at the endpoints of an interval.
The couplings are  simple, such as: $\Tr \phi_1 \Phi_m + \Tr \phi_2 \Phi_m$ with $\phi_{1,2}$ (matrix) fields of the boundary theories and $\Phi_m$ a (matrix) field of the messenger theory. We find that the Euclidean system is perfectly well-defined, but its two natural analytic continuations in Lorentzian signature (along the additional messenger or boundary dimensions), sometimes lead to the existence of a non-positive spectral weight and/or the presence of tachyonic instabilities. In the case of analytic continuation along the messenger dimension, the two boundary QFT's remain Euclidean (but cross-interacting), a setup that provides an interesting twist to the dS/CFT proposal~\cite{Hull:1998vg,Strominger:2001pn,Witten:2001kn,Maldacena:2002vr,Anninos:2011ui}.

It is natural then to ponder whether the peculiarities of the Lorentzian tripartite systems we study are in correspondence with the physical properties of gravitational backgrounds such as Big-bang/Big-crunch type of universes, or Lorentzian traversable wormholes, which as gravitational solutions demand a violation of the bulk energy conditions\footnote{This is in contrast with \emph{Euclidean} signature wormholes, for which negative ``energy" can be easily obtained in the presence of additional non-trivial bulk fields and fluxes~\cite{worm}.}. We are still missing a crisp - UV complete higher dimensional example, where we thoroughly understand the gravitational dual of such tripartite systems, but we initiate an analysis of self-interacting tripartite models (that could perhaps be rendered UV complete with some adjustments), in sections~\ref{boundaryselfinteractions} and~\ref{freefieldmatrix}.

We also provide several appendices with further examples and details on our calculations. Most importantly appendix~\ref{miscellaneousmatrix} contains various novel examples of $0d - 1d$ models, where two matrix models are coupled via a messenger MQM. In addition in appendix~\ref{orbifold}, we show that the seemingly unrelated microscopic model of MQM on the $S_1/\mathbb{Z}_2$ orbifold, studied in~\cite{Betzios:2016lne} as a model of a two-dimensional bang-crunch universe, can also be rewritten as a model of the general type studied in this paper. This is an additional example, that might give further credence to the idea that Euclidean wormholes and Lorentzian bang-crunch types of geometries are intimately related and can arise in our setup~\cite{Maldacena:2004rf,Betzios:2016lne,Betzios:2017krj,worm,VanRaamsdonk:2020tlr}.

Finally in section~\ref{conclusions}, we summarise and conclude with the main lessons of this work, as well as list some interesting avenues for future investigations.

\section{Matrix models}\label{Matrixmodels}

We may consider various systems of cross-coupled matrix models or matrix field theories. The lowest-dimensional cases of $0d$ matrix models coupled to $1d$ matrix quantum mechanics (MQM), are discussed and analysed in appendix~\ref{miscellaneousmatrix}. Even though we find that such models are interesting on their own right, we would like to have a less trivial example with a non-trivial space-time dependence. Therefore,  in the main text we shall focus on a concrete case of cross coupled $1d$ matrix quantum mechanics through an intermediary two-dimensional theory of the Yang-Mills or BF type (quasi-topological)\footnote{In section~\ref{higherdim} we shall generalise the idea of having a ``messenger" intermediate (quasi)-topological theory to higher dimensions.}. Such a model has the advantage of being both amenable to a detailed analysis, as well as presenting interesting characteristics that are shared by higher-dimensional examples such as those we discuss in section~\ref{higherdim}.

\subsection{MQM coupled to generalised YM ($D_0/D_1$ system)}\label{D0D1}

We shall now analyse a tripartite system of two one-dimensional MQM models coupled via a generalised YM-theory (gYM). The simplest possibility is that this gYM theory is only indirectly coupled to the MQM's through the asymptotic value of the gauge field $A$. Using a form notation, the gYM part of the action is \cite{Douglas:1994pq,Blau:1993hj} ($A$ is the one form gauge field, $F = d A +   A \wedge  A = d_A A$ and $d \mu$ is the volume form)
\be\label{gYMaction}
S_{gYM} = \frac{1}{g_{YM}^2} \int_\Sigma  \Tr B  F \, + \, \frac{\theta}{g_{YM}^2} \int_\Sigma \Tr B  \, d \mu - \frac{1}{2 g^2_{YM}} \int_\Sigma \Tr \Phi(B) \, d \mu \, .
\ee
In this action $B$ is an auxiliary zero form, and $\Phi(B)$ is a potential that specifies the gYM theory we are considering\footnote{Unitarity imposes that this potential should be bounded from below.}. We have also explicitly indicated the possibility of introducing a $\theta$-angle via the second term.
Considering a concrete example, if we set $\Phi(B)=B^2$ and use its equations of motion, we recover the usual 2d YM theory. Another special potential is $\Phi(B)=0$, that results in a purely topological theory (pure BF model). An additional possibility is that the field $B$ is a compact field, in which case one finds a q-deformed version of the gYM theory.

The equations of motion of \eqref{gYMaction} are ($\star$ is the Hodge dual)
\be
d_{A}  B= d B + A \wedge B = 0\,,\quad \star F =    \frac{\delta \Phi(B)}{\delta B} - \theta  \, .
\ee
The variational principle yields also a boundary term that is
\be
\delta S_{\partial \Sigma} = \frac{1}{g_{YM}^2} \int_{\partial \Sigma} \Tr B \delta A \, .
\ee
This variational principle is well defined in the case of Dirichlet boundary conditions for which $\delta A = 0$ at the boundary/ies. It is also possible to add a boundary term
\be - \frac{1}{g_{YM}^2} \int_{\partial \Sigma} \Tr B  A = S_{\partial \Sigma}\;,\ee
 if we wish to fix $\delta B = 0$ at the boundaries (``electric" vs. ``magnetic" type of boundary conditions).

In the rest, when using explicit indices instead of abstract forms, we shall denote the two dimensions by $(0,1) = (\tau, z)$. We introduce two boundaries at $z=\pm L$ and identify
$$A_\tau (\tau, z=L) = A^1_\tau (\tau)\sp A_\tau (\tau, z=-L) = A^2_\tau (\tau)$$
 as the two asymptotic gauge fields living along the one-dimensional boundaries. From now on we set $\Phi(B) = B^2$ and focus on 2d YM for concreteness. It is a simple exercise to generalise the discussion to any of the gYM theories.

We now enrich further the dynamics of the system by introducing additional one-dimensional dynamical fields on each boundary. In particular, one can introduce a $U(N)$-gauged matrix quantum mechanics (MQM) model of (Euclidean) time dependent Hermitian matrices $M_{1,2}(\tau)$ on each boundary, through the action
\be\label{MQMaction}
S_{MQM_{1,2}} = \int d \tau \Tr \left( \half  (D_\tau M_{1,2})^2  \, - V(M_{1,2}) \right) \, , \quad D_\tau M_{1,2} =  \partial_\tau M_{1,2} + i [A^{1,2}_\tau , M_{1,2}] \, .
\ee
An important point here is that both MQM's transform under the same $U(N)$ group due to their coupling to the asymptotic values of a common two dimensional gauge field.  We can relax this condition, by introducing other types of (bi-fundamental) matter fields and such an extended example will be analysed in the next section~\ref{bifundmodel}.

For each boundary MQM we can diagonalise the matrix with time dependent rotations
\be
M_{1,2} = U_{1,2} \Lambda_{1,2} {U_{1,2}}^\dagger \sp  J_{MQM_{1,2}} = \frac{\delta S_{MQM_{1,2}}}{\delta A^{1,2}_\t} = U_{1,2} K_{1,2} {U_{1,2}}^\dagger\;.
\ee
Dropping the $1,2$ boundary superscripts for compactness, the Hamiltonian for the eigenvalues of each MQM, is expressed as
\be\label{HamiltonianR}
\hat{H}_{MQM} = \left[ - \half \sum_i \left(\frac{\partial^2}{\partial \lambda_i^2} + V(\lambda_i) \right)+ \half \sum_{i<j} \frac{K_{i j}^R K_{i j}^R}{(\lambda_i - \lambda_j)^2} \right]  \, ,
\ee
and acts on wave-functions $\Psi(\lambda) = \prod_{i < j}(\lambda_i - \lambda_j) \tilde{\Psi}(\lambda)$ transforming in the $U(N)$ representation\footnote{More precisely, one actually needs to consider $SU(N)$ representations, since the trace of the gauge field does not appear in the action. On top of that, one has to project to
representations admitting zero weight states (see~\cite{Gross:1990md,Klebanov:1991qa,Boulatov:1991xz} and appendix~\ref{partitions}), since the diagonal part of the gauge field matrix (Cartan generators) does not couple to the eigenvalues $\lambda_i$. The complete $U(N)$ can become relevant if we introduce a 1d ``Chern-Simons" term $k \int d \tau \Tr A_\tau $~\cite{Betzios:2017yms}. We shall explain when this is possible in the next section~\ref{bifundmodel}.} $R$. For a single gauged MQM model on an infinite domain $\tau$, the non-dynamical gauge field $A_\tau$ can be set to zero, but one has to impose its equations of motion as a (Gauss's law) constraint i.e.
\be
J_{MQM} = K = 0 \, .
\ee
This would project each individual MQM into the singlet sector, but now the two MQM models are coupled through the gauge field of the 2d YM theory, so that we expect non-trivial representations to contribute.

In order to understand in more detail the induced cross correlations between the two MQM's, one can integrate out the 2d YM for an effective action coupling the two MQM's and we would like to understand the precise form of the induced coupling. We shall treat the cases of compact and non-compact $\tau$ independently, the second case corresponding to the partition function of the system of MQM's (cylinder topology), where $A_\tau$ has non trivial winding modes.

\subsubsection{Non compact $\tau$}\label{noncompacttau}

In the case that $\tau$ is of infinite extend, there exist two possible interpretations of the path integral, and quantisations of the 2d Yang-Mills theory. In the first case one can interpret $\tau$  as the Euclidean ``time", that is also shared by the two boundary MQM's. The other option is to interpret $z$ as the Euclidean ``time", the path integral now being a transition amplitude for the 2d Yang-Mills , where the two boundary states at $z = \pm L$ are coupled to the two MQM's. If we pass to a Hamiltonian formulation of the Yang-Mills theory, we find that $A_0$ (that can be either $A_\tau$ or $A_z$ respectively) acts as a Lagrange multiplier enforcing the Gauss-law constraint. This means that we can set $A_\tau = 0$ or $A_z = 0$ (no winding modes), but we should then impose the corresponding equation as a constraint acting on all the physical states. Both gauges preserve the boundary symmetries and make manifest different physical aspects of the system.

Let us first consider the gauge $A_z = 0$.
In this case one finds $F_{\tau z} = - \partial_z A_\tau$, as the only non-vanishing component of the field strength. The equations of motion read
\be\label{EOMzgauge}
\frac{1}{g_{YM}^2}\partial_z F^{z \tau} = \delta(z-L) J_{MQM_{1}}^\tau(\tau) + \delta(z+L) J_{MQM_{2}}^\tau(\tau)  \, , \qquad \partial_\tau F^{\tau z} + i [A_\tau, F^{\tau z}] = 0 \, ,
\ee
with  the MQM currents on the boundaries
\be\label{MQMcurrents}
J^{MQM_{1,2}}_\tau = \frac{\delta S_{MQM_{1,2}}}{\delta A^{1,2}_\tau} =   i [M_{1,2}, D_\tau M_{1,2}] \, ,
\ee
playing the role of sources that are defined only on the boundary surfaces $z = -L,L$ where the two MQM models reside.
The second equation of \eqref{EOMzgauge} is the constraint equation and can also be written as
\be\label{PB2}
\partial_\tau \partial_z A_\t  + i [A_\t , \, \partial_z A_\t ] = 0  \, .
\ee
The homogeneous part of the first equation of \eqref{EOMzgauge} is readily solved
\be\label{PB1}
\partial_z^2 A^\tau = 0 \quad \Rightarrow \quad A_\tau(\tau, z) = f(\tau) + z g(\tau) \, ,
\ee
with $f,g$ Hermitean matrices.
Setting $z = \pm L$, we can identify $2 f(\tau) = A^1_\tau (\tau) + A^2_\tau (\tau) $ and $2 L g(\tau) = A^1_\tau (\tau) - A^2_\tau (\tau)$. At this point we should mention that the axial-gauge is not a complete gauge and there is further residual gauge invariance~\cite{Frishman:2010tc}. In particular one can perform further gauge transformations that depend only on $\tau$, to set $f(\tau) = 0\,  \Rightarrow  \, A^1_\tau (\tau) = -  A^2_\tau (\tau) $. On the other hand as shown in appendix~\ref{Decouplingghosts} the ghosts do decouple from the path integral in such axial gauge choices.

To complete the classical analysis, one needs to solve a Green's function problem with the two boundary current sources\footnote{It is actually more similar to a propagator in the presence of two branes at $z = \pm L$.}. In particular, we demand continuity at $z= \pm L$ and that the derivative jumps with a jump proportional to the sources $J_{MQM_{1,2}}(\tau)$. With these conditions, one finds\footnote{We can introduce an infinitesimal regulator $\epsilon$ in order to define two additional small regions near the boundaries.}
\be\label{cases}
\partial_z A_\tau = g(\tau) +  g_{YM}^2 \Theta(z-L+ \epsilon)J_{MQM_1}(\tau) + g_{YM}^2 \Theta (z+L - \epsilon) J_{MQM_2}(\tau)  \, , \quad z \in [-L, L] \, ,
\ee
with $\Theta(x)$ the step function.
This solution for the gauge field has the correct behaviour, being continuous and the normal jump of its derivative inducing the boundary current sources. We then analyse the constraint equation \eqref{PB2}, in the three regions defined by the split $z \in [-L , -L + \epsilon) \cup (-L + \epsilon, L - \epsilon) \cup (L - \epsilon, L]$ to find
\be\label{sln1}
\partial_\tau g = 0 \, , \quad \partial_\t J_{MQM_{1,2}} \pm i L [g , \, J_{MQM_{1,2}}] = D_\tau^{1,2} J_{MQM_{1,2}}  = 0 \, .
\ee
In the last equation the $+$ sign is for $J_{MQM_{1}}$ and the $-$ sign is for $J_{MQM_{2}}$ and the result is simply the covariant conservation law of the two MQM currents. The solution of eqns. \eqref{sln1} is
\be\label{sln2}
g = C  \, , \qquad J_{MQM_{1,2}}(\tau) = e^{ \pm i \tau L C} J_{1,2}(0) e^{ \mp i \tau L C}    \, ,
\ee
in terms of an arbitrary constant matrix $C$.

We conclude  that the complete set of EOM's in the axial gauge leads to
\be
f=0  \Rightarrow A^{1}_\tau = - A^{2}_\tau = L C  \;,
\ee
 so that
the conservation laws for the two MQM's contain the same constant gauge field, $C$,  (up to a sign). Since $\tau$ is non-compact, there is no classical finite (YM) action configuration with a classical constant non-zero gauge field~\cite{Frishman:2010tc}, due to the divergence
\be
 S_{YM} \sim \int d \tau \Tr C^2\;.
  \ee
  Similar to the case of a single gauged MQM, one can then set the zero mode $C=0$, but needs to impose as a constraint $\delta S_{total}/\delta C = 0$. This gives the condition $J_{total} = J_{MQM_{1}} - J_{MQM_{2}} = 0$, or in other words a constraint between the allowed representations for the two MQM models\footnote{Including 1d C.S. terms on the boundaries $k_{1,2} \int d \tau \Tr A^{1,2}$, would lead to $J_{total} = (k_1 - k_2) I$ with $I$ the identity matrix and $k_{1,2}$ the C.S. levels. The theory is non anomalous only when $k_1 = k_2$, since the MQM currents \eqref{MQMcurrents} are traceless.}. If we wish to derive this result using the path integral formalism, we can compute the gauge fixed path integral for the 2d YM coupled to sources (up to an overall normalisation)
\be
Z \sim \int d   C ~e^{ - \frac{1}{4 g_{YM}^2} L_\tau \Tr C^2 + L L_\tau \Tr C (J_{1}(0) - J_2(0))  } \, .
\ee
In order to regulate this expression, we assumed that the $\tau$ direction has  a finite extend $L_\tau$. Upon performing the integral, we find
\be
Z \sim e^{- g^2_{YM} L^2 L_\tau \Tr (J_1 - J_2)^2} (\pi L_\tau )^{N^2/2} \, .
\ee
Taking the limit $L_\tau \rightarrow \infty$ we recover a matrix version of the delta function as expected
\be\label{constraint1}
\lim_{L_\t \rightarrow \infty} Z \sim \delta^{N^2} (J_1 - J_2) \, .
\ee
This is in contrast with the case where the 2d theory would have dynamical degrees of freedom, in which we generally expect an induced direct cross interaction of the $\Tr (J_1 J_2)$ type.

One can also find a complementary (but more technically involved) analysis of the model in the $A_\tau = 0$ gauge in appendix~\ref{alternategauge}. The upshot is equation \eqref{currentsmatch}
\be\label{currentsmatchmain}
\frac{1}{ 2 g_{YM}^2   L} J_{YM}^\tau = \frac{1}{ 2 g_{YM}^2   L} [W^{-1} , \, \partial^\t W] =   J_{MQM_{1}}^\tau -   J_{MQM_{2}}^\tau  \,  , \quad W = \mathcal{P} \exp \left( \int_{-L}^L dz  A_z \right) \, ,
\ee
that holds non-perturbatively as a constraint and generalises the condition $J_{MQM_{1}} - J_{MQM_{2}} = 0$, beyond the analysis of the $A_z = 0$ gauge. In this formula, by $W$ we denote an open Wilson line, with end-points at $z = \pm L$.
In this case there is a non-trivial YM current $ J_{YM}^\tau$, related to the two MQM currents/charges by the constraint. Again this results in selection rules for the allowed representations, and the Hilbert space of the system, in the presence of open Wilson lines stretching across the two boundaries.

We shall now move forward to the case when $\tau$ is compact (an $S^1$), so that physical zero modes of $A_\tau$ become relevant.

\subsubsection{Compact $\tau$ (circle)}\label{compacttau}

In this case, we can only work in the $A_z = 0$ gauge, since $A_\tau$ has physical zero modes. We consider a compact $\tau$ direction with size $\beta$, ($\tau \sim \tau + \beta$), so that fields are periodic and the gauge field has a zero mode. One can now gauge fix in eqn.\eqref{PB1} $f = C_f$ to be a constant matrix. Equations \eqref{sln1} and \eqref{sln2} are now replaced by
\bea
g = C_g \, , \quad \partial_\t J_{MQM_{1,2}} + i [ C_f \pm  L C_g , \, J_{MQM_{1,2}}] = 0 \, . \nn \\
 J_{MQM_{1,2}}(\tau) = e^{ i \tau ( C_f  \pm  L C_g)} C_{1,2} e^{- i \tau ( C_f  \pm  L C_g)} \, .
\eea
This analysis makes clear that one does not have a unique common zero mode since $A_\tau = C_f + z C_g$, hence there are two independent holonomies/zero modes at the ends of the cylinder ($z = \pm L$). A further discussion of Wilson loops and the Hamiltonian $\hat{H}_z^{YM}$ of 2d YM in the case of compact $\tau$ can be found in appendix~\ref{compacttauwilson}.

\subsubsection{The partition function}\label{thermalpf}

If we wish to study the thermal partition function of the combined system, then $\tau$ is compactified into an $S^1$, so that the $2d$ geometry is a cylinder (in this section we take the two boundary $S^1$'s to have equal size $\beta$, in section~\ref{HallLittlewood} we lift this restriction). In this case, we have a $\tau$-periodic gauge field with two independent  non-trivial holonomies (winding modes) on the boundary $S^1$'s.  The path integral for 2d YM on the cylinder with area $A = \beta L$ is expressed in terms of the two boundary holonomies as a sum over $U(N)$ representations (denoted by $R$)~\cite{Rusakov:1990rs}
\be\label{2dYMpf}
Z^{YM}_{cyl}(U_1,U_2 ) = \sum_R  \chi_R(U_1) \chi_R(U_2)e^{- L \frac{g_{YM}^2}{N}  C_R^{(2)} + i \theta  C_R^{(1)}}  \, .
\ee
In this formula $L$ is the cylinder length and $\chi_R(U)$ are $U(N)$ characters. The result for the path integral depends on the holonomy/zero modes of the gauge fields on the boundaries which can be expressed as
 \be
 U_{1,2} =  \exp i \beta A_\tau(z= \pm L) = \exp i \beta A^{1,2}_\tau\;,
  \ee
  and is also interpreted as a propagator/heat kernel (along $z$), with the quadratic Casimir $C_R^{(2)}$ playing the role of the Hamiltonian and boundary states defined by the representation characters. For more details on this interpretation, see appendix~\ref{compacttauwilson}. One should then couple this expression to the two MQM thermal partition functions through the boundary holonomies. In particular the matrices of MQM are periodic up to a twist
\be
M_{1,2}(\beta) = U_{1,2} M_{1,2}(0) U_{1,2}^\dagger
 \ee
 dictated by the gauge field zero mode on each boundary. If the MQM model is Gaussian (as for $c=1$ MQM), one can integrate $M^{1,2}$ out and derive an explicit result for the twisted thermal partition function that is~\cite{Boulatov:1991xz,Betzios:2016lne,Betzios:2017yms}
\be\label{twistedMQM}
Z^{MQM}_{1,2} (U_{1,2})  =  \int \mathcal{D} M_{1,2} \, \langle M_{1,2}(0) |  U_{1,2} M_{1,2}(0) U_{1,2}^\dagger \rangle \, =
\ee
$$
= \det \frac{1}{q_{1,2}^{-1/2} U_{1,2} \otimes I  - q_{1,2}^{ \half} I \otimes  U_{1,2} } \, ,
$$
with
\be
  q_{1,2} = e^{- \omega_{1,2} \beta} \qquad \,
\ee
In these formulae, one should set $\omega = i$ for the inverted oscillator relevant for the $c=1$ MQM dual to Liouville theory and
\be
\langle M_{1,2}(0) |  U_{1,2} M_{1,2}(0) U_{1,2}^\dagger \rangle\ee
 is then the transition amplitude of each inverted oscillator with the twisted periodic bc's\footnote{It is also possible to pass to an eigenvalue basis where $U_{i j} = e^{i \theta_i} \delta_{i j}$, see eqn.~\eqref{inner}.}.

The total partition function of the system is
\bea\label{pf1}
Z(\beta) &=& \sum_{R} \int D U_1 \int DU_2 \, \chi_R (U_1) Z^{MQM}_1 (U_1)  \chi_R (U_2) Z^{MQM}_2 (U_2) e^{- L \frac{g_{YM}^2}{N}  C_R^{(2)} + i \theta  C_R^{(1)}}  \, \nn \\
&=&  \sum_{R}  e^{- L \frac{g_{YM}^2}{N}  C_R^{(2)} + i \theta  C_R^{(1)}} Z_{R}^1(\beta) Z_{R}^2(\beta)  \, .
\eea
In this expression $Z^{MQM}_{1,2} (U)$ corresponds to the twisted MQM thermal partition for each matrix model, and
\be
Z^{1,2}_R(\beta) = \int D U \chi_R(U) Z_{1,2}^{MQM}(U) \, ,
 \ee
corresponds to the finite temperature MQM partition function in the $U(N)$ representation $R$. In particular one can also write it as a trace over the fixed $R$-representation Hilbert space $Z_R(\beta) = \Tr_R e^{- \beta \hat{H}_R}   $ with the Hamiltonian given by eqn.~\eqref{HamiltonianR}. It admits an explicit description in terms of the highest weights of the representation $R$, see  eqn.\eqref{MQMrep} of appendix~\ref{characterexpansion}. The result for the partition function implicitly depends on $N$, since the reps contributing to the sum are $U(N)$ reps, whose highest weights correspond to Young diagrams having a maximum number of $N$ non-empty rows. To be more precice, as described below eqn. \eqref{HamiltonianR} and in~\cite{Gross:1990md,Klebanov:1991qa,Boulatov:1991xz}, actually only zero weight states of $SU(N)$ reps contribute to the expression~\ref{pf1}, unless one introduces one-dimensional Chern-Simons terms, see section~\ref{bifundmodel} for this possibility.

We therefore find that the two MQM partition functions are coupled by having common representations appearing, that are additionally weighted by the Casimirs of $U(N)$. For the 2d YM model this is the quadratic Casimir $C^{(2)}_R$, while for gYM theories one also finds higher Casimirs~\cite{Frishman:2010tc}. As mentioned in the introduction, this form for the partition function is expected to be relevant in the path integral description of systems dual to Euclidean wormholes~\cite{Maldacena:2004rf} and has recently being discussed in~\cite{VanRaamsdonk:2020tlr}\footnote{This structure of the model is also reminiscent of~\cite{Dijkgraaf:2005bp}.}.

The physical interpretation is the following:
We have two MQMs in arbitrary reps without explicit local cross-interaction terms in their Hamiltonians. Nevertheless, they are effectively coupled by ``weighted selection rules" in their allowed states due to a constraint eqn. \eqref{currentsmatchmain}, arising from the ``messenger" quasi-topological $2d$ YM theory. This is affecting the combined partition function, so that it does not factorise. On the other hand, there is factorisation for each individual representation $R$ reminiscent to the holomorphic factorisation property of WZW models~\cite{Witten:1991mm}. The effective cross-interaction is therefore as indirect as possible, without the presence of an explicit local interaction term in the Hamiltonian of the combined system. Additionally, the role of the Casimirs $C_R^{(n)}$ of $U(N)$, is to provide a weighting in the averaging procedure over different representations\footnote{This is reminiscent to the idea of wormholes being related to averaged systems, but is now implemented in a unitary quantum mechanical setup.}. It is clear that one could also define similar models in higher dimensions, and we shall describe some examples in section~\ref{higherdim}.

The next step is to try to find out what is the leading representation appearing in the partition function (saddle point). This can be achieved by considering the large-N limit, and discussing large representations (continuous limit of the Tableaux). We expect the different leading saddles to dictate the form of the dual geometric background. In particular for a single MQM, this is the singlet representation dual to the linear dilaton background of $c=1$ Liouville theory\footnote{Higher representations exhibit a logarithmically large gap in the string coupling with respect to the singlet~\cite{Gross:1990md}. A single gauged MQM model also has all the non-singlet representations projected out of its spectrum by construction.}. Non singlets are related to long-strings and different backgrounds such as black holes~\cite{Maldacena:2005hi,Gaiotto:2005gd,Aniceto:2006rr,Betzios:2017yms}, but unfortunately there exist several missing gaps in this extended dictionary. A straightforward analysis of the large representation limit using fusion of $U(N)$ characters and Littlewood-Richardson coefficients is presented in appendix~\ref{characterexpansion}. Due to mathematical difficulties in interpreting the large-$N$ limit of the Littlewood-Richardson coefficients, in section~\ref{HallLittlewood} of the main text, we provide an analysis using Hall-Littlewood polynomials, that proved to be more tractable.

\subsection{Liberating messenger and MQM ranks via bi-fundamental fields}\label{bifundmodel}

We could also consider the case of coupling the two gauged MQM theories to the 2d gauge field using bi-fundamental fields, so that the (quasi) topological messenger and MQM gauge groups are independent. The motivation for studying this more elaborate construction is that one has at hand an additional tunable parameter (the ratio of ranks) that proved to be useful in the models of~\cite{Bachas:2017rch,VanRaamsdonk:2020tlr}. We shall then take the two sets of bi-fundamental fields to be charged both under the messenger $U(n)$ gauge group, and the $U(N_{1,2})$ gauge groups of each MQM. The simplest case to analyse is when the bifundamentals only couple to the associated gauge fields and not to $M^{1,2}$ (it offers also the least direct interaction between the two MQM theories). Another posibility is that the bifundamentals live on the full $2d$ geometry, as in models of 2d QCD~\cite{Gross:1993hu,Gross:1993yt} coupled to two MQMs, where the flavor group is gauged with respect to the boundary MQM gauge field. Since this situation is more complicated to analyse and introduces further propagating degrees of freedom, we shall focus on the simplest case.

The action on the first boundary is comprised out of a $U(N_1)$ gauged MQM action as in eqn. \eqref{MQMaction} together with the bifundamental field action
\be\label{bifundaction1}
S_{bif.}^1 =  \sum_{\a=1}^{n} \sum_{i = 1}^{N_1} \int d \tau   \psi_{\a i}^\dagger \left( \delta_{i j} \delta_{\a \b} (\partial_\tau + m_{1}) - i q_B A^{gYM}_{\a \b}(z=L) \delta_{i j} - i q_{bd_1} A^{bd_1}_{i j} \delta_{\a \b}  \right) \psi_{\b j}  \, .
\ee
In this equation $i,j$ are $U(N_1)$ indices, while $\a , \beta$ are $U(n)$ indices.
We also introduce a similar action for the second boundary and the BF-type action eqn. \eqref{gYMaction} with the $U(n)$ gauge field $A^{gYM}_{\a \b}(z, \tau)$ living in the 2d geometry. Notice that Latin indices shall be reserved for the $U(N_{1,2})$ gauge groups of the two MQM's, while Greek indices for the $U(n)$ gauge group and fields. The bifundamental fields can be either bosonic or fermionic\footnote{In one dimension it is possible to have a first derivative action for a bosonic field.}. In the rest we set unit charges in order not to carry many parameters in the expressions that follow.

The last type of terms we shall consider are the 1d Chern-Simons terms of the form $S_{C.S.} = k_{1,2} \int d \tau  \Tr  A^{bd_{1,2}} $. By tuning them, one can enrich the class of admissible representations for the combined model, beyond those that admit states with zero weight. In order to make this more explicit, we have now three constraints analogous to eqn.~\eqref{currentsmatchmain}, associated to the three different gauge groups
\bea\label{currentsmatchmainb}
\frac{1}{ 2 g_{YM}^2   L} \left(J_{YM}^\tau\right)_{\a \b} =    \sum_{i = 1}^{N_1} \psi_{\a i}^{\dagger \, (1)}  \psi_{\b i}^{ \, (1)}  - \sum_{j = 1}^{N_2} \psi_{\a j}^{\dagger \, (2)}  \psi_{\b j}^{ \, (2)} \nn \\
(J_{MQM_{1}}^\tau)_{i j} = k_1 \delta_{i j} - \sum_{\a = 1}^n   \psi_{\a i}^{\dagger \, (1)}  \psi_{\a j}^{ \, (1)}  \,  , \nn \\
(J_{MQM_{2}}^\tau)_{i j}  =  k_2 \delta_{i j} - \sum_{\a = 1}^n   \psi_{\a i}^{\dagger \, (2)}  \psi_{\a j}^{ \, (2)}  \, .
\eea
In these expressions the first is the $U(n)$ constraint, while the other two correspond to the $U(N_{1,2})$ constraints. Taking the trace of the expressions in \eqref{currentsmatchmainb}, we find the consistency conditions  (see also~\cite{Betzios:2017yms} for normal ordering the quantum version of such expressions)
\be\label{restrictionlevels}
\sum_{\a = 1}^n \sum_{i = 1}^{N_1} \psi_{\a i}^{\dagger \, (1)}  \psi_{\b i}^{ \, (1)} \, = \, k_1 N_1   = \sum_{\a =1}^n \sum_{j = 1}^{N_2} \psi_{\a j}^{\dagger \, (2)}  \psi_{\b j}^{ \, (2)} \, = \, k_2 N_2 \, .
\ee
The condition $k_1 N_1 = k_2 N_2$ shall be reproduced through properties of the Kostka polynomials appearing in the analysis of the partition function in section~\ref{HallLittlewood}.

We shall now describe the thermal partition function of the combined system including the bi-fundamentals using a direct character expansion (setting $k_{1,2} = 0$ momentarily). For a compact $\tau$, one needs to consider the zero modes of the various gauge fields. In particular we set $\Omega_{1,2} = e^{i \beta A^{gYM}(z= \pm L)}$ and $U_{1,2} =  e^{i \beta A_{bd}^{1,2}}$ as the unitary matrices arising from exponentiation of the zero modes of each gauge field at the boundaries $z = \pm L$.
If we integrate out the bi-fundamentals, they give a partition function~\cite{Betzios:2017yms} ($m$ is their mass)
\be
Z^1_b(U_1, \Omega) = \sum_{R_1} e^{- \beta m_1 C_{R_1}^{(1)}} \chi_{R_1} (U_1)  \chi_{R_1} (\Omega_1^\dagger) \, ,
\ee
and similarly for the second boundary (the representations $R_1$ run over those of the smallest group). The MQM matter fields can be integrated as before. Using the orthogonality relation for characters
\be
\int D \Omega \, \chi_{R}(\Omega) \, \chi_{R'}(\Omega^\dagger) \, = \, \delta_{R R'} \, ,
\ee
we arrive at the following expression for the combined partition function
\bea
Z &= \sum_{R, R_1, R_2}   \int D U_1 \int DU_2 \, \chi_{R_1} (U_1) Z^{MQM}_1 (U_1)  \chi_{R_2} (U_2) Z^{MQM}_2 (U_2) \nn \\
&\times e^{- \beta m_1 C^{(1)}_{R_1} - \beta m_2 C^{(1)}_{R_2} } \, \delta_{R R_1} \delta_{R R_2} \, e^{- L \frac{ g_{YM}^2}{N}  C_R^{(2)} + i \theta  C_R^{(1)}} = \nn \\
&= \sum_{R}  \, Z_{R}^{MQM_1} Z_{R}^{MQM_2}  \,  e^{- \beta (m_1 + m_2) C^{(1)}_{R} }  \, e^{- L \frac{ g_{YM}^2}{N}  C_R^{(2)} + i \theta  C_R^{(1)}} \, . \nn \\
\eea
This might look qualitatively similar to the expression encountered in eqn.~\eqref{pf1}, but the difference lies in the fact that $R$ is now a $U(n)$ representation with $n$-rows (we assume from now on $n \leq N_{1,2}$). Using the formula \eqref{MQMrep} of appendix~\ref{characterexpansion}, one can also write this expression in terms of integers labelling the highest weights of three representations (see appendix~\ref{partitions} for details on partitions and highest weights), and two Littlewood Richardson coefficients $C_{R R_{1,2}}^{R_{1,2}}$, where $R$ is a $U(n)$ representation and $R_{1,2}$ is a $U(N_{1,2})$ representation respectively. The Littlewood Richardson coefficients capture the multiplicity that $R_{1,2}$ appears in the irrep decomposition of the tensor product $R \otimes R_{1,2}$.

It is clear that even in the case of a very small rank for the messenger theory $n \ll N_{1,2}$, the exact partition function does not factorise, but is expressed as a factorised sum over various ``sectors". Of course in this case, in the large $N_{1,2}$ limit, the leading saddle is expected to be determined solely from each individual MQM and if there is any connection between the two almost factorised dual geometries it is extremely weak (``microscopic quantum wormhole") and not a semi-classical saddle\footnote{Such extremely weak links and microscopic wormholes are relevant for the $\a$ parameter story~\cite{BKP}.}. If we wish to find a leading wormhole saddle, one should therefore perform a more thorough analysis of the limit of large representations. Since it is not easy to analyse the large representation asymptotics of the Littlewood-Richardson coefficients $C_{R R_{1,2}}^{R_{1,2}}$, we shall resort to using the technique of appropriate orthogonal polynomials, that bypasses this problem.

\subsection{The partition function via Hall-Littlewood polynomials}\label{HallLittlewood}

In order to analyse in detail the partition functions we found in the previous sections, it is convenient to use the formalism of Hall-Littlewood and Schur symmetric polynomials~\cite{Macdonald}, that was applied in similar models in~\cite{Dorey:2016hoj,Betzios:2017yms,Barns-Graham:2017zpv}. In particular we shall exploit the relation between representations $R$ of $U(n)$ and partitions $\lambda : (\lambda_1 , ... \lambda_n)$, in terms of Young diagrams with $\lambda_i$ boxes on the $i$'th row and with length $\ell(\lambda) = n$ (number of rows). The order of the partition is the number of boxes $\sum_{i=1}^{\ell} \lambda_i = |\lambda| $. See appendix~\ref{partitions} for details on partitions and fig~\ref{partitionexample} for a simple example.

In the rest we shall also denote by $s_\l(Z)$ the Schur polynomials, $K_{\l \m}(q)$ the Kostka polynomials and by $P_{\l}(Z ; q), Q_{\l}(Z ; q) $ the Hall-Littlewood/Milne polynomials. The definitions of these polynomials as well as various technical details and formulae can be found in appendix~\ref{HallLittlewoodappendix} and the book~\cite{Macdonald}. Additional details regarding algebras, representations and branching functions that we shall later use, can be found in appendix~\ref{modulesbranching}.

Putting into use the formalism of appendix~\ref{HallLittlewoodappendix}, it is possible to integrate over the $U_{1,2}$ gauge group zero modes with eigenvalues $z_i^{(1,2)} = e^{i \theta_i^{(1,2)}}$. The part of the partition function from the first gauge group contains a contribution from the MQM and the bifundamental field on the first boundary, depends also on the $U(n)$ holonomy $\Omega_1$ with eigenvalues $\omega_\a^{(1)}$, and reads\footnote{For MQM this part of the partition function for $\Omega_1 = I$ was analysed in~\cite{Betzios:2017yms} and in~\cite{Dorey:2016hoj} for the matrix model related to the Quantum-Hall effect.}
\bea\label{singleboundarypf}
\mathcal{Z}_{N_1}^{k_1}(\Omega_1) &=& \frac{q_1^{N_1^2/2}}{N_1!} \left(\prod_{i=1}^{N_1} \frac{1}{2 \pi i} \oint \frac{d z_i}{z_i^{k_1} } \right) \frac{\prod_{i \neq j} (z_i - z_j)}{\prod_{i,  j} (z_i - q_1 z_j) }  \prod_{{{a}}=1}^{n} \prod_{i=1}^{N_1} \frac{1}{(1- e^{- \beta m_1} \omega^{(1)}_\a z_i )} \, \nn \\
&=& \frac{q_1^{N_1^2/2}}{(1-q_1)^{N_1}} \sum_{\l \, : \, \ell(\lambda) \leq \text{min} \lbrace n, N_1 \rbrace  }   e^{- \beta m_1 |\lambda|} Q'_\l(\Omega_1 ; q_1)    \langle P_\l  \, , P_{(k_1^{N_1})}   \rangle_{q_1}  \,  \nn \\
&=&  \frac{q_1^{N_1^2/2}}{\phi_{N_1}(q_1)}  \sum_{\l \, : \, \ell(\lambda) \leq n}   e^{- \beta m_1 |\lambda|}  K_{\lambda , (k_1^{N_1})}(q_1) s_\l (\Omega_1) \, , \qquad q_1 = e^{- \omega_1 \beta_1} \, . \nn \\
\eea
We should note at this point that $\lambda$ is a partition corresponding to the highest weights of the $U(n)$ representation (see appendix~\ref{modulesbranching}) and $(k_1^{N_1})$ is the partition with $ N_1$ rows each having $k_1$ boxes. The definition of the Hall inner-product $\langle P_\l  \, , P_{(k_1^{N_1})}   \rangle_{q_1}$ is given in eqn. \eqref{inner} of the appendix and the function $\phi_{N_1}(q_1)  = \prod_{j = 1}^{N_1} (1- q_1^j)$. We have assumed that $n \leq N_1$ and hence the representations have maximum length $\ell(\lambda) = n$.
For fermionic bifundamentals, one can use the dual Cauchy identity eqn. \eqref{idenntitiespolynomials}, and follow similar steps to find ($T$ denotes the transpose partition, reflecting along the diagonal, see~\ref{partitions})
\bea
\mathcal{Z}_{N_1}^{k_1} =    \sum_\l \frac{q_1^{N_1^2/2} e^{- \beta m_1 |\lambda|}}{(1-q_1)^{N_1}}  P_{\l^T}(\Omega_1 ; q_1) \langle P_\l  \, , P_{(k_1^{N_1})}   \rangle_{q_1}  = \frac{q_1^{N_1^2/2}}{\phi_{N_1}(q_1)}    e^{- \beta m_1 |(k_1^{N_1})|} P_{(N_1^{k_1})}(\Omega_1 ; q_1)  \, . \nn \\
\eea
Similar formulae hold for the second boundary as well. The final step is to combine the pieces from the boundary contributions together with that of the 2d YM
\be\label{2dYMpropag}
Z^{YM}_{cyl}(\Omega_1,\Omega_2 ) = \sum_{\nu \, : \, \ell(\nu) \leq n}  s_\nu(\Omega_1)   s_\nu(\Omega_2) e^{- L \frac{g_{YM}^2}{n}  C_\nu^{(2)} + i \theta  |\nu |}  \, ,
\ee
where the quadratic and linear Casimirs are expressed in terms of the partition $\nu$ as
\be\label{Casimirs}
 C_\nu^{(2)} = \sum_{i=1}^n  \nu_i (\n_i - 2 i + n + 1) \, , \quad   |\nu | = C_\nu^{(1)} = \sum_{i=1}^n \nu_i \, ,
\ee
and integrate over the zero modes of the $U(n)$ gauge field holonomies $(\Omega_{1,2})$ at the two boundaries. Using the simple orthogonality relation of Schur polynomials, we find in the case of bosons
\be\label{bifundpartition}
\mathcal{Z} = \frac{q_1^{N_1^2/2} q_2^{ N_2^2/2}}{\phi_{N_1}(q_1) \phi_{N_2}(q_2)}  \sum_{ \nu \, : \, \ell(\nu) \leq n } \, e^{- \beta (m_1+ m_2) |\nu|} K_{\nu , (k_1^{N_1})}(q_1) K_{\nu , (k_2^{N_2})}(q_2) e^{- L \frac{g_{YM}^2}{n}  C_\nu^{(2)} + i \theta  |\nu |}  \, ,
\ee
with a similar expression in the case of fermions involving the transpose representations. This is the main general form of the partition function as a sum over Kostka polynomials $K_{\m, \n}(q)$. We readily observe that the prefactor describes a factorised singlet partition function for the two MQM models corresponding to the canonical partition function of $N_{1,2}$ free fermions in the oscillator potential
\be
Z_{singlet} = Z_{singlet}^{(1)} Z_{singlet}^{(2)} = \frac{q_1^{N_1^2/2} q_2^{  N_2^2/2}}{\phi_{N_1}(q_1) \phi_{N_2}(q_2)} \, .
\ee
The complete partition function does not factorise and one has to find the leading saddle governing the dual geometry.

The next thing to notice is a consistency condition $k_1 N_1 = k_2 N_2$ for the C.S. levels and the gauge groups, due to the properties of the Kostka polynomials that are non-zero only for $|\nu| = |k_1^{N_1}| = k_1 N_1 = k_2 N_2 = |k_2^{N_2}|  $, matching eqn.\eqref{restrictionlevels}\footnote{This is analogous to an anomaly matching condition.}. The model of common YM and MQM groups of section~\ref{thermalpf} is simply obtained by specialising $N_1 = N_2 = n$ (and $m_{1,2} = 0$) that leads to $k_1 = k_2 = k$. We also observe that had we picked $k_1 = k_2 = 0$, the only states that would contribute would be zero-weight states (inside a highest weight module) as was first noticed in~\cite{Gross:1990md,Klebanov:1991qa,Boulatov:1991xz} and described in section~\ref{thermalpf} and appendix~\ref{modulesbranching}.

For the normal oscillator, the lowest energy state for each MQM is determined as one sends $|q| \ll 1$ and derived from properties of the Kostka polynomials~\cite{Dorey:2016hoj}. This corresponds to a representation consisting of two rectangular boxes (we parametrize $N_1 = n L_1 + C_1$)
\be
\nu_{ground} = \left((L_1+1)^{C_1}, \, L_1^{n - C_1} \right) \, ,
\ee
This determines the ground state representation in the case of the normal oscillator. On the other hand, for the inverted oscillator $q= e^{i \beta}$ (a pure phase). Therefore, this limit is not enough  to determine the leading saddle of the system. What one should do instead is to take the limit of large representations in an appropriate manner.

This limit depends on how we scale $N_1, N_2 , n$. A case that has been discussed in~\cite{Dorey:2016hoj,Betzios:2017yms} is the limit when
\be\label{limit1}
N_1 = L_1 n + C_1 \, , \quad  N_2 = L_2 n + C_2 \, , \quad  \text{with} \quad  L_{1,2} \rightarrow \infty \, ,  \quad n \quad \text{finite} \, .
\ee
For the present model, this is the limit in which the intermediate gauge group has relatively small rank, and is analogous to the analysis of models related to ``wormbranes", where one considers a quiver gauge theory with two sets of nodes with large rank connected via a node with small rank~\cite{Bachas:2017rch}.

To actually perform this limit, we shall use two important facts. First the Kostka polynomial $K_{\l (k^N)}(q)$ is non zero only when $|\l| = k N = k L n + k C$ in the above decomposition. Second given such a partition $\lambda$ with length $\ell(\lambda) = n$  there is a unique partition $\tilde{\lambda} \, , \, \ell (\tilde{\lambda}) < n$ with $|\tilde{\lambda}| = k C \,$ mod $ n $ such that~\cite{Dorey:2016hoj}
\be
\lambda = \tilde{\lambda} + (Q^n) \, , \quad \lambda_i = \tilde\lambda_i + Q \, , \quad Q = \text{max} \lbrace 0,  k L  - \frac{|\tilde{\lambda}| - k C}{n} \rbrace \, ,
\ee
with $(Q)^n$ the representation of $Q$ boxes in each of the $n$ rows.
This provides a way to relate representations of $U(n) \simeq SU(n) \times U(1)$ to those of $SU(n)$ (tilde's). We then consider the irreps of the complexified $\mathfrak{sl}_n$ ($A_{n-1}$) Lie-algebra as well as those of the affine  $\hat{\mathfrak{sl}}_n$ ($\hat{A}_{n-1}$) Lie-algebra (at level $k$). For notation and conventions see Appendix~\ref{modulesbranching}. We denote  with $\Lambda = \Lambda(\lambda)$ the dominant integral weights that dictate the $\mathfrak{sl}_n$ ($A_{n-1}$) irrep and with $R_{\Lambda}$ the Cartan subalgebra generators in the Chevalley basis and with similar hatted symbols those of the affine $\hat{\mathfrak{sl}}_n$ ($\hat{A}_{n-1}$) Lie-algebra.
In~\cite{Nakayashiki:1995bi,Hatayama:1998eeq} it is shown that the large $N$ - fixed $n$ limit as described in \eqref{limit1}, each of the the Kostka polynomials appearing in eqn.~\eqref{bifundpartition}, simplifies into an appropriate branching function\footnote{with $\Lambda \equiv \Lambda(\lambda)$ we denote the highest weight Dynkin labels that are in $1-1$ correspondence with the partition $\lambda$, see appendix~\ref{modulesbranching}.}
\bea\label{Kostkabranch}
K_{\lambda(\Lambda) , (k^{N})}(q) \big|_{L \rightarrow \infty} = q^{E_0} b_\Lambda^{\hat{\Lambda}}(q) = q^{E_0} \Tr_{M^{\hat\Lambda}_\Lambda} q^{ R_{\hat{\Lambda}}(L_0)}   \, , \nn \\
\text{with} \quad \hat{\Lambda} = k \Lambda_{(C)} \, , \qquad E_0 = \frac{n k L(L-1)}{2} + k L C  \, . \quad
\eea
In this formula,  hats denote objects in the affine-Lie algebra and the coefficients $b_\Lambda^{\hat{\Lambda}}(q)$ are also known as the branching functions for the embedding of $A_{n-1}$ into $\hat{A}_{n-1}$ or the coset $\hat{A}_{n-1}/ {A}_{n-1}$ Virasoro submodule $M^{\hat{\Lambda}}_\Lambda$. The trace is therefore over $\hat{A}_{n-1}$ modules, but restricted to $M^{\hat{\Lambda}}_\Lambda$, which is the integrable highest weight $\hat{A}_{n-1}$ - module with respect to the global generators belonging to $\Lambda$. The parameter $k$ corresponds to the level of the associated $\hat{A}_{n-1}$ affine Lie algebra~\cite{Dorey:2016hoj,Betzios:2017yms}. These works also make intuitively clear the appearance of the Kac-Moody algebra in the limit \eqref{limit1}, by constructing ``(quasi) current algebra" operators in a Schwinger-Wigner representation that involves bi-linears of the bifundamental fields $\psi_{\a i}$ of the action \eqref{bifundaction1}, together with powers of the MQM matrix creation/annihilation operators $A_{i j}, A^\dagger_{i j}$. In the limit \eqref{limit1}, one finds that the algebra indeed closes, a central extension term appears, and the states are then organised according to the affine current algebra. For the expression \eqref{Kostkabranch}, the relevant representation that has the lowest energy (lowest $L_0$ eigenvalue - primary states) is the $k$-fold symmetrization  of the $C^{th}$ antisymmetric power of the fundamental representation. This representation has congruence class $k C  \, \text{mod} \, n$ and all the rest of the dominant integral weights fall into the same congruence class. Additional details of this restricted module and the associated branching functions are presented in appendix~\ref{modulesbranching}. In particular, they can also be expressed in terms of $q$-independent branching coefficients/multiplicities as
\be\label{branchingfunct1}
b_{\Lambda}^{\hat{\Lambda}} (q) \equiv \sum_{m=0}^\infty \dim M^{\hat{\Lambda}}_{\Lambda - m \delta} \, q^m \, ,
\ee
with  $\delta$ the additional imaginary root of the affine Lie algebra.

An important property of the multiplicities $\dim M^{\hat{\Lambda}}_{\Lambda - m \delta}$ in \eqref{branchingfunct1}, is that they exhibit a fast universal Cardy-like growth for high excitation number $m \rightarrow \infty$, for fixed highest weights $\Lambda$. In particular, for $\hat{A}_{n-1}$~\cite{Klimyk}
\be\label{asymptoticrestricted}
\dim M^{\hat{\Lambda}}_{\Lambda - m \delta}  \rightarrow_{ m \rightarrow \infty}  (c/6)^{n/4} b m^{-(n+ 2)/4} e^{\pi \sqrt{2 c m /3  }} \, ,
\ee
with $c$ the central charge of the affine algebra $\hat{A}_{n-1}$ at level $k$ and $b$ a numerical coefficient that can also be computed explicitly depending on the specific algebra and the level $k$. This Cardy type of growth is also present in the simpler non-singlet MQM model of~\cite{Betzios:2017yms}, proposed to describe a two-dimensional black hole. We provide some further comments for the connection between the present model, Euclidean wormholes and black holes in the conclusions~\ref{conclusions}.

Using the simplification of the Kostka polynomials in terms of branching functions, one then finds that the partition function for the combined system \eqref{bifundpartition} is expressed as $(\hat{\Lambda}_{1,2} = k_{1,2} \Lambda_{(C_{1,2})} $ with $k_1 C_1 = k_2 C_2$)
\be
\mathcal{Z} = q_1^{E_0^{(1)}} q_2^{ E_0^{(2)}} \prod_{j=1}^{N_1} \frac{1}{1- q_1^j} \prod_{j=1}^{N_2} \frac{1}{1- q_2^j} \sum_{\Lambda(\lambda)} \,  b_{\Lambda}^{\hat{\Lambda}_1} (q_1) b_{\Lambda}^{\hat{\Lambda}_2} (q_2)   e^{- L    \frac{g_{YM}^2}{n}  C_{\Lambda}^{(2)} + i \theta  |\Lambda |} \, .
\ee
One could make this expression more concrete for small algebras such as $\hat{A}_1$, using the formulae in appendix~\ref{modulesbranching}. The qualitative properties are transparent though, without such a detailed analysis. The piece that does not factorise, is not dependent on $N_{1,2}$. This means that strictly speaking in the limit~\eqref{limit1} of the canonical partition, the leading saddle is factorised and of order $O(N_{1,2}^2)$ while the non-factorised contributions are subleading $O(1)$ corrections (only ``quantum microscopic wormholes" are possible). We therefore find that in contrast to the analogous model of wormbranes where two quivers are coupled with a small rank node~\cite{Bachas:2017rch}\footnote{The model of wormbranes has a single asymptotic boundary.}, our cross coupling between the two MQM theories via the topological messenger theory is much weaker when $N_{1,2} \rightarrow \infty$ with $n$-finite\footnote{It might be worthwhile to consider also the limit where both $N_{1,2}, n \rightarrow \infty$ with $N_{1,2} \gg n$.}.

Nevertheless, even in the case when $n$ is finite, there do exist $U(n)$ representations with $O(N)$ boxes. In other words, even if the length of the partition $\ell(\lambda)= n \ll N$, there do exist partitions such that $|\lambda| \sim O(N)$, the quadratic Casimir for these representations being of order $O(N^2)$. These correspond to very elongated thin Young diagrams and are the leading configurations at large-$N$ that induce a weak cross communication between the otherwise decoupled MQM's. As we discussed in the introduction, in $c=1$ Liouville string theory non-trivial representations with few boxes are related to the presence of long string excitations on the linear dilaton background (whose free energy without the long strings is captured by the singlet prefactor). At finite temperature these long strings correspond to winding modes that modify the free energy of the system~\cite{Maldacena:2005hi,Betzios:2017yms}. Now at large $N$ finite $n$, we have two such linear dilaton geometric saddles at finite temperature (described by the factorised part of the partition function) and in addition the presence of very large and thin Tableaux describing the collection of such long strings/winding modes that are correlated. For a example with a single row we have a totally symmetric object with bosonic permutation statistics. Intuitively one could expect that the condensation of these modes, would at least mildly deform the background, creating a (weak and subleading in this case) connection between the two originally decoupled linear dilaton backgrounds\footnote{It would be interesting if one could find a subleading or perhaps complex saddle in the dual Liouville theory similar to the double cone of~\cite{Saad:2018bqo}.}.

We shall now turn our attention to some specific values of the levels $k_{1,2}$, for which one can obtain more explicit results for the partition function, and then analyse a limit where all $N_{1,2}, \, n$ are sent to infinity (limit of continuous Young diagrams). In this case, even the leading saddle of the path integral is directly affected by the non-factorised piece of the partition function.

\subsection{The non-factorised large $N,n$ limit}\label{LargeNn}

If we wish to study the limit where both $N_{1,2}, n$ are large, it is possible to analyse further the partition function \eqref{bifundpartition}, if we specialise the Chern-Simons parameters $k_{1,2}$ of the model. In this case one can use explicit exact formulae for the Kostka polynomials $K_{\lambda, (k)^N}(q)$.

The first case that can be analysed is the case when $k_{1,2} = 0$ so that $N_1 = N_2 = N$. This corresponds to our original model analysed in section \ref{thermalpf}, where only states with zero weights (belonging to each highest weight module) contribute to the partition function~\eqref{bifundpartition}. The Kostka polynomial $K_{\lambda, 0}(q)$ is then a generating series for generalised exponents~\cite{Kostant,Broer}, generalising the dimension of the zero weight submodule of a highest weight module/partition $\Lambda(\lambda) $, see also appendix~\ref{modulesbranching}. Fixing the total weight $|\lambda| = \sum_i \lambda_i = m $, one can exactly map the zero weight sub-modules to permutation group $S_m$ modules~\cite{Broer,Matsuzawa}. Using this correspondence, we can use an explicit formula by Kirillov~\cite{Kirillov1} (see also~\cite{Olshanetsky} for a related quantum mechanical model) that reads
\be
K_{\lambda , 0}(q)  \, = \, \prod_{x \in \lambda} \frac{q^{f(x)} }{ (1- q^{h(x)})} \, , \quad h(x) = \lambda_i + \lambda^T_j - i - j + 1 \, , \quad i \in [1, \ell(\lambda)] \, ,
\ee
in terms of the hook length\footnote{This is the sum of horizontal cells right from $x$ and lower vertical cells including $x$.} $h(x)$ corresponding to the box $x = (i,j) \in \lambda $, and the foot/leg length of the hook  $f(x) =  \lambda^T_j - j $ defined as the sum of lower vertical cells when projecting the hook to the vertical axis. The total foot/sum of leg lengths over all cells is\footnote{For more details and definitions, see~\cite{Macdonald}.}
\be
f(\lambda) = \sum_{x \in \lambda} f(x) =  \sum_{i \geq 1}^{\ell(\l)} (i-1)\lambda_i  = \sum_{i \geq 1}^{\ell(\lambda)} \begin{pmatrix}
\lambda_i^T \\ 2
\end{pmatrix}  \, .
\ee
Another useful formula for the product over hook lengths is
\be
\prod_{x \in \lambda} (1- q^{h(x)}) = \frac{\prod_{i \geq 1}^{\ell(\nu)} \prod_{j = 1}^{\nu_i} (1- q^j)}{\prod_{i < j}^{\ell(\nu)} (1 - q^{\nu_i - \nu_j})} \, , \qquad \nu_i = \lambda_i - i + \ell(\lambda) \, ,
\ee
in terms of the strictly decreasing shifted weights $\nu_i$ labelling the same partition (with $\ell(\nu) = \ell(\lambda)$). 
Finally there is the relation $\sum_{x \in \lambda} h(x) = f(\lambda^T) + f (\lambda) + |\lambda|$ to transition between the sums over lengths of a partition and its transpose.

Using these formulae, we find the following explicit form for the complete partition function as a sum of partitions labelled by the strictly decreasing integers $\nu_i$
\bea\label{bifundpartition0}
\mathcal{Z} &=& \frac{q_1^{N^2/2} q_2^{N^2/2}}{\phi_N(q_1) \phi_N(q_2)}  \sum_{ \nu_i \in \nu \, : \, \ell(\nu) \leq n }  \, e^{- \beta (m_1+ m_2) |\nu|} \times  \nn \\
 &\times&  \frac{(q_1 q_2)^{ f(\nu)} \prod_{i < j} (1 - q_1^{\n_i - \n_j}) (1 - q_2^{\n_i - \n_j}) }{ \prod_{i \geq 1}^n \prod_{j = 1}^{\nu_i} (1- q_1^j) (1- q_2^j)}  e^{- L \frac{g_{YM}^2}{n}  C_\nu^{(2)} + i \theta  |\nu |}  \, , \nn \\
\eea
with $q_{1,2} = e^{- \omega_{1,2} \beta_{1,2}}$, $T$ the transpose partition and the quadratic and linear Casimirs given by equation \eqref{Casimirs}.

Another example we can analyse in detail is the case $k_1 = k_2 = 1$ (so that $N_1 = N_2 = N$). One can then use the following explicit formula for the Kostka polynomials~\cite{Macdonald}
\be
K_{\lambda \, , (1^N)} (q) \, = \, \frac{q^{f(\lambda^T)} \phi_N(q)}{\prod_{x \in \lambda} (1- q^{h(x)})} \, , \quad h(x) = \lambda_i + \lambda^T_j - i - j + 1
\ee
in terms of the hook length $h(x)$ and where $\phi_N(q) = \prod_{i=1}^{N}(1- q^i)$. In this case the complete partition function is expressed as
\bea\label{bifundpartition2}
\mathcal{Z} &=& q_1^{N^2/2} q_2^{N^2/2} \sum_{ \nu_i \in \nu \, : \, \ell(\nu) \leq n }   \, e^{- \beta (m_1+ m_2) |\nu|}   \times  \nn \\
 &\times& \frac{(q_1 q_2)^{  f(\nu^T)} \prod_{i < j}^n (1 - q_1^{\n_i - \n_j}) (1 - q_2^{\n_i - \n_j}) }{ \prod_{i \geq 1}^n \prod_{j = 1}^{\nu_i} (1- q_1^j) (1- q_2^j)}  e^{- L \frac{g_{YM}^2}{n}  C_\nu^{(2)} + i \theta  |\nu |}  \, . \nn \\
\eea
In both cases, the form of the exact partition function resembles that of q-deformed\footnote{Now the q-deformation is related to the inverse temperature $\beta \sim 1/T$ and not to 2d YM parameters.} YM theories and partition functions appearing in the context of topological strings.

\paragraph{Large $N,n$ limit} - Since the exact formulae \eqref{bifundpartition0} and \eqref{bifundpartition2} are now expressed only in terms of the partition integers $\nu_i$, one can directly study the large $n$-limit (so that both $N,n$ are scaled to infinity and $N \sim n$) using the continuous variables described in appendix~\ref{characterexpansion}. This is a limit in which both sides of the Tableaux become large, so that it acquires a continuous shape if we scale it properly~\cite{Douglas:1993iia}. In particular one finds the leading saddle to be determined by the following effective action (from now on we set $q_1 = q_2 = q$ for simplicity)
\bea
\mathcal{Z} &=& Z_{singlet}^{(1)} Z_{singlet}^{(2)} \int D h ~e^{- n^2 S_{eff}(h)} \, , \nn \\
S_{eff}(h) &=& L g_{YM}^2  \int_0^1 d x h^2(x) + \left(\beta (m_1 + m_2) -i \theta \right) \int_0^1 d x h(x)    \nn \\
&-& \int_0^1 d x \int_0^1 d y \log \left(1 - q^{|h(x) - h(y)|}  \right) + 2  \int_0^1 d x \int_0^{h(x)} d y \log ( 1 - q^{-y}) \nn \\
\eea
Introducing the normalised density of boxes $\rho(h)$, the saddle point is determined by
\be
\omega \beta \int {d h' \rho(h')} \coth \left( \frac{\omega \beta (h-h')}{2}  \right) = 2  \log(1 - q^{-h}) + 2 L g_{YM}^2  h +  \left(\beta (m_1 + m_2) -i \theta \right) \, .
\ee
It is useful to redefine variables $U = q^{h}$ ($q = e^{- \omega \beta}$) and the new density of boxes
\be
\bar{\rho}(U) = \frac{1}{\log q} \rho \left( h = \frac{1}{\log q} \log U \right) \, , \quad \int d U \frac{\bar{\rho}(U)}{U} = 1 \, ,
\ee
with respect to which the saddle point equation becomes
\be\label{saddlefinal}
\mathcal{P} \int \frac{d V \bar{\rho}(V)}{U - V} = \frac{1}{\omega \beta} \left( \frac{L g_{YM}^2}{\omega \beta} +  1 \right) \log U  - \frac{1}{\omega \beta} \log (U-1)    - \frac{\left( \beta (m_1 + m_2) -i \theta \right)}{2 \omega \beta}  \, .
\ee
This is a common form of saddle point equations\footnote{Similar equations appear for example in the analysis of chiral YM and q-deformed YM~\cite{Crescimanno:1994eg,Kostov:1997bn}.} that can be readily analysed (for real $\omega$). The term on the left hand side provides a mutual repulsion between the eigenvalues while the ``force" from the potential term on the righthand side tries to clump them. The $\theta$-term makes the action and the force complex and one needs to consider an appropriate ``instanton" expansion to analyse the model~\cite{Arsiwalla:2005jb,Jafferis:2005jd}. In particular the form of the ``force" and the potential is such that we expect the presence of various phases. There exist two possible different one-cut phases (with or without saturation of the eigenvalue density near the endpoint of its support). There also exist two different two-cut phases (with one or two saturation regions). In the low temperature limit $\beta \rightarrow 0$, the saddle point equations reduce to those of the 2d YM having two possible saddles. This is the qualitative phase diagram for the saddles in the case of the normal oscillator  (real $\omega$)\footnote{The usual matrix oscillator is relevant for the description of $1/2$ BPS states in $\mathcal{N}=4$ SYM~and the dual LLM geometries~\cite{Berenstein:2004kk,Lin:2004nb,Berenstein:2017abm}.}. Unfortunately for the inverted oscillator relevant to $c=1$ Liouville string theory, the analysis is much more intricate, since $\omega $ is imaginary\footnote{In this case one has also to work in the grand-canonical ensemble at fixed chemical potential $\mu$, dual to $N$, since $1/\mu \sim g_{st}$ is the parameter most directly related to the string coupling.}.  This means that we actually need to perform a very careful ab-initio analysis to solve the combined saddle point equations, and not merely rely on a simple analytic continuation in $\omega$.

We conclude that the class of coupled MQM - 2d YM models we analysed, can admit various possible leading semi-classical saddles that are different from the decoupled gauged MQM singlet state (in the limit where both the gYM and MQM gauge groups are comparably large). Moreover the leading saddle depends both on how we perform the large $N,n$-limit as well as on the various other parameters such as the temperature and $L g_{YM}^2$, $\theta$,  leading to the possibility of obtaining a quite interesting and rich phase diagram. While we did not find explicitly the description of the leading dual geometric saddles for the relevant case of the inverted oscillator, we expect that some of them will correspond to a connected geometry with two asymptotic regions, because in the Hamiltonian description the two MQM models are not directly cross coupled via an explicit term in the action\footnote{As long as the asymptotic regions are governed by the inverted oscillator potentials of $\lambda^{1,2}$, there are two such regions.}. Even though deriving the explicit dual geometry (bulk reconstruction) is a hard problem, perhaps some crude properties such as the number of boundaries and the connectedness properties of the manifold are somehow directly encoded in the density of boxes in the Young diagram and the solution of the matrix model saddle point equations (see~\cite{deMelloKoch:2008hen} for a related analysis in the context of LLM geometries~\cite{Lin:2004nb}). A logical possibility is that the structure shall remain similar to the one of the normal oscillator, and depending on the parameters, some saddles will correspond to a connected wormhole geometry, while others to two-factorised geometries and/or geometries connected by microscopic wormholes. This is an important problem for the future that should be tackled in parallel from the geometric gravitational side (we do not yet know explicit target space wormhole solutions in the bosonic or $0A$/$0B$ Liouville string theory).

\subsection{Two- and four-point correlation functions}\label{MQMcorrelators}

Using the technique of Hall-Littlewood polynomials, it is also possible to study some simple two- and four-point correlation functions. In particular, at finite temperature, one finds the following result for the Wick contractions of two Matrix operators obeying the twisted periodicity condition $M(\beta) = U M(0) U^\dagger$~\cite{Brigante:2005bq}
\bea\label{wickmain}
\wick{\c{M}_{i j}(\tau) \c{M}_{k l}(0)} = \frac{1}{N} \sum_{m = - \infty}^\infty G_s(\tau - m \beta ; \omega) U_{i l}^{-m} U_{k j}^m \, , \nn \\
G_s(\tau  ; \omega) =  \frac{1}{2 \omega} e^{- \omega |\tau|} \, .
\eea
This equation holds independently for each of the two MQM theories, so there is an implicit $1,2$ index. We emphasize that there are no cross-MQM Wick contractions. Armed with this result, we can proceed to integrate over the Gaussian MQM action to derive once more an effective action depending on the two holonomies $U_{1,2}$ as shown in eqn. \eqref{twistedMQM}. The resulting integral over the unitary matrices is hard to compute if the indices are uncontracted, since it involves a string of unitary matrices $U_{i j}$. On the other hand, for gauge invariant expectation values with contracted indices we obtain  a product of traces, that can also be expressed using characters or Hall-Littlewood polynomials.
Some simple examples are
\bea\label{wickexamples}
\langle \Tr M^2(0) \rangle_U &=& \frac{1}{N} \sum_{m = - \infty}^\infty G_s( - m \beta ; \omega) \langle \Tr U^m \Tr U^{-m}  \rangle_U \, , \nn \\
\langle \Tr M^2(\tau) \Tr M^2(0) \rangle_U &=&	 \frac{2}{N^2} \sum_{m,  n  =  - \infty}^\infty G_s(\tau - m \beta ; \omega) G_s(\tau - n \beta ; \omega) \langle \Tr U^{m-n} \Tr U^{n-m}   \rangle_U  \nn \\
\eea
where with $\langle A \rangle_U$ we denote the unitary average that contains the part from the MQM effective action (and any other $U(N)$ characters that come from integrating the 2d BF theory and the bifundamental messengers). The first example is the one point function, that is found to be time independent (but non-zero) as expected from translational invariance on the circle.

The computation of the correlators now parallels the discussion of section~\ref{HallLittlewood}, where the averages over $U_{1,2}$ contain additional insertions of traces of the unitary matrices. The most efficient manipulation is a direct expansion of power sum symmetric polynomials in terms of Hall Littlewood polynomials~\cite{Macdonald}
\be
p_\rho(Z) = \sum_\lambda X^\l_\r (q) P_\l (Z ; q) \, ,
\ee
in terms of Green's polynomials  $ X^\l_\r (q)$ that for $q=0$ reduce to the irreducible character $\chi^\l$ of $S_n$ at elements of cycle type $\rho$. In the examples of eqn. \eqref{wickexamples} that only involve traces, the expression simplifies into
\be
p_j(Z) = \Tr Z^j = \sum_{|\lambda| = j} q^{n(\lambda)} \prod_{i = 1}^{\ell(\lambda)-1} (1- q^{-i}) P_\lambda(Z ; q) \, .
\ee
Now using the description of section~\ref{HallLittlewood} that expresses all the pieces of the integral over unitary matrices in terms of inner products of Hall-Littlewood polynomials, the single boundary part analogue of eqn.\eqref{singleboundarypf} for the correlators is
\bea\label{correlators}
&\langle \Tr M_1^2(0) \rangle_1 = \frac{2 q_1^{N_1^2/2}}{N_1 (1-q_1)^{N_1}} \sum_{m =1}^\infty G_s( - m \beta ; \omega)   \sum_{\l}   e^{- \beta m_1 |\lambda|} Q'_\l(\Omega_1 ; q_1)      \,  \nn \\
& \sum_{ \rho , \mu }^{ |\rho| = |\mu| =  m }  q_1^{n(\m) + n(\rho)} \prod_{i = 1}^{\ell(\m)-1} (1- q_1^{-i}) \prod_{k = 1}^{\ell(\rho)-1} (1- q_1^{-k})   \langle P_\l P_\rho  \, , P_{(k_1^{N_1})} P_\mu   \rangle_{q_1}  \, , \nn \\
\eea
\bea
\langle \Tr M_1^2(\tau) \Tr M_1^2(0) \rangle_1 =	 \frac{4 q_1^{N_1^2/2}}{N_1^2 (1-q_1)^{N_1}} \sum_{n = -\infty}^\infty \sum_{m = 1}^\infty G_s(\tau - (m+n) \beta ; \omega) \times \, \nn \\
\times G_s(\tau - (n-m) \beta ; \omega)  \sum_{\l}   e^{- \beta m_1 |\lambda|}  Q'_\l(\Omega_1 ; q_1) \sum_{ \rho , \mu }^{ |\rho| = |\mu| =  m } q_1^{n(\m) + n(\rho)} \times     \,  \nn \\
 \times \prod_{i = 1}^{\ell(\m)-1} (1- q_1^{-i}) \prod_{k = 1}^{\ell(\rho)-1} (1- q_1^{-k})   \langle P_\l P_\rho  \, , P_{(k_1^{N_1})} P_\mu   \rangle_{q_1}  \, , \nn \\
\eea
We can further manipulate these expressions via fusion of two Hall-Littlewood functions
\be
P_\m(Z; q) P_\n(Z ; q) = \sum_\lambda f^\l_{\m \n}(q) P_\l(Z ; q) \, ,
\ee
with structure constants/polynomials denoted by $f^\l_{\m \n}(q)$ (obeying the condition $f^\l_{\m \n}(q=0) =  C^\l_{\mu \n}$). After the dust settles, we obtain an expression in terms of skew Kostka polynomials $ K_{\n/\rho , \l}(q)$ and partitions
\bea\label{correlators2}
&\langle \Tr M_1^2(0) \rangle_1 = \sum_{m = 1}^\infty G_s( - m \beta ; \omega)  F_1  \, , \nn \\
&\langle \Tr M_1^2(\tau) \Tr M_1^2(0) \rangle_1 =	\sum_{n = -\infty}^\infty \sum_{m = 1}^\infty G_s(\tau - (m+n) \beta ; \omega) G_s(\tau - (n-m) \beta ; \omega) \frac{ 2 F_1}{N_1} \nn \\
&F_1 = F_1(q_1, m_1, k_1, N_1 ; m ; \Omega_1 ) =  \frac{2 q_1^{N_1^2/2}}{N_1 \phi_{N_1}(q_1)}  \sum_{\n , \l}  e^{- \beta m_1 (|\n| - |\rho|)} s_\l(\Omega_1) \times \nn \\
&\times \sum_{ \rho , \mu  \, ; \, |\rho| = |\mu| =  m }  K_{\n/\rho , \l}(q_1) q_1^{n(\m) + n(\rho)} \prod_{i = 1}^{\ell(\m)-1} (1- q^{-i}) \prod_{k = 1}^{\ell(\rho)-1} (1- q^{-k}) f_{\m (k_1^{N_1})}^\n (q_1) \nn \\
\eea
One then has to couple this expression to a similar expression coming from the second MQM model and integrate over $\Omega_{1,2}$ that are unitaries coupling them to the propagator of the 2d YM theory on the cylinder eqn. \eqref{2dYMpropag}. This has the effect of identifying a common representation (index $\lambda$ in $F_1$ of eqn. \eqref{correlators2}) that couples the two correlators forming a cross-correlator of the general form ($1,2$ sub-indices  and tildes, distinguish different operators of the two MQM's)
\bea\label{generalcorrelatorMQM}
\langle O_{i_1}(\tau_{i_1}) \, ... \, \tilde{O}_{i_2}(\tau_{i_2}) \, ... \rangle = \sum_{\lambda \, : \ell(\lambda) \leq n} \langle O_{i_1}(\tau_{i_1}) \, ... \rangle^\lambda_1 \langle  \tilde{O}_{i_2}(\tau_{i_2}) \, ... \rangle^\lambda_2 \,  e^{- L \frac{g_{YM}^2}{n}  C_\l^{(2)} + i \theta  |\l |}  \, \, . \nn \\
\eea
This correlator generically only depends separately on the differences $\tau_{i_1} - \tau_{j_1}$ and $\tau_{i_2} - \tau_{j_2}$ and not on time differences that mix the $1,2$ sub-indices or normal with tilde operators.

Some concrete examples of the individual pieces inside the $\lambda$-summation are those of eqn. \eqref{correlators2}.
The result hence is a similar (but more complicated version) of the equation describing the partition function \eqref{bifundpartition}, now involving skew Kostka polynomials\footnote{Their definitions along with their properties can be found in~\cite{Kirillov2}.} and additional summations over the integer images due to the MQM thermal propagators. This computation again shows that cross correlators remain non-trivial as long as the leading saddle effectively couples the two MQM partition functions (through a common representation/partition index $\lambda$). Again this cross-communication arises at a leading saddle point level when we take both $N,n \rightarrow \infty$, while when $N \rightarrow \infty$, but $n$ finite the leading saddle is factorised and there is a subleading cross-communication from very long thin partitions with $O(N)$ boxes.

We conclude this section by noting once more that the two-point cross correlators of simple traces of the matrices $M_{1,2}(\tau)$ are simply constants (translational invariance on the circle is preserved for any representation), and there are also no short distance effects in higher point cross correlators for operators inserted on opposite MQM's. It would be interesting to analyse the OTOC limit of our four point cross correlation function using the second equation of \eqref{correlators2}. Using the formalism of Hall-Littlewood polynomials, it is also possible to compute a two-point cross correlator with non trivial time dependence, using more complicated operators for each matrix model,
such as loop operators $\Tr e^{M(\tau)}$ (loop-loop cross correlator~\cite{worm}), or bilocal operators $\Tr M(\tau) M(0)$, whose importance in reconstructing the dual space-time in the case of the SYK model was emphasised in~\cite{Das:2017wae}.

\section{Higher-dimensional examples}\label{higherdim}

It is possible to generalise the construction of the previous section~\ref{D0D1}, in higher dimensions. In particular we can consider a $d+1$-dimensional (quasi) topological theory that plays the role of a ``messenger" coupling two systems in $d$-dimensions. In this section we shall list various examples, but analyse in some detail the case where the boundary theories are in two dimensions (coupled to the $3d$ Chern-Simons messenger). We leave a more thorough quantitative study of higher-dimensional examples for the future~\cite{BP}.

\subsection{Chern-Simons messenger}

A simple example of a higher-dimensional sandwich setup that can be analysed in detail, is in terms of two $2d$ gauged BCFT's coupled through a topological field theory like Chern-Simons theory living in $3d$. The most basic example of Chern-Simons theory\footnote{Examples with two boundaries appear in \emph{ constriction} setups in the condensed matter literature.} with induced dynamics on the boundaries, has been studied quite thorougly in the literature and in particular in physical problems related to Quantum Hall effect. In the presence of manifolds with the topology $\Sigma \times R$ where $\Sigma$ contains boundaries and $R$ is interpreted as time, one finds chiral edge modes (or a chiral WZW model) propagating on them.

In the seminal papers~\cite{Elitzur:1989nr,Wen:1990se} a geometry of topology $Cyl \times R$ was studied. The two boundaries are the edges of the finite size cylinder where a dynamical chiral boson propagates. Naively one would expect the Hilbert space of states to factorise as $\mathcal{H} = \mathcal{H}_{edge}^1 \otimes \mathcal{H}_{edge}^2$, with $\mathcal{H}_{edge} = \mathcal{H}_{KM} \otimes \mathcal{H}_p$, with the last two factors representing a Kac-Moody Hilbert space and a charge Hilbert space generated by a Kac-Moody current $J^+$ and a charge operator $\psi^\dagger$ creating ``electrons" in the fermionic formulation (in the description in terms of a chiral boson these are solitonic sectors). Nevertheless, the true Hilbert space does not factorise as such, since in the particular example bulk flux can make ``electrons" to tunnel quantum mechanically through the bulk. The Hilbert space can then be shown to be constructed from a non-chiral boson and takes the form $\mathcal{H} = \mathcal{H}_{KM}^1 \otimes \mathcal{H}_{KM}^2 \otimes \mathcal{H}_{p_1 p_2}$ with $\mathcal{H}_{p_1p_2}$ spanned by
states $| I_1, I_2, M \rangle$ with $I_{1,2}$ two integers and $M = 0, 1... l$ with $l$ the maximum flux allowed.

Similarly, for the non-abelian model, the Hilbert space corresponds to that of a full WZW model when both left/right movers are taken into account. In this case, the partition function of the system is holomorphically factorisable~\cite{Witten:1991mm}, but one can consider even more general setups that contain indecomposable representations which do not have holomorphically factorisable partition functions.

The relevance of the real time setup of topology $Cyl \times R$, for Euclidean wormholes can be understood upon  analytic continuation of the real line $t = - i \tau$, with a subsequent periodic identification $\tau \sim \tau + \beta$. The Euclidean topology then corresponds to $\Sigma \times I$, with $\Sigma = T^2$ a two-torus and $I$ an interval (the former length of the cylinder or annulus). It is then possible to associate two different complex structures $\tau_{1,2}$ on the two tori at the endpoints of the interval. Defining $q_{1,2} = e^{2 \pi i \tau_{1,2}}$
the partition function takes the form of a trace over the Hilbert space of the two boundary WZW models
\be
Z_{T^2 \times I} = \Tr_{WZW} \left( q_1^{L_0} q_2^{\bar{L}_0} \right) \, ,
\ee
that also enjoys similar modular invariance properties as the usual torus partition function of a single WZW model~\cite{Cotler:2020ugk}.

We would like now to modify and generalise this setup so that we can have two general $2d$ gauged BCFT's\footnote{In principle we would like them to be holographic CFT's, but our setup is expected to be interesting also for non-holographic systems and condensed matter applications.} coupled to the $3d$ Chern-Simons theory as a definition of our higher-dimensional tripartite field theory system.

The Chern-Simons (CS) action with group $H$ is\footnote{We took this to be $SU(N)_k$, but we can also consider $U(N)_{k, k'}$, with different CS levels for the the $SU(N)$ and $U(1)$ part.}
\be
S_{CS} = \frac{k_{CS}}{4 \pi} \int_{\mathcal{M}} d^3 X \epsilon^{MN \Lambda} \Tr \left( A_M \partial_N A_\Lambda + \frac{2}{3} A_M A_N A_\Lambda \right) \, .
\ee
The variation of the CS action gives
\be
\delta S_{C.S.} = \frac{k_{CS}}{4 \pi} \int_{\mathcal{M}} d^3 X  \epsilon^{MN \Lambda} \Tr (\delta A_M F_{N \Lambda} ) \, + \, \frac{k_{CS}}{4 \pi} \int_{\mathcal{M}} d^3 X  \epsilon^{MN \Lambda} \partial_N  \Tr (  A_M \delta A_{ \Lambda} ) \, .
\ee
The first term gives the bulk equations of motion for a flat connection and the last boundary term can vanish either by demanding that the variation of the gauge field at the endpoints of the interval is zero, or by inducing an additional $WZW$ action at level $k_{CS}$ at the boundary.

We shall consider an analogous construction to the model of section~\ref{D0D1}, where the boundary values of the CS gauge field correspond to the gauge fields that are coupled to the two boundary theories $A_\m (r = \pm L) = A^{1,2}_\m $ via a usual minimal coupling/gauging procedure for the two boundary theories. If we wish to have different CS and boundary gauge groups, we can introduce additional bifundamental fields as in section~\ref{bifundmodel}, or consider gauging a diagonal subgroup of the boundary symmetry group $G$ (considering for example a pair of diagonally gauged WZW models with $H \subset G$).

Due to the topology being $\Sigma \times I$, a convenient gauge choice to analyse such models is the radial gauge $A_r = 0$. This gauge choice can be obtained
using gauge transformations that kill the (matrix) scalar degree of freedom \be
\phi(z,\bar{z}) = \int_{-L}^L d r A_r\;.
\ee
In this gauge, the Chern-Simons action simplifies and becomes quadratic
\be
S_{CS} = \frac{k_{CS}}{4 \pi} \int d r \int_{\Sigma} d^2 x \epsilon^{\m \n} \Tr \left( A_\m \partial_r A_\n \right) \, ,
\ee
the non-linearity in the non-abelian case remains due to the constraint that one needs to impose
\be
\frac{\delta S_{CS}}{\delta A_r} = 0 \, \quad \Rightarrow \quad \epsilon^{\m \n} F_{\m \n} = 0
\ee
Since the action is linear in derivatives, the Poisson brackets are
\be
\lbrace A^a_\m(x) \, , \, A^b_\n(y)  \rbrace = \frac{4 \pi}{k} \delta^{ab} \epsilon_{\m \n} \delta^{(2)}(x-y) \, .
\ee
Now there exist various choices of canonical pairs one can take or in other words various choices of polarisations leading to different quantisations. This is important especially when we introduce the boundaries and matter fields coupled to the boundary values of the gauge fields. For example one can choose to fix $A_1$ and $A_2$ to play the role of the conjugate momentum, or to use holomorphic quantisation going to the variables $A_z = \half \left(A_1 - i A_2 \right)$ and fix
$A_z$ or $A_{\bar{z}}$. In this latter case the quantum operators obey
\be
[ A^a_z(z) \, , \, A^b_{\bar{w}}(w)  ] = \frac{2 \pi}{k} \delta^{ab} \epsilon_{\m \n} \delta^{(2)}(z - w) \, ,
\ee
and $A^a_z$ can be realised as functional derivatives to wavefunctions that depend only on $A_{\bar{z}}$. The appropriate inner
product is expressed as a generalisation of the Bargmann coherent state inner product
\be
\langle \Psi \, | \, \Phi \rangle \, = \, \int \mathcal{D}(A_z, A_{\bar{z}}) \exp \left( - \frac{k}{2 \pi} \int d^2 z \Tr A_z A_{\bar{z}} \right) \overline{\Psi(A_{\bar{z}})} \Phi(A_{\bar{z}}) \, .
\ee
The transition between the two bases can be performed using the overlap
\be
\langle A_1 \, | \, A_{z} \rangle \, = \, C \exp \left( - \frac{k}{2 \pi} \int_{\Sigma} d^2 x A_z^2 + \frac{k}{\pi} \int_{\Sigma} d^2 x A_z A_1 -  \frac{k}{4 \pi} \int_{\Sigma} d^2 x A_1^2 \right) \, ,
\ee
with $C$ an overall normalisation for the vacuum amplitude.

The object we need to study that couples the two BCFT's, is the transition amplitude between the two boundary states for the gauge field that can be written as
\be
\langle A\big|_{\Sigma_1} \, \big| A\big|_{\Sigma_2} \rangle \, = \, \mathcal{N} \, \int_{A (r = \pm L) = A\big|_{\Sigma_{1,2}} } \mathcal{D} A \, e^{- S_{CS} - S_{\Sigma_1} - S_{\Sigma_2} } \, ,
\ee
where $S_{\Sigma_{1,2}}$ are appropriately chosen boundary terms that render the combined model gauge invariant. Remarkably this amplitude does not appear to have been explicitly computed in the literature\footnote{A useful paper discussing similar amplitudes is~\cite{Porrati:2021sdc}.}. In radial quantisation, using the radial Hamiltonian $H_r$ as an evolution operator, we can also write schematically
\be
\langle A\big|_{\Sigma_1} \, \big| A\big|_{\Sigma_2} \rangle = \sum_n  \Psi_n( A\big|_{\Sigma_1}) \overline{\Psi_n(A\big|_{\Sigma_2} )} e^{- L {E}^r_n} \, .
\ee
Since the Hamiltonian of a topological theory is trivial the result exhibits the expected factorisation into sectors labelling the wave-functions (degeneracy). This is also related to the holomorphic factorisation of the WZW model partition function (sum of squares of conformal blocks/wavefunctions)~\cite{Witten:1991mm}.

Two comments are in order here. Since one needs to make a choice of which components of the gauge field to fix on the two boundary surfaces, the overall normalisation $\mathcal{N}$ might also depend on the complex structure. The second is that one will have to introduce appropriate boundary terms $S_{\Sigma_{1,2}}$ to render the total action gauge invariant and these terms depend on this choice. Once this transition amplitude is defined and computed for a particular system, one can then derive an ``effective action" for the 2d BCFT's, by integrating over the 3d gauge field in the path integral.

We would like now to sketch the main features of such models in a simple example, leaving a more thorough and complete analysis for the future~\cite{BP}.  We shall therefore concentrate in the simplest case of a two torus boundary surface $\Sigma_{1,2} = T_2$ and for an abelian CS gauge group $H=U(1)$. Using holomorphic quantisation on the torus one can decompose
\be\label{decomp}
A_{\bar{z}} = \partial_{\bar{z}} \chi + i \pi \tau_2^{- 1} \alpha \, ,
\ee
with $\chi$ a complex function and $\a$ a complex group valued vector (in the abelian case it is simply a $U(1)$ compact complex function).
In the $A_r = 0$ gauge the C.S. Lagrangian decomposes into
\be
L_{C.S.} = \frac{i k \pi^2}{\tau_2} (\dot{\alpha} \alpha^* - \dot{\alpha}^* \alpha) + i k \int_{T^2} d^2 x \left( \partial_{\bar{z}} \dot{\chi} \partial_z \chi^* - \partial_z \dot{\chi}^* \partial_{\bar{z}} \chi \right) \, .
\ee
and the path integral measure factorises
\be
\mathcal{D} (\partial_z \chi^* , \partial_{\bar{z}} \chi) \mathcal{D} (\alpha, \alpha^*) e^{- \frac{k}{\pi} \int d^2 z \partial_z \chi^* \partial_{\bar{z}} \chi} e^{- k \pi a^* \tau_2^{-1} \alpha } \, .
\ee
Since the manifold is a two torus, there exist large gauge transformations that affect the $\alpha$'s i.e.
\be
\alpha \rightarrow \alpha + p + q \tau \, , \quad p, q \in \mathbb{Z} \, ,
\ee
and the system behaves as being in the background of a magnetic field (torus magnetic translations). The flux quantisation condition translates to $2 \pi k$ being an integer. In this case one has finite-dimensional representations of the group of magnetic translations\footnote{This case will lead to rational CFT's with finite number of conformal blocks as the relevant wave-functions.}, else one would have to consider infinite-dimensional representations.

Using holomorphic quantisation, the physical wave-functions on the torus $T^2$ (analogous to the characters in the example of the 2d gYM theory), can be explicitly expressed in terms of $\vartheta$-functions, \cite{Bos:1989kn}
\be\label{toruswavefunctions}
\Psi_m(\chi, \alpha) =  \frac{e^{\frac{k \pi}{2 } \alpha  \tau_2^{-1} \alpha } }{\eta(\tau)} \, e^{ \frac{k}{2 \pi} \int_{\Sigma} \partial_{\bar{z}} \chi \partial_z \chi } \,
\vartheta \left[ \begin{matrix}
\frac{m}{k} \\ 0
\end{matrix} \right] \left(k \alpha , k \tau \right)  \, , \quad m \in [0, k-1] \, .
\ee
For $\tau \in \mathfrak{h}_g$ (the Siegel upper half plane), the $\vartheta$-functions appearing in the wave-functions, form a basis of the Hilbert space of $\mathcal{L}_\tau$ quasi-periodic functions of weight $k$. Using the fact that the Hamiltonian in radial quantisation is zero, the result for the $\mathcal{M}_3 = T_2 \times I$ transition amplitude in the abelian case is
\bea
\langle A_{\bar{z}}^1 | A_{\bar{z}}^2 \rangle =  \frac{e^{ \alpha_1 \frac{k \pi}{2  \tau_2^{(1)}}  \alpha_1 + \alpha_2 \frac{k \pi}{2 \tau^{(2)}_2 }  \alpha_2 + \frac{k}{2 \pi} ( \int_{\Sigma_1} \partial_{\bar{z}} \chi \partial_z \chi + \int_{\Sigma_2} \partial_{\bar{z}} \chi \partial_z \chi)}}{\eta(\tau^{(1)}) \eta(\tau^{(2)})} \, T(\a_1 , \t_1 ; \a_2 , \t_2) \, \nn \\
 T(\a_1 , \t_1 ; \a_2 , \t_2)  =  \sum_{m = 0}^{k-1}
\vartheta \left[ \begin{matrix}
\frac{m}{k} \\ 0
\end{matrix} \right] \left(k \alpha_1 , k \tau^{(1)} \right) \vartheta \left[ \begin{matrix}
\frac{m}{k} \\ 0
\end{matrix} \right] \left(k \alpha_2 , k \tau^{(2)} \right) \, ,
\eea
where we used the decomposition \eqref{decomp} and the wavefunctions/conformal blocks \eqref{toruswavefunctions}. This sum is reminiscent of an addition theorem for the $\vartheta$-functions/wave-functions. It is clear that in this example, the ``sector" label $S$ of eqn. \eqref{maineqn} is simply the integer $m$ labelling the wave-functions and that the combined source functional takes the form
\be\label{sourceCS}
\mathcal{Z} = \sum_{m=0}^{k-1} Z^{(1)}_m(J_1) Z^{(2)}_m(J_2)  \, ,
\ee
\be
 Z^{(i)}_m(J_i) = \int \mathcal{D} \left(  \chi_i, \alpha_i   \right) Z_{BCFT}^{(i)} \left(\chi_i,  \alpha_i  ; J_{i}  \right) \Psi^{(i)}_m(\chi_i, \alpha_i) \, ,
\ee
with $Z_{BCFT}^{(i)} \left( \chi_i , \alpha_i  ; J_{i}  \right)$ being each BCFT partition function in the presence of external sources\footnote{This is analogous to the twisted partition function of the MQM, eqn. \eqref{twistedMQM}.} and $\Psi^{(i)}_m(\chi_i, \alpha_i)$ the Chern-Simons boundary wave-functions \eqref{toruswavefunctions}, both evaluated on a torus. We  observe that the resulting two-point cross-correlators cannot exhibit short distance singularities being sums of products of one-point functions.

We close this section with some comments on the non-abelian version of the model. For the non-abelian Chern-Simons theory, the wave-functions on the boundary slices are Weyl-Kac characters $\chi_{\mu, k}(u, \tau)$ where $\mu$ labels the representation, $k$ is the level of the affine algebra and $A_{\bar{z}} = i u \tau_2^{-1}$~\cite{Elitzur:1989nr}. These characters are analogous to the ones we used in the study of the non-abelian matrix model of section~\ref{HallLittlewood} and the sectors are again enriched into integrable representations/partitions of the affine Lie algebra of the group $H$. The resulting sums are again reminiscent of addition theorems. The conclusions about the general form \eqref{maineqn} of the partition function and the UV softness of cross correlation functions remain unchanged.

\subsection{Other higher-dimensional examples}

Moving one dimension further up, a promising setup is to consider two holographic three-dimensional gauge theories coupled via a ``messenger" four-dimensional theory of the BF-type~\cite{Blau:1989bq,Horowitz:1989ng} in analogy to the simple 2d BF 1d MQM model. In condensed matter physics, such BF-theories may describe the low energy physics of topological insulators (see~\cite{Cho:2010rk,Qi:2011zya} and references within). Therefore, such two-boundary models could also be of interest for more practical applications of bi-layered materials.

For concreteness, we consider a pair of holographic 3d field theories (for example of ABJM type) and the 4d BF theory on a space with the topology of a manifold times an interval $\mathcal{M} = \Sigma \times I$ and symmetry group $G$
\be
S_{BF} = \int_{\mathcal{M}}  \Tr F \wedge B + \frac{1}{1 2 g} \Tr B \wedge B \, ,
\ee
where we again define the restriction of the gauge field on the two boundaries $A\vert_{\Sigma_{1,2}} = A_{1,2}$ corresponding to the two three-dimensional gauge fields\footnote{This case is more intricate, since the gauge field can have non-trivial dynamics on the three dimensional boundaries.}. It is clear that the list of examples does not stop here, but one may consider similar setups in any number of dimensions. The next example is that of two four-dimensional theories of the YM type (like $\mathcal{N}=4$ SYM) coupled by an intermediate five-dimensional Chern-Simons theory and so on. For this latter case, see also section~\ref{freefieldmatrix}.

\section{Correlators in a simple model of cross-coupled QFTs}\label{crosscoupledsimple}

\subsection{The Euclidean theory}

In this section we slightly modify our setup and consider the case of a non-self interacting messenger theory coupled to fields that belong to the two $d$-dimensional boundary theories. The difference with the previous topological messenger theories, is that now the messengers do have physical propagating degrees of freedom.

The basic question we would like to address is regarding the form of the combined source functional (analogue of \eqref{maineqn}) and as a consequence whether the cross correlators remain UV soft when the messenger theory has non-trivial propagating states. This theory realizes the idea of \cite{worm} as well as the higher-dimensional messenger theory suggestion of \cite{VanRaamsdonk:2020tlr} in a weakly coupled context.
Our motivation is to settle the issue of the source functional and the nature of the cross correlators in an explicit example\footnote{We choose to study the case of the most direct cross interactions between the various sector, should this model exhibit the desired features, then less direct interactions involving bifundamental fields etc. are guaranteed to do so.}. We shall first analyse the case of free $d$-dimensional boundary theories and relegate the discussion of self-interacting boundary theories in section~\ref{boundaryselfinteractions}.

Consider two scalars $\phi_{1,2}$ propagating in $d$ dimensions interacting with a $(d+n)$-dimensional scalar $\Phi$ via bilinear interactions\footnote{This model can be easily generalised in the case where the fields are matrices as we explain in the end of this section and section~\ref{freefieldmatrix}.}. We keep the field $\Phi$ massive to avoid any possible issue with IR divergences. We shall denote by $x$ a point in the $d$ dimensions and by $y$ a point in the $n$ extra dimensions. The scalar $\phi_1$ is localised at a point $y_1$ in the extra $n$ transverse directions while the scalar  $\phi_2$ is localised at a point $y_2$ in the extra $n$ transverse directions so that their distance is $|y_1-y_2|=L$.

The relevant action is
\be\label{action1}
S=S_1(\phi_1)+S_2(\phi_2) + S_3(\Phi) +S_{cross-int}
\ee
with
\be
S_{1,2}=-{1\over 2}\int d^d x ~\phi_{1,2}(x)\left(\square_d-  m^2\right)\phi_{1,2}(x)
\ee
\be
S_3=-{1\over 2}\int d^d xd^n y ~\Phi(x,y)\left(\square_{d+n}-  M^2\right)\Phi(x,y)
\ee
\be\label{intpart}
S_{int}=g\int d^dx ~\phi_1(x)\Phi(x,y_1)+g \int d^dx ~\phi_2(x)\Phi(x,y_2)
\ee
Passing to momentum space and integrating out the scalar $\Phi$ we obtain
\be
S=\int{d^dp\over (2\pi)^d}\left[\phi_1(p)(p^2+ m^2)\phi_1(-p)+\phi_2(p)(p^2+ m^2)\phi_2(-p)
+\right.
\ee
\be
+\left.
g^2\int {d^n q\over (2\pi)^n}{  \phi_1(p) \phi_1(-p)+ \phi_2(p) \phi_2(-p)+ \phi_1(p) \phi_2(-p)[e^{iq\cdot (y_1-y_2)}+cc]\over
p^2+q^2+ M^2}\right]
\label{a9}\ee
There are two types of terms. Those that correct each individual theory and those that couple the two theories. The terms that correct each individual theory are finite when $n=1$ and contain UV divergences when $n\geq 2$. We henceforth assume $n=1$ to continue and find
\be
S=\int{d^dp\over (2\pi)^d}\left[\phi_1(p)\left(p^2+ m^2+ g^2 \Sigma_{11}(p) \right)\phi_1(-p)+
\right.
\ee
\be
+\phi_2(p)\left(p^2+ m^2+ g^2  \Sigma_{22}(p) \right)\phi_2(-p)
+\left.
{g^2} \phi_1(p)\phi_2(-p) \Sigma_{12}(p)   \right] \, ,
\label{a10}\ee
where we can use the following integrals to determine the self-energies
\be\label{self1}
\int_{-\infty}^{+\infty}{dq\over 2\pi}{1\over q^2+p^2+ M^2}={1\over 2\sqrt{p^2+ M^2}} \equiv \Sigma_{11}(p) \equiv \Sigma_{22}(p)
\ee
\be\label{self12}
\int_{-\infty}^{+\infty}{dq\over 2\pi}{[e^{iq\cdot (y_1-y_2)}+cc]\over q^2+p^2+ M^2}={e^{-L\sqrt{p^2+ M^2}}\over \sqrt{p^2+ M^2}} \equiv \Sigma_{12}(p)
\ee
This is the result when $y$ is non-compact and the $d+1$-dim field $\Phi$ has an infinite $y$-space to propagate.

On the other hand, if $y$ is a compact interval of length $L$, and $\Phi$ can only propagate inside the interval\footnote{This case can be further generalised by considering induced dynamics on the boundary ``branes" for the scalar $\Phi(x,y)$. Then one has to consider Neumann bc's on the two boundaries.} we find instead the self-energies (these results can be also reproduced upon analytically continuing the transition amplitude computation of appendix~\ref{freetrans} into Euclidean signature - eqn. \eqref{Dirichleteffective})
\be\label{intervalself}
 \frac{1}{2 L(p^2 + M^2)} + \frac{1}{L}\sum_{n=1}^\infty {1\over \left(\frac{ \pi n}{L} \right)^2 +p^2+ M^2}= {\coth (L \sqrt{p^2 + M^2}) \over 2 \sqrt{p^2+ M^2}} \equiv \Sigma^I_{11}(p) \equiv \Sigma^I_{22}(p)
\ee
\be\label{intervalself12}
 \frac{1}{2 L(p^2 + M^2)} + \frac{1}{L}\sum_{n=1}^\infty {(-1)^n \over \left(\frac{ \pi n}{L} \right)^2 +p^2+ M^2}={\sinh^{-1} (L \sqrt{p^2 + M^2}) \over 2  \sqrt{p^2+ M^2}}  \equiv \Sigma^I_{12}(p) \, .
\ee
We observe that the self-energies for the interval reduce to the ones of the infinite $y$-space as $L \rightarrow \infty$ and that their qualitative features are very similar for any value of $L$. Nevertheless, there do exist important subtle differences between these two cases. We shall discuss and analyse them in the places when they are relevant. In particular, such differences are very important when trying to implement two possible analytic continuations into Lorentzian signature, in section~\ref{anboundary} and~\ref{anrad}.

We shall now prove that this setup has all the expected properties for the correlation functions of $\phi_{1,2}$.
As a first step, we rediagonalize this action to cast it in the form
\be
S=\int{d^dp\over (2\pi)^d}\left[\phi_+(p)D_+(p)\phi_+(-p)+\phi_-(p)D_-(p)\phi_-(-p)\right]
\label{a11}\ee
with
\be
\phi_{\pm}={\phi_1\pm\phi_2\over \sqrt{2}}\sp D_{\pm}(p)=p^2+ m^2+{g^2~\left(1\pm e^{-L\sqrt{p^2+ M^2}}\right)\over 2\sqrt{p^2+ M^2}}
\label{a12}\ee
We would like now to arrange that the Euclidean theory is well defined and for this we must have that
$D_{\pm}(p)$ is regular for all physical Euclidean momenta ($p^2\geq 0$), and that the position space propagators are reflection positive, see also appendix~\ref{reflcpos}. Here we observe that $D_{\pm}(p)>0$ and regular for all values of $g,L$. In this case, the Euclidean propagators $D_{\pm}^{-1}$ are well defined and finite everywhere. In the IR we find
\be\label{a12b}
D_\pm(p) \sim  m^2 + {g^2~\left(1\pm e^{-L M}\right)\over 2 M} + O(p^2)  \, ,
\ee
while in the UV
\be\label{a12c}
D_\pm(p) \sim p^2 + m^2 + {g^2 \over 2 p} + O(p^{-2}) \, .
\ee
This is a common asymptotic behaviour consistent with dispersion relations and indicates that the propagators could admit a spectral representation. For more details on the relation between the spectral representation and reflection positivity, see appendix \ref{reflcpos}.

For the field correlators, we obtain
\be
\langle \phi_1\phi_1\rangle(p)\equiv G_{11}(p)=G_{22}(p)\equiv \langle \phi_2\phi_2\rangle(p)={D_{+}^{-1}(p)+D_{-}^{-1}(p)\over 2}=
\label{a14}\ee
$$
={p^2+ m^2+{g^2\over \sqrt{p^2+ M^2}}\over
\left(p^2+ m^2+{g^2~\left(1+e^{-L\sqrt{p^2+ M^2}}\right)\over 2\sqrt{p^2+ M^2}}\right)\left(p^2+ m^2+{g^2~\left(1- e^{-L\sqrt{p^2+ M^2}}\right)\over 2\sqrt{p^2+ M^2}}\right)}
$$
Therefore $G_{11}$ at short distances behaves as in the original theory
\be
G_{11}={1\over p^2+m^2+{\cal O}(p^{-4})}\sp p\to \infty
\label{a16}\ee
The same applies to large distances but the effective mass $\hat m^2$ is different
\be
G_{11}^{-1}=Z(p^2+\hat m^2)+{\cal O}(p^4)\sp p\to 0
\label{a25}\ee
with
\be
Z={(g^2+2m^2M)(6g^2M^3+4m^2M^4-g^4)+g^4\left(g^2(2LM+1)+2M(M^2+m^2(1+LM))\right)\over 8M^3(m^2M+g^2)^2 e^{2LM}}
\ee
\be
\hat m^2={2M^2 e^{2LM}(g^2+m^2M)((g^2+2m^2M)^2-e^{-2LM}g^4)\over (g^2+2m^2M)(6g^2M^3+4m^2M^4-g^4)+g^4\left(g^2(2LM+1)+2M(M^2+m^2(1+LM))\right)}
\ee
This is a signal of the IR-relevance of the interaction.

On the other hand,  the cross correlator of the original scalar operators
\be
\langle \phi_1\phi_2\rangle(p)\equiv G_{12}(p)={D_{+}^{-1}(p)-D_{-}^{-1}(p)\over 2}=
\label{a15}\ee
$$
=-{{g^2\over \sqrt{p^2+M^2}}e^{-L\sqrt{p^2+M^2}}\over
\left(p^2+ m^2+{g^2~\left(1+e^{-L\sqrt{p^2+ M^2}}\right)\over 2\sqrt{p^2+ M^2}}\right)\left(p^2+ m^2+{g^2~\left(1- e^{-L\sqrt{p^2+ M^2}}\right)\over 2\sqrt{p^2+ M^2}}\right)}\sim {e^{-Lp}\over p^5}
$$
is exponentially suppressed at short distances. This implies that in real space, $G_{12}(x-y)$ asymptotes to a constant as $x\to y$. This is similar to what was found in the holographic examples and the simple (non-local) model of~\cite{worm}. In addition,  we can also consider arbitrary composite operators of the form
\be
O^{1,2}_{m}\equiv :\phi_{1,2}^m: \, ,
\ee
and find that their mixed two-point functions
\be
\langle O^1_{m}O^2_{m}\rangle=\langle \left({\phi_++\phi_-\over \sqrt{2}}\right)^m\left({\phi_+-\phi_-\over \sqrt{2}}\right)^m\rangle={m!\over 2^m}\left(D^{-1}_{+}-D^{-1}_{-}\right)^m \, ,
\ee
are also very soft in the UV. The same applies to all similar local operators containing a finite number of derivatives.
When the messenger theory is defined on an interval, the mixed two-point functions again exhibit the same features.

What remains to be discussed are the reflection positivity properties of the various correlators (the two possible analytic continuations into Lorentzian signature are discussed further in section~\ref{Lorentziancase}), using the spectral representation of the two-point functions~\ref{reflcpos}.

\begin{figure}[t]
\begin{center}
\includegraphics[width=0.45\textwidth]{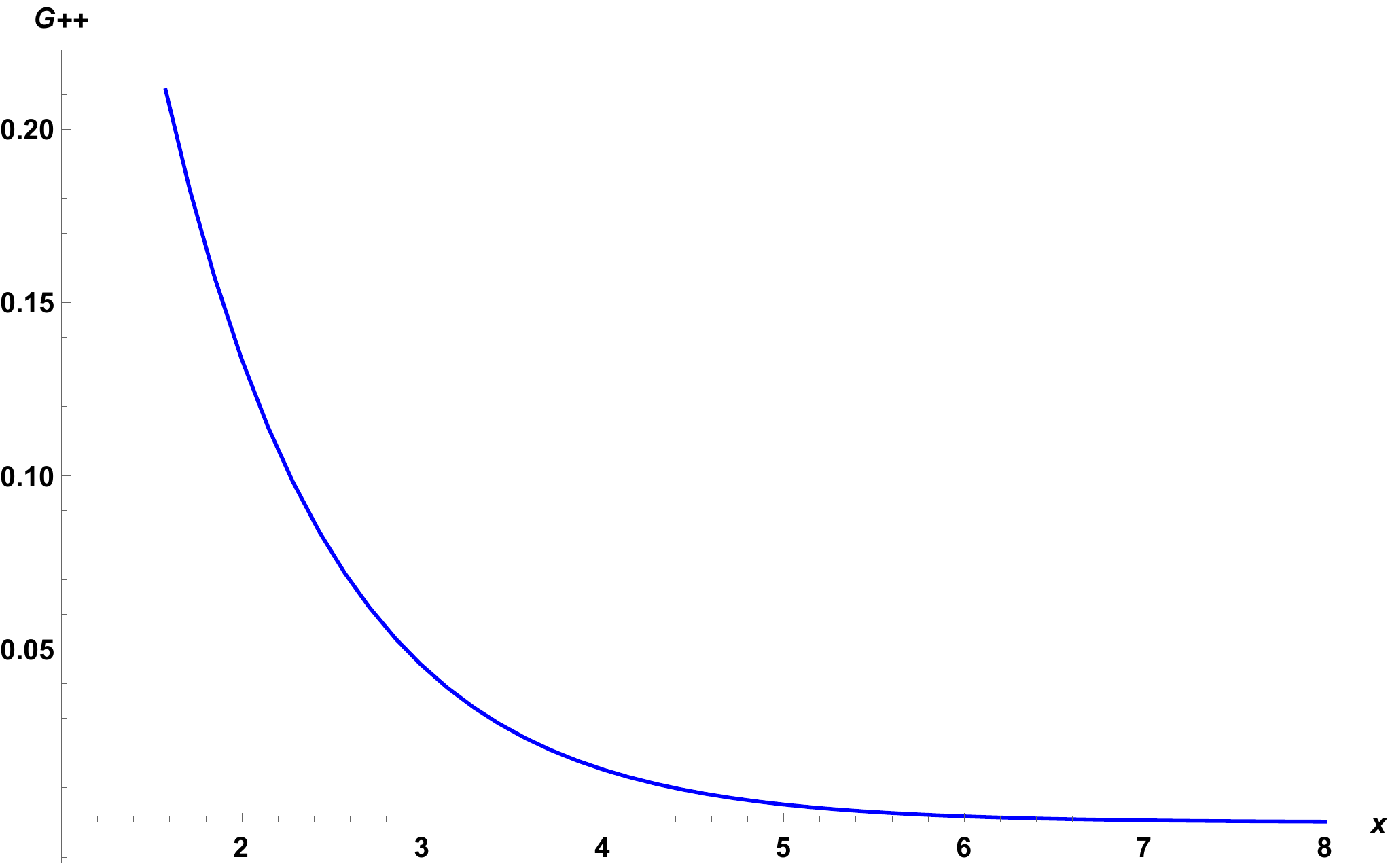} \hspace{0.1em} \hspace{0.1em} \includegraphics[width=0.45\textwidth]{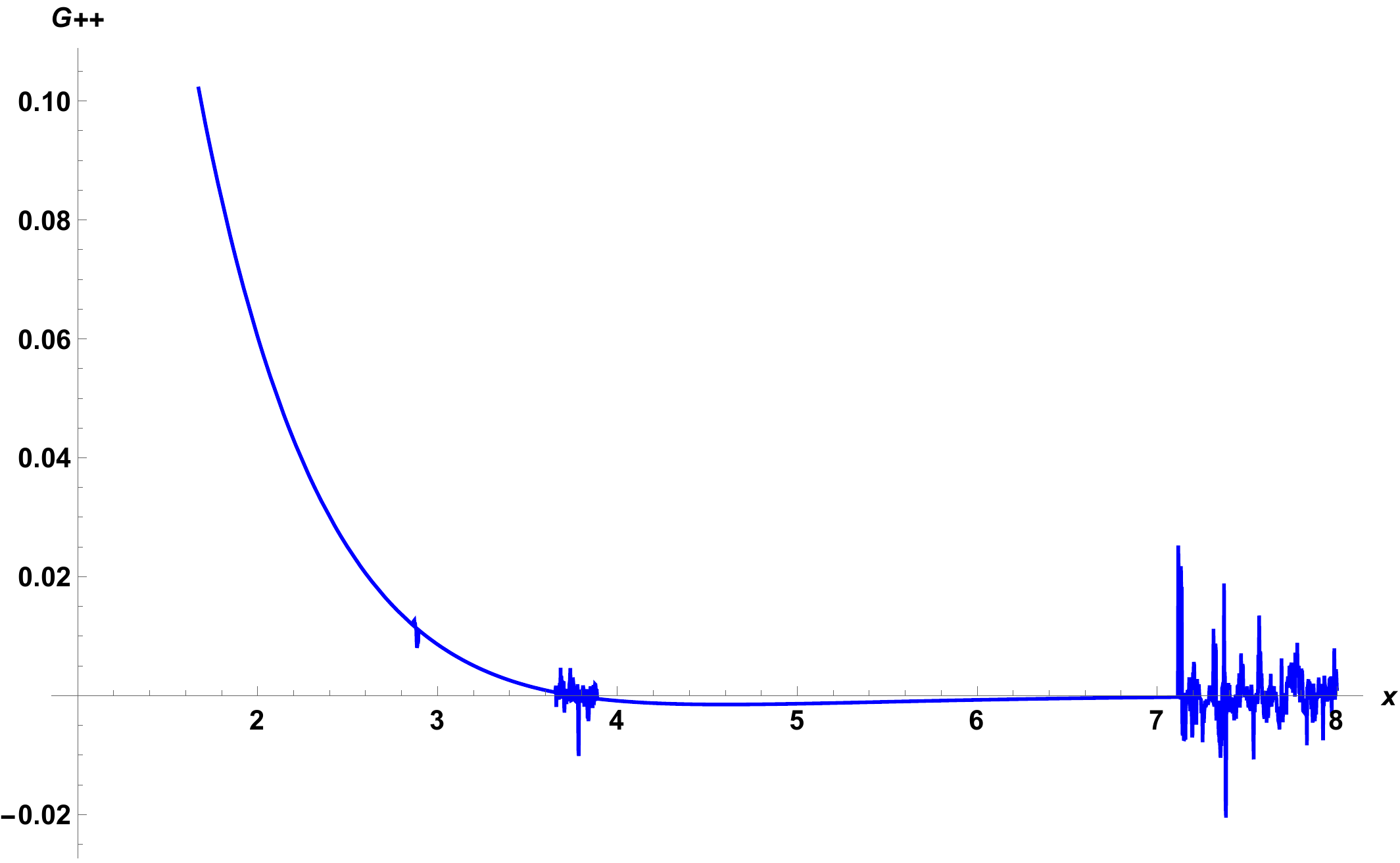}  \vspace{1em}
\includegraphics[width=0.45\textwidth]{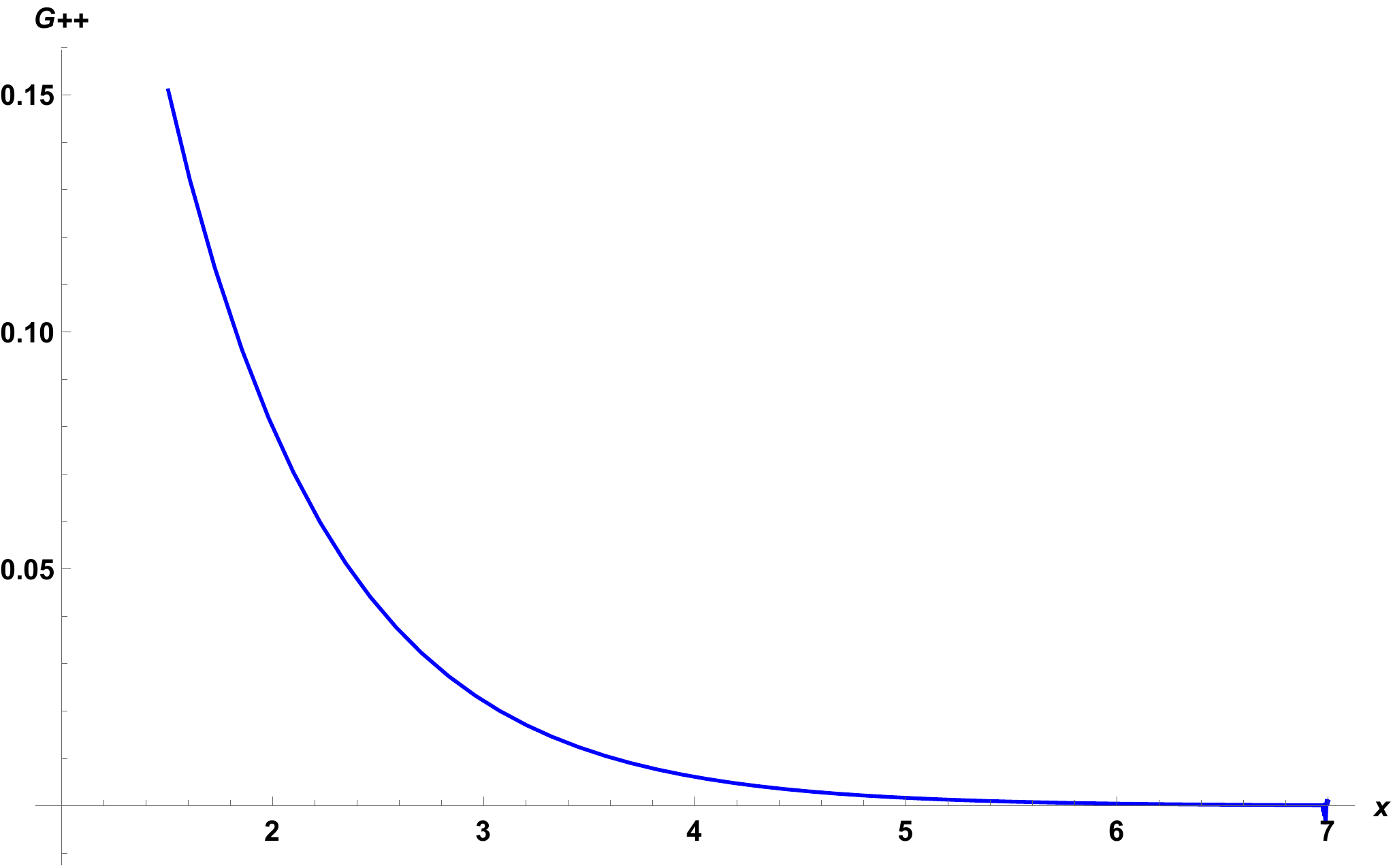} \hspace{0.1em} \hspace{0.1em} \includegraphics[width=0.45\textwidth]{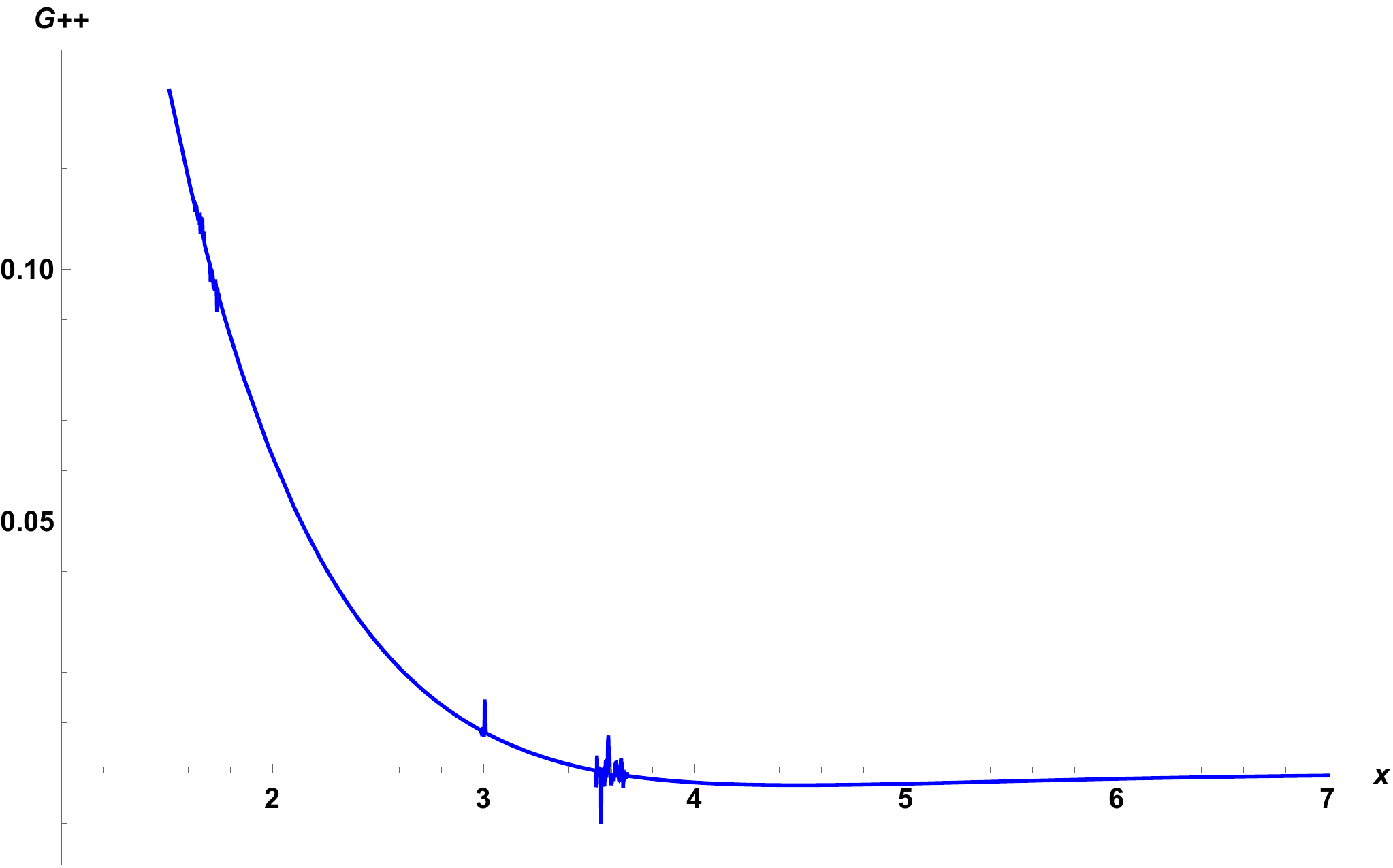}
\end{center}
\caption{The plots on the left are for $g = 1, m=1, M=10$. The plots on the right are for $g = 1, m=1, M=1$. The upper plots depict $G_{++}(x) = D_{+}^{-1}(x)$ (extra dimension is an infinite line), while the lower plots are for $G_{++}(x) = (D^I_{+})^{-1}(x)$ (extra dimension is a finite interval). The noise on the right side plots is in regions of large numerical error, but we can safely trust the complementary region, where the graph is smooth.
We observe a violation of reflection positivity in the regime $m \simeq M$ (and for values of $x$ for which we have acceptable numerical error). This violation is also dependent on the parameter $g$ and increases for large $g$. The $G_{--}(x)$ correlators are found to be reflection positive in the regime of parameters that we could numerically analyse. The results are similar to the analytic results of the non-local model~\cite{worm}.}
\label{fig:reflectionpositivity}
\end{figure}

The operators $\phi_1, \, \phi_2$ are good primary operators of the tensor product of the two decoupled theories $ \mathcal{H}_1 \times \mathcal{H}_2$. After including the cross interactions, a good basis is that constructed out of the $\phi_+ , \, \phi_-$ operators and the original factorisation does not hold anymore. In particular, the cross-correlator $\langle \phi_1\phi_2\rangle(p)$ eqn. \eqref{a15}, due to its exponential vanishing in the UV, does not admit a K\"allen-Lehmann spectral representation as shown in appendix~\ref{reflcpos}, and the propagating states after the interaction are the $+$ and $-$, that could admit such a representation, since $D_\pm(p)>0 $ (no zeros) for $p^2 >0 $, and with asymptotics \eqref{a12b}, \eqref{a12c}, that are consistent with dispersion relations. On the other hand, these conditions alone are not sufficient to guarantee reflection positivity as was exemplified in the non-local example of~\cite{worm}. This motivates a more careful study of the real space propagators in order to actually check reflection positivity. Unfortunately, it is possible to fourier transform the propagators only numerically. The features of the real space Euclidean propagators are depicted in figure~\ref{fig:reflectionpositivity}.

We find that the $D^{-1}_-(x)$ and $(D_-^I)^{-1}(x)$ propagators are indeed reflection positive, for various choices of parameters $m,M,g$ that we checked (these are measured in units of $L$ to form three independent dimensionless parameters). Remarkably there does exist a non-trivial parameter regime (when $m \simeq M$) in which $D_+(x)$ and $D^I_+(x)$ are not reflection positive\footnote{This behaviour for the $+$ propagator was also observed in the model of~\cite{worm}.}, as can be seen in the right-hand side of figure~\ref{fig:reflectionpositivity}. So the Euclidean theories we consider are consistent and satisfy the O.S. axioms~\cite{Osterwalder:1973dx} only for a subset of the parameters that define the model\footnote{Notice that as we increase the messenger mass $M$, or send $g \rightarrow 0$ the models are always well defined, since the cross correlation becomes increasingly weaker.}.

It would also be very interesting if one could construct a strongly coupled system, where one of the $\phi_+ , \, \phi_-$ operators gets ``confined" in the IR, ceases to propagate and does not appear in the effective IR Hilbert space. In this case one forms a bound confined state that ``liberates" at high energies (similar to a QCD-like bound state). The natural state to ``confine" in our example is $\phi_+$ in the regime where it does not admit a Kallen-Lehmann spectral representation. One could then make it disappear from the IR correlation functions and perhaps make sense of the parameter regime in which the free model is not reflection positive.\footnote{We remind the reader that in \cite{worm} we gave calculations and generic arguments that in wormholes the physics is always confining.}

We conclude this section by noting that this class of models can be readily upgraded into cross interacting matrix field theories by promoting the fields $\phi_{1,2}$ into $N$-dimensional hermitian matrices. If we keep the linear couplings of \eqref{intpart}, the rank of $\Phi$ is fixed to be $N$ as well, and hence all the quadratic terms scale in the same way\footnote{All the two point functions can be made to be of $O(1)$ upon rescaling the fields.}. Higher point interactions are then suppressed in an $1/N$ expansion as per usual. A richer possibility of inequivalent ranks can be obtained by promoting the matrix model of appendix~\eqref{bifundmatrix} into a matrix field theory (equivalently by adding ``messenger" bi-fundamental fields into the action \eqref{action1}). In the model of the appendix, the matrix $M(\tau, x)$ would then be a ($d+1$)-dimensional field, whilst the rest of the matrices (bi-fundamentals and boundary fields) are $d$-dimensional fields dependent only on $x$. It might be interesting to analyse in more detail this class of bi-fundamental models with propagating messengers especially in the limit $N \rightarrow \infty$, $n$-finite (the large $N,n$-limit has a similar behaviour to the model \eqref{action1}). This is useful as a comparison to the topological messenger model of section~\ref{HallLittlewood}, for which non trivial cross correlators only appear in subleading order in this limit (naive $n,N$ counting suggests a similar behaviour for the propagating messenger model as well).

\subsection{The Lorentzian Theory}\label{Lorentziancase}

As we mentioned in the introduction, one can consider two possible analytic continuations of the Euclidean theory. One  is along the boundary directions and the other is along the extra dimension in which only the messengers can propagate. We are after two types of behaviour, negative residues near poles of the propagator (negative spectral weight or ghost-like ``pathological" behaviour) and the presence of complex poles or ``tachyonic" type of instabilities (which could nevertheless have positive residues). Similar comments apply for branch-cuts that describe a continuum of states.

\subsubsection{Analytic continuation along the boundary directions}\label{anboundary}

In this case, we can analytically continue $D_{\pm}(p)$ and $D^I_{\pm}(p)$ from eq\eqref{a12} to find the Lorentzian inverse propagators by sending $p_{0}\rightarrow i \omega$. We find (we define $s = \omega^2 - \vec{p}^2$ that can be both positive and negative)
\bea
D_{\pm}(s)=- s + m^2+{g^2~\left(1\pm e^{-L\sqrt{- s + M^2}}\right)\over 2\sqrt{-s+ M^2}} \nn \\
D^I_{+}(s)=-s+ m^2+ g^2 \left(  {\coth (L \sqrt{-s + M^2}/2) \over 2 \sqrt{-s+ M^2}} \right) \nn \\
D^I_{-}(s)=-s+ m^2+ g^2 \left(  {\tanh (L \sqrt{-s + M^2}/2) \over 2 \sqrt{-s+ M^2}}  \right)  \nn \\
\label{DLor}\eea
We observe that for $D_\pm(s)$ the point $s = M^2$ is a branch/threshold point. Crossing the branch cut we usually move to a different sheet of the $s$-plane (denoted as the ``unphysical sheet").

Unfortunately,  this branch point is transcendental, in the sense that the expansion of the exponential yields an infinite collection of terms with branch points of all possible orders. It is therefore impossible to determine the spectral weight using the discontinuity formula, even in the parameter regime where we expect spectral positivity due to reflection positivity of the Euclidean propagator. In addition,  for $s = \omega^2 - \vec{p}^2 \geq M^2 $, the exponential terms become oscillatory. For masses in the range where the Euclidean model is well defined, we find one pole in the propagator that exists in the real axis
with positive residue, validating the O.S. theorem of the possibility of analytically continuing reflection positive theories~\cite{Osterwalder:1973dx}.
In the regime $m \simeq M$, where reflection positivity fails, the poles lie on the complex plane and have complex residues, so the theory has ghosts. We conclude that the main trouble with the Lorentzian version of this model (for all the parameter regimes) is the transcendental cut having no obvious Lorentzian interpretation.

On the other hand,  for $D_\pm^I(s)$, the discontinuity due to the square root is cancelled. Hence the interval propagator admits a Laurent series expansion in the complex $s$ plane. This means that we only need to understand the structure of poles/singularities of the propagator and whether they admit positive residues. Once more, in the parameter regime where the Euclidean theory was found to be well defined and reflection positive, the poles are on the positive real axis and with positive residue, (the case when the Euclidean model is not reflection positive is similarly sick).  Hence we can make sense of the model on the interval, even in Lorentzian signature due to the cancellation of the transcendental branch cuts and the normal behaviour of poles in the propagator.

We conclude that the ``sickness" of the boundary Lorentzian theory in the model on the infinite extra dimension, has to do with the fact that there seems to be a  huge continuum of modes (that condense in a transcendental branch cut), induced on the lower-dimensional Lorentzian theories from the messenger that can propagate on an infinite dimension\footnote{Though perhaps at strong coupling one can make such branch cuts to disappear.}. On the other hand,  for the theory on the finite interval there is no such issue, since the extra-dimensional modes are bound on an interval and hence discrete, the resulting Lorentzian theory having a propagator with healthy poles only.

 Additionally, in the case of the interval, one expects the presence of a natural negative vacuum energy due to the Casimir effect. All these properties make more probable the construction of traversable wormholes in similar finite interval field theoretic setups, rather than in the case where the extra dimension is of infinite extend. Some related comments on the possibility of creating traversable wormholes by exploiting the Casimir effect in a similar fashion can be found in~\cite{May:2021xhz}.

\subsubsection{Analytic continuation along the messenger direction}\label{anrad}

The other possibility is to analytically continue along the additional radial direction on which only the messenger field can propagate. In this case one has to perform the continuation in the original description of the model eqn. \eqref{a9}, by rotating $q = i \omega$ (and using Lorentzian/real time propagators for the messenger theory).
This analytic continuation corresponds to a Lorentzian messenger theory, coupled to two Euclidean boundary theories at two distinct points in time.

The result takes the form of a Lorentzian transition amplitude, in the presence of Euclidean sources (the fields $\phi_{1,2}$) at the endpoints in time (we define $\Delta T = (t_2 - t_1)$). Of course there are various choices on how to perform this analytic continuation and boundary choices for the messenger field $\Phi(x, t)$. The boundary theories of the fields $\phi_{1,2}(x)$ nevertheless do remain Euclidean, which is a quite interesting generalisation of the ideas revolving around the dS/CFT proposal~\cite{Hull:1998vg,Strominger:2001pn,Witten:2001kn,Maldacena:2002vr,Anninos:2011ui}, since in our setup the two Euclidean boundary QFT's are cross-interacting\footnote{It is also possible to introduce further self interactions for each Euclidean QFT as shown in section~\ref{boundaryselfinteractions}}.

The real time free messenger transition amplitude in the presence of sources is written as
\be\label{free1}
\left\langle\Phi(t_b)\vert\Phi(t_a)\right\rangle^{J} = \int_{\Phi(t_a)}^{\Phi(t_b)} \mathcal{D}\Phi \,  \exp\left[i\int_{t_a}^{t_b}dt\int d^{d}x \left(\frac{1}{2}\left(\partial_{\mu}\Phi\partial^{\mu}\Phi- M^2 \Phi\right)+J\Phi\right)\right] \, .
\ee
The details of the computation of this transition amplitude are given in appendix~\ref{freetrans}. We shall use Dirichlet boundary conditions that fix the values of the field $\Phi(t_a), \Phi(t_b)$ at the end-points in time, see \eqref{DirichletGreen} for the relevant Green's function.

\begin{figure}[t]
\begin{center}
\includegraphics[width=0.45\textwidth]{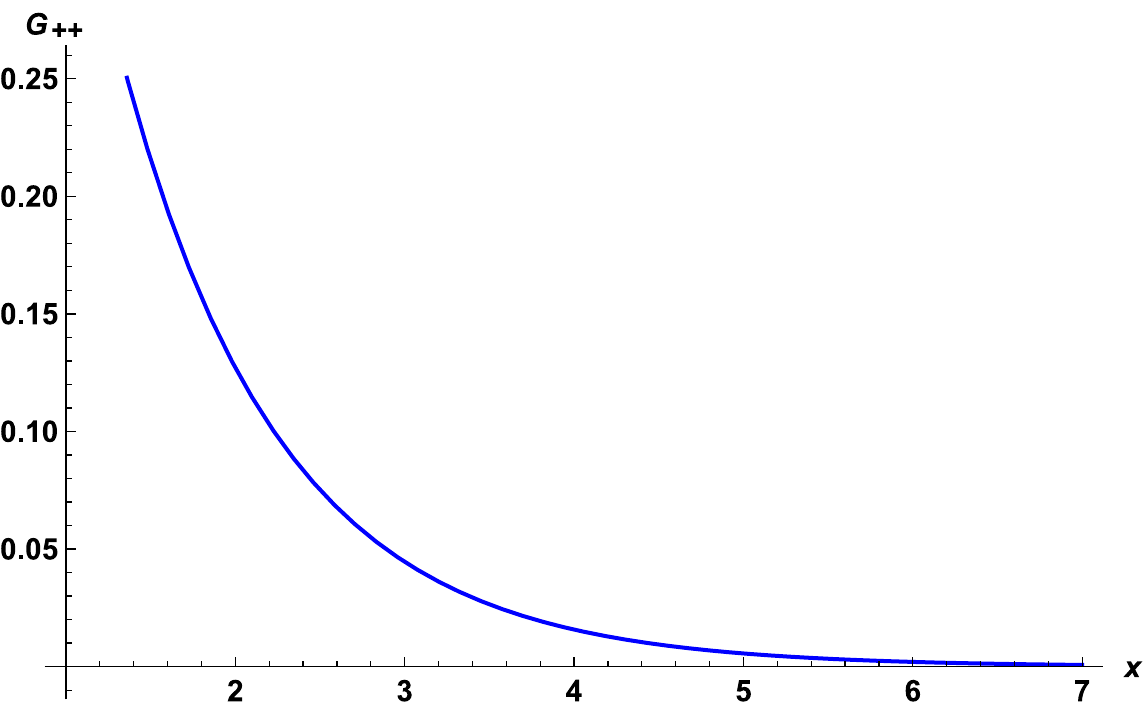} \hspace{0.1em} \hspace{0.1em} \includegraphics[width=0.45\textwidth]{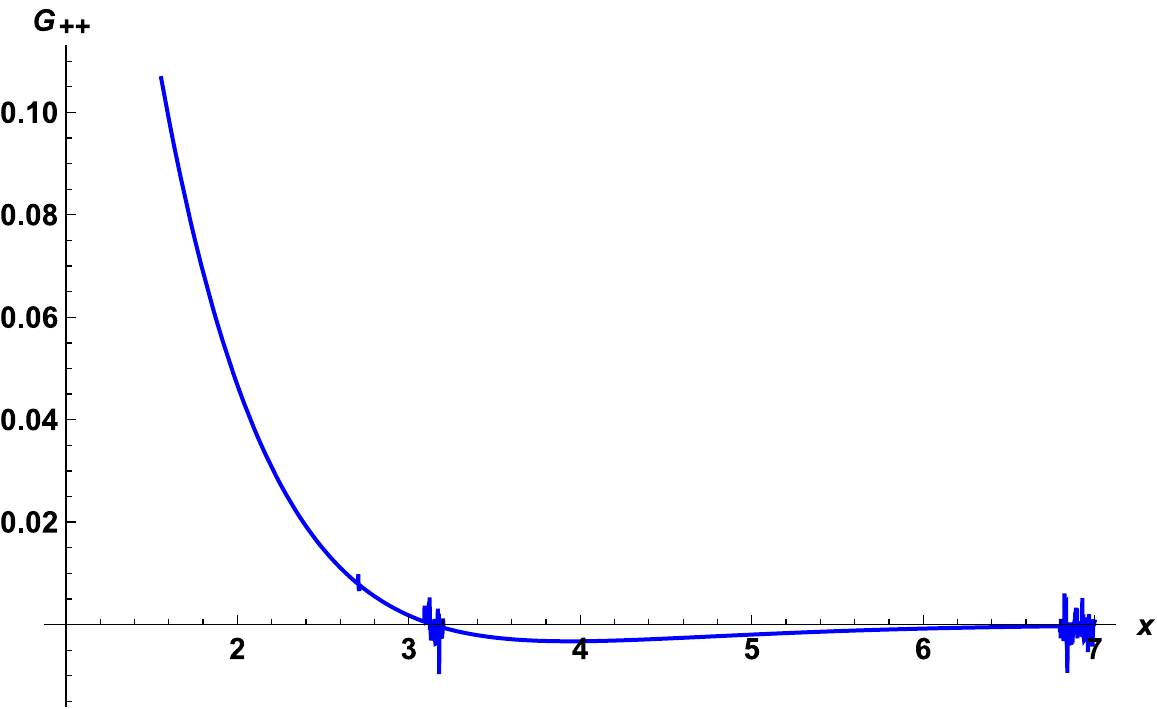}  \vspace{1em}
\includegraphics[width=0.45\textwidth]{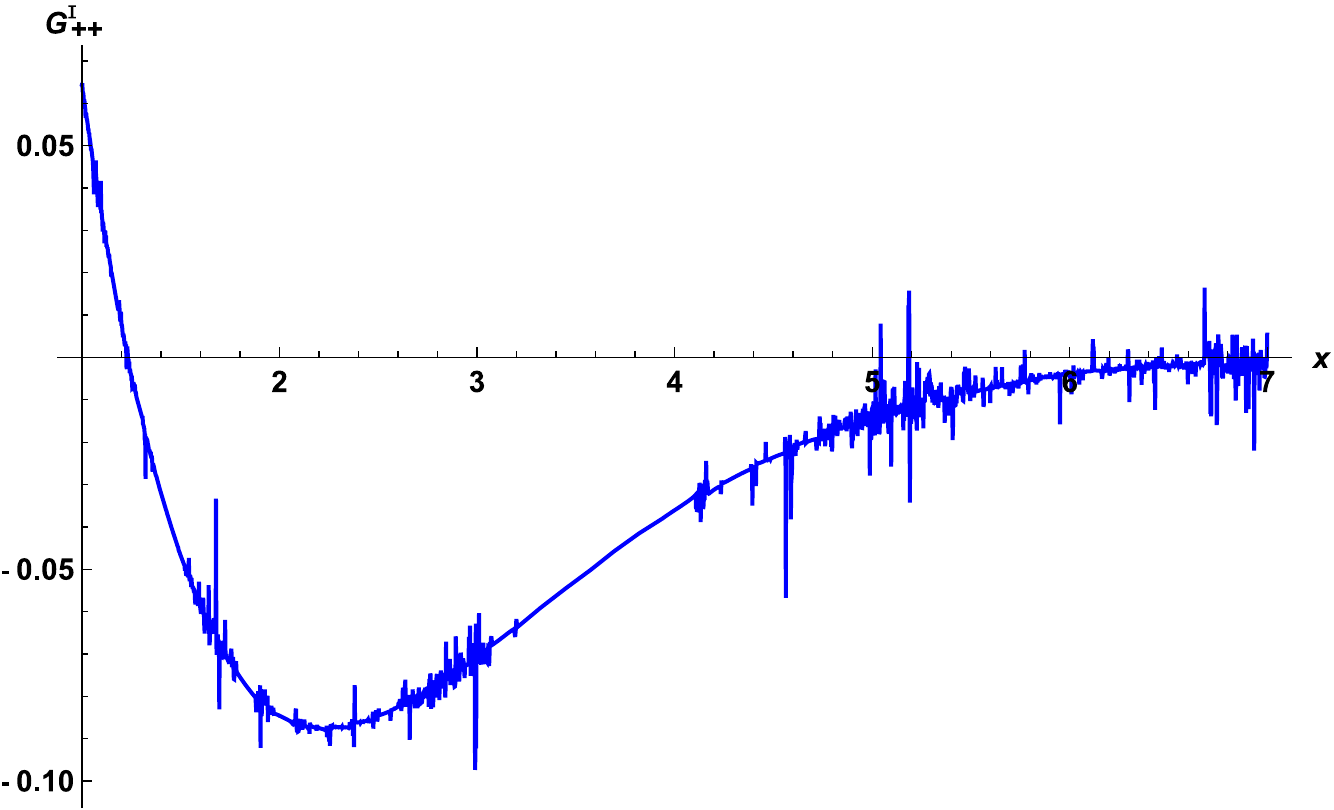} \hspace{0.1em} \hspace{0.1em} \includegraphics[width=0.45\textwidth]{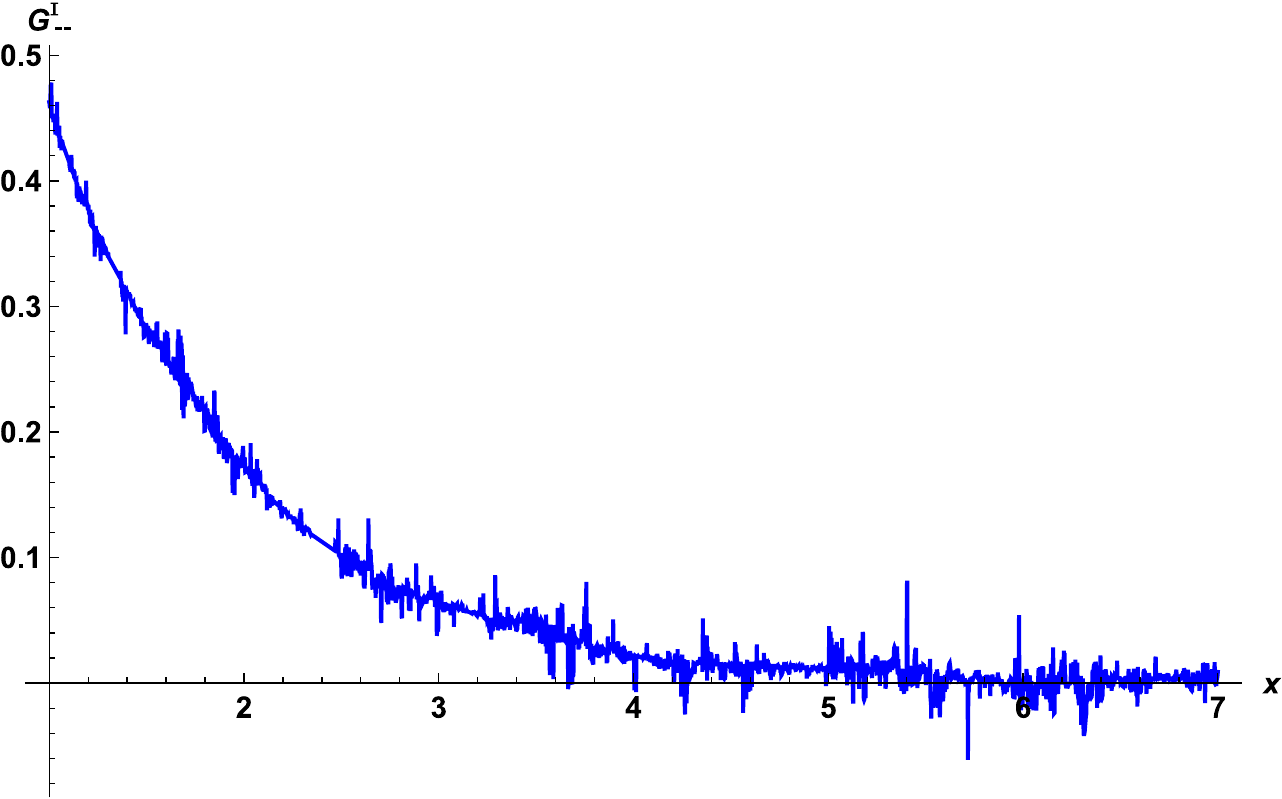}
\end{center}
\caption{The upper plots depict $G_{++}(x) = D_{+}^{-1}(x)$ (extra dimension is an infinite line). The upper left figure is for $g = 1, m=1, M=10$, showing the reflection positivity of the propagator, while the upper right is for $g = 1, m=1, M=1$, where it is not reflection positive. The lower left plot is for $G^I_{++}(x) = (D^I_{+})^{-1}(x)$ (extra dimension is a finite interval), while the lower right plot is for $G^I_{--}(x) = (D^I_{-})^{-1}(x)$ both for $g = 1, m=1, M=10$, where they used to be reflection positive in the case of Euclidean messenger propagation. We find relatively large noise and numerical error, since the integrand is highly oscillatory near the regions in which it diverges. We observe a clear violation of reflection positivity for the  $G^I_{++}(x)$ propagator, in the interval where the curve is smooth, while in the case of the $G^I_{--}(x)$ the ratio error/value is large and we cannot safely draw a conclusion.
}
\label{fig:reflectionpositivityL}
\end{figure}

Once the messenger theory is integrated out, instead of eqn. \eqref{a11}, one finds the diagonalised action
\be
S=\int{d^dp\over (2\pi)^d}\left[\phi_+(p)D_+(p)\phi_+(-p)+\phi_-(p)D_-(p)\phi_-(-p)\right]
\label{a11L}\ee
with
\bea
D_{\pm}(p)&=& p^2+ m^2+{g^2~\left(1\pm \cos ( \Delta T \sqrt{p^2+ M^2})\right)\over 2\sqrt{p^2+ M^2}} \, , \nn \\
D^I_{+}(p) &=& p^2+ m^2+ g^2 \left(  {\cot (\Delta T \sqrt{p^2 + M^2}/2) \over 2 \sqrt{p^2+ M^2}} \right) \, , \nn \\
D^I_{-}(p) &=& p^2+ m^2+ g^2 \left(  {\tan (\Delta T \sqrt{p^2 + M^2}/2) \over 2 \sqrt{p^2+ M^2}}  \right) \, .
\label{a12L}\eea

One common property of these inverse propagators is that the UV and IR behaviour \eqref{a12b}, \eqref{a12c}, is only changed in subleading terms and the asymptotic conditions for dispersion relations still hold. Nevertheless, we once more observe qualitative differences between the model with an infinite extend and finite interval.
The difference that is always present is the branch cut of the infinite extend model. Again this branch cut does not affect the Euclidean propagators in the physical regime $p^2 >0$, since it still starts at $p^2 = - M^2$. In addition in the first case of \eqref{a11L}, the Euclidean propagators remain positive definite, but oscillatory as one varies $p$ in the physical region $p^2 > 0$, while in the second finite interval case we loose this positivity property for $D^I_\pm(p)$ completely. This means that the Euclidean propagators (for the interval) diverge at points belonging to the physical regime $p^2 > 0$. It can be checked that the divergences  do not to correspond to simple poles, indicating a pathology of the interval theory. Upon performing a numerical fourier transform in position space,  we find that the theory on the infinite line is reflection positive for a subset of parameters, while the one on the interval is not (in particular the $(D^I_{+})^{-1}(x)$ propagator), even in the regime where it used to be. The results are summarised in the plot ~\ref{fig:reflectionpositivityL}.

We shall close this section with a speculative comment. Contemplating the fact that the analytic continuation of the Euclidean wormhole geometries gives rise to Bang/Crunch type of Cosmological manifolds, for which the energy conditions cease to hold it is natural to expect that holographic duals of such geometries would indeed suffer from some kind of ``pathology". If the duals of such geometries are Euclidean as in the dS/CFT correspondence, then it is natural to expect that these Euclidean theories would violate the usual O.S. axioms for Euclidean QFT's~\cite{Osterwalder:1973dx}.
Perhaps the violation of reflection positivity we observe in the interval model with Lorentzian messenger (whose cross correlators are sufficiently UV soft when the messenger theory is Euclidean - consistent with holographic cross correlators on Euclidean wormhole geometries), is an indication that this specific analytic continuation along the messenger dimension is related to the analytic continuation in the bulk geometry. In order to actually check this idea one would have to study a top down example of the tripartite systems, with a known holographic dual. In the next section we shall improve on our toy model by including the effects of self interactions for each boundary QFT.

\subsection{Including self interactions}\label{boundaryselfinteractions}

So far we assumed that both the messenger and the two boundary theories are non-self interacting. We would like to see what happens to the cross-correlators once we turn on self-interactions for the two boundary theories. This is especially important if we wish to eventually construct  models of strongly coupled (while weakly cross-coupled in the UV) QFT's with a potential semi-classical geometric dual admitting a wormhole saddle, see section~\ref{freefieldmatrix}. As can be seen from eqn. \eqref{a9} and \eqref{a10}, the effects of the messenger theory,  once integrated out in the path integral,  is to correct the propagators of the individual theories and to induce a UV soft cross interaction. These actions now have to be supplemented with additional self-interaction terms for each individual boundary theory, that introduce loop integrals, so one has to ascertain the effects of loop corrections to the cross interaction.

Before treating this problem in full generality, we assume simple cubic vertices for $\phi_{1,2}$ with couplings $g_3^{(1,2)}$. We then find the first correction to the two-point cross correlator to be determined from a one loop diagram that evaluates to
\be\label{cubicint}
 \langle \phi_1(p) \phi_2(-p) \rangle_{1} =    g_3^{(1)} g_3^{(2)} \int \frac{d^d q}{(2 \pi)^d} D^{-1}_{12}(q) D^{-1}_{12}(p-q) \, ,
\ee
with $D^{-1}_{12}$ read from \eqref{a14}. We would like to understand the UV asymptotics of the cross correlator as $p \rightarrow \infty$. In this case we find that the integrand is exponentially suppressed both in the large-$q$ and large-$p$ regions due to the properties of the cross-propagators $D^{-1}_{12}$. This means on one hand that the UV of the loop integral (large-$q$) does not give any divergence, rendering the result of the integration finite (no need for renormalisation). This means that we can simply consider the case $p \rightarrow \infty$ inside the integrand, to find that this particular one-loop correction to the cross-correlator is again exponentially suppressed at large $p$ similarly to the tree level one.

We can also make a qualitative remark to all orders. Since the cross communication is only induced via the cross-propagator, all possible loop integrals that contribute to cross-correlators inevitably contain a string of such cross-propagators, that vanish exponentially in the UV. We can restrict our discussion to graphs that are 1PI with respect to the cross propagator, else the exponential in the UV falloff is trivial and follows from the tree level computation. If the cross propagators are part of a subgraph containing a loop integral, we can safely obtain a finite result for this loop integral (due the exponential suppression in the UV that swamps any power law divergences\footnote{Assuming no IR divergences (a non zero messenger mass ensures this).}) that depends on the external momenta\footnote{Subgraphs that do not contain the cross propagators will exhibit UV divergences, but we assume that these are renormalised in the usual fashion for each individual boundary QFT.}. For consistency, the cross propagators can only appear in even pairs in loop integrals, so there will always be at least one additional external momentum for which the subgraph will exhibit exponential UV suppression (and due to momentum conservation this affects at least two legs of the subgraph). For the two-point cross-correlation function, all the subgraphs are glued together in a way that in the end only two external lines appear, each belonging to a different boundary theory. This means that inevitably, all the internal lines will in the end exhibit an exponential suppression in the UV, rendering the cross correlator sufficiently soft. This argument indicates that the result we found for the specialised interaction \eqref{cubicint} generalises to all loop orders at least for the two-point cross correlator.

We shall now formally recover this result, using generating functionals. Assuming that the messenger theory is not self interacting, we can simply integrate it out in the path integral to obtain an effective source functional for the boundary theories of the form
\be
Z[J_1(x_1), J_2(x_2)] = \int  \mathcal{D} \phi_1 \mathcal{D} \phi_2 \, e^{- S_1 - S_2 - S_{int} - \int_1 d^d x_1 \phi_1(x_1) J_1(x_1) -  \int_2 d^d x_2 \phi_2(x_2) J_2(x_2) }    \, .
\ee
In this case $S_{1,2}(\phi_{1,2})$ corresponds to a general interacting action for the field on the first/second boundary (with a self energy shift given by \eqref{self1} or \eqref{intervalself}). The only cross communication is induced by the bilinear cross interaction term
\be
S_{int} = g^2 \int \frac{d^d p}{(2 \pi)^d} \, \phi_1(p) \Sigma_{12}(p)  \phi_2(-p) \, ,
\ee
with $\Sigma_{12}(p)$ given by \eqref{self12} or \eqref{intervalself12} (in both cases it falls exponentially in the UV). This term can be decoupled using the Hubbard-Stratonovich trick (or delta function insertions) to yield the following expression for the source functional
\be
Z[J_1(x_1), J_2(x_2)] = \int  \mathcal{D} \zeta_1 \mathcal{D} \zeta_2 \, \, Z_1 [J_1 + i \zeta_1]  Z_2 [J_2 + i  \zeta_2 ]  \, e^{ - \frac{1}{g^2} \int \frac{d^d p}{(2 \pi)^d} \, \zeta_1(p) \, \Sigma_{12}^{-1}(p) \, \zeta_2(-p) }    \, ,
\ee
which is the equation that replaces the main result eqn.~\eqref{maineqn} that holds for topological messengers.

Any cross correlator is computed as an average of two individual pieces in the presence of ``dynamical" sources $\zeta_{1,2}$
\be\label{crosscorrelatorfreemess}
\langle O_{i_1} O_{j_1} \, ... \, \tilde{O}_{i_2} \tilde{O}_{j_2} \, ... \rangle = \int  \mathcal{D} \zeta_1 \mathcal{D} \zeta_2 \, \, \langle O_{i_1} O_{j_1} \, ... \rangle_{i \zeta_1} \langle  \tilde{O}_{i_2} \tilde{O}_{j_2} \, ... \rangle_{i \zeta_2} \, e^{ -\frac{1}{g^2} \int \frac{d^d p}{(2 \pi)^d} \, \zeta_1(p) \, \Sigma_{12}^{-1}(p) \, \zeta_2(-p) }
\ee
This means in particular that $\langle O_{i_1} O_{j_1} \, ... \rangle_{i \zeta_1}$ and $\langle  \tilde{O}_{i_2} \tilde{O}_{j_2} \, ... \rangle_{i \zeta_2}$, contain only external source legs $\zeta_1$ and $\zeta_2$ respectively. These legs are fused pairwise in a $1-1$ fashion, due to the nature of the single term in the exponent, and always involve an exponentially suppressed (in the UV) propagation for any momentum flow between the two theories.

 The individual pieces $\langle O_{i_1} O_{j_1} \, ... \rangle_{i \zeta_1}$ and $\langle  \tilde{O}_{i_2} \tilde{O}_{j_2} \, ... \rangle_{i \zeta_2}$, contain the usual short distance singularities, since as we observe from \eqref{self1} or \eqref{intervalself}, the self energies do not affect the UV properties of the propagators on each boundary theory (but are IR relevant and introduce an effective mass). On the other hand as we move the operator $O_{i_1}(x_1)$ from the first theory close to the operator  $\tilde{O}_{i_2}(x_2)$ of the second theory, we can instead choose to fourier transform with respect to $x_1-x_2$ and send the corresponding momentum $p$ to infinity. The cross interaction and the momentum flow inevitably involves a number of $\zeta_{1,2}$ propagators which as we mentioned above are all exponentially soft in the UV. All the loop integrals that involve these momenta are then finite (and at least one of them involves the momentum $p$). Assuming that the divergences on each individual boundary theory are regularised and renormalised, this means that the result is an expression that falls-off exponentially in large $p$, rendering the cross correlator UV soft.

We conclude that arbitrary self interactions on each individual boundary theory do not spoil the UV softness of cross correlators.

\subsection{A matrix version of the model}\label{freefieldmatrix}

While the model described in this section~\ref{crosscoupledsimple} does not have a simple holographic dual interpretation, it can be easily generalised to incorporate two boundary (matrix) theories that could have such a dual\footnote{This also relies on the fact that quantum corrections on each individually interacting theory continue to preserve the softness of the cross-correlators, see section~\ref{boundaryselfinteractions}, for a proof of this.}. For example, one could consider two
$\mathcal{N}=4$ SYM $U(N)$ gauge theories in $d=4$ (or any set of holographic theories with matrix field degrees of freedom) with a matter action of the form ($I = 1,...6$ in $\mathcal{N}=4$)
\be\label{SYMbos}
S_{1,2} =  \half \int d^4 x \Tr \left(D_\m X^I_{1,2} D^\m X^I_{1,2} + g^2_{1,2} [X^I_{1,2}, X^J_{1,2}]^2 \right) \, ,
\ee
and couple them with a simple five-dimensional matrix messenger field\footnote{Here we consider only couplings with the bosonic sector of the four-dimensional YM-theory, but this can be easily generalised to account for the fermions.} with an action of the form
\be
S_{m}= {1\over 2}\int d^4 x d y ~ \Tr \left( \Phi^I(x,y)\left(\square_{d+1}-\tilde M^2\right)\Phi^I(x,y) \right)
\ee
\be\label{intpart1}
S_{int}=g\int d^4 x ~  \Tr \left( X^I_1(x)\Phi^I(x,y_1) \right) + g \int d^4 x ~ \Tr \left( X^I_2(x)\Phi^I(x,y_2) \right) \, ,
\ee
where the two four-dimensional theories are placed at $y = y_{1,2}$. In this case the analysis is almost equivalent to the one presented in the previous section~\ref{crosscoupledsimple}, the only difference is keeping track of factors of $N$ due to the matrix structure and the presence of self interactions on each boundary. In particular equation \eqref{crosscorrelatorfreemess}, is generalised into
\bea\label{crosscorrelatorSYM}
\langle O_{i_1} O_{j_1} \, ... \, \tilde{O}_{i_2} \tilde{O}_{j_2} \, ... \rangle = \int  \mathcal{D}  \zeta^I_1 \mathcal{D} \zeta^I_2 \, \, \langle O_{i_1} O_{j_1} \, ... \rangle_{i \zeta_1} \langle  \tilde{O}_{i_2} \tilde{O}_{j_2} \, ... \rangle_{i \zeta_2} \, e^{ - S(\zeta_{1,2}) }  \, , \nn \\
 S(\zeta_{1,2}) =  \frac{1}{g^2}\int  \frac{d^d p}{(2 \pi)^d} \, \Tr \zeta^I_1(p) \, \Sigma_{12}^{-1}(p) \, \zeta^I_2(-p)
\eea
In this expression $\zeta^I_{1,2}$ are $N \times N$ matrix fields and $\Sigma_{12}^{-1}(p) = (\Sigma_{12}^{-1})_{i j  , k l}(p)$ is an appropriate matrix inverse propagator with UV soft properties. The individual averages should be computed in each SYM, in the presence of source terms for the elementary matrix fields $S_{1,2} = i \int d^4 x \Tr  X^I_{1,2}(x) \zeta^I_{1,2}(x) $. We observe that as expected,  the cross interaction appears to leading order in $N$, on the same footing with the other terms in the action \eqref{SYMbos}, since the coupling \eqref{intpart1} involves gauge group indices and not products of traces. The property of UV softness for the cross correlation functions follows according to the discussion of section~\ref{boundaryselfinteractions}. One should also take care, since the sources do not couple to color singlets and hence it is not possible at this stage to replace each individual expectation value with a supergravity dual description even at the 't Hooft limit.

We should mention a final possibility, that is to couple not only the matter fields $X^I_{1,2}(x)$ with the messengers $\Phi^I(x,y)$, but consider the addition of a five-dimensional Chern-Simons term for a gauge field $A_M(x,y)$, whose boundary values at $y_{1,2}$ correspond to the gauge field of the $4d$ $\mathcal{N}=4$ SYM theories\footnote{These higher dimensional examples are more intricate to define, since the gauge field has propagating degrees of freedom on the boundary. Since we wish to keep the boundary gauge field fluctuating and dynamical, we might need to add additional boundary terms to define a Neumann boundary value problem~\cite{Compere:2008us}.}. Such an extended model blends features of the models in this section and the topological messenger theories we considered in the first part of this work, nevertheless it is similarly not clear if it can admit a weakly coupled holographic dual.

\section{Conclusions}\label{conclusions}

As discussed in the introduction~\ref{introduction}, there are various possible avenues that one can explore with respect to the factorisation problem and the holographic interpretation of Euclidean wormhole solutions. It seems  that there are several possible ``resolutions", depending on the precise setup and assumptions one is imposing (number of dimensions, worldsheet vs. target space vs. replica wormholes, microscopic holography  vs. statistical/averaged computations). Taking also into account that we do not have access to a generic non-perturbative formulation of string theory, except indirectly through holography, we do believe that a solution that deviates from the very sharp and well studied paradigm of $\mathcal{N}=4$ as little as possible, and has the property of generalisability in several number of dimensions and setups, provides the most satisfactory resolution of this paradox.

In this work, we chose the approach that requires the most minimal modifications to the well-studied paradigm of AdS/CFT, in terms of a system of cross coupled QFT's~\cite{worm,VanRaamsdonk:2020tlr}, the coupling being mediated by a higher-dimensional theory of (quasi)-topological (or non self interacting) character. Remarkably, we found that these types of systems satisfy two non-trivial holographic requirements. The first is the form of the source functional of the combined system as an average over ``sectors" of individual source functionals (eqn. \eqref{maineqn}), which in turn were identified with group representations in the (quasi) topological examples we studied. The second is the UV softness in the behaviour of the two-point cross correlators, computed as averages over one-point correlators of each individual subsystem. An interesting feature of our (quasi)-topological proposal is that while being a unitary construction, at the same time exhibits some features analogous to those that appear when considering averages over boundary theories. On the other hand it is also crucially different, in the sense that the sum over representations is not (a priori) related to an average over couplings multiplying products of gauge invariant operators, but involves the gauge group structure and the constraints of the system\footnote{This also helps to evade issues with finding marginal couplings to average over.}.

While we do not know if some variants of the theories we analyse in this work can admit an explicit weakly coupled semi-classical holographic dual, with small $\a'/L^2_{AdS}$ ratio, we do believe that they deserve further exploration from a holographic perspective. In particular a D-brane picture would help to search for UV complete models, with better holographic control.

A possible D-brane construction proposed in~\cite{VanRaamsdonk:2020tlr}, involves the study of D-brane systems of the Gaiotto-Witten type~\cite{Gaiotto:2008ak}, where the role of the messengers is played by a stack of $D3$ branes connecting two widely separated stacks of $NS5$ and $D5$ branes. This system describes two $3d$ BCFT's coupled to a $4d$ $\mathcal{N}=4$ messenger theory on an interval~\cite{Nishioka:2011dq} and hence falls in the general category of systems that we consider in this work. This model is hard to study, especially in the proposed case where the widely separated stacks of $D5$'s and $NS5$'s have opposite orientation~\cite{VanRaamsdonk:2020tlr}, so that SUSY is broken by boundary conditions\footnote{A motivation behind these boundary conditions, was that the system would flow to a confining theory in the IR, so that the dual gravitational end of the world branes would bend and reconnect in the bulk, forming a wormhole. We provide below some additional discussion on wormholes and confinement.}. On the other hand a great simplification would be, if the low energy theory of the connecting $D3$'s could be replaced instead by a (quasi) topological theory. Perhaps it is possible to achieve this using ideas like the ones in~\cite{Bershadsky:1995qy,Kapustin:2006pk,Brunner:2021tfl}.

In this work we did not study at all the interesting setup of the ``gas of microscopic wormholes" and $\a$-parameters. We find that in that setup and in the derivation of the $\a$-parameter effective action, there do exist various assumptions, whose validity should be re-examined in detail~\cite{BKP}. On the other hand, in the bifundamental model~\ref{bifundmodel}, we found that in the limit $n \ll N_{1,2}$ where the messenger rank is much smaller to that of the boundary ranks, the two source functionals are only connected via subleading contributions that could have a dual interpretation as ``microscopic wormholes". Nevertheless there is no notion of $\a$-parameters appearing.

Another remark concerns the two-dimensional quantum gravity models that have attracted most of the recent attention due to their solvability. As described in~\cite{Betzios:2020nry,McNamara:2020uza}, these models can also be thought of as string theory models, the two-dimensional geometry corresponding to the worldsheet of the string. We therefore think that it is of utmost importance to analyse microscopic models of strings propagating on \emph{target space wormhole} geometries, so that one has a double topological expansion at hand. This was in part our motivation for studying the model of section~\ref{D0D1}, with which we can possibly achieve this feat (analysing further the grand canonical ensemble for the \emph{inverted oscillator}). So far our analysis mostly relied on the matrix model side, but we think that it is very important to revisit the gravitational equations of motion arising from the $c=1$ Liouville string, to see if they do admit more general wormhole types of solutions. In addition at the moment we do not know what the decompositions eqn. \eqref{maineqn}, and the explicit \eqref{pf1} and \eqref{bifundpartition} in terms of representations, mean from a worldsheet point of view. Other promising worldsheet string theory models (exept the $c=1$ string), for which it is straightforward to understand the target space geometry are WZW models (which are known to be able to incorporate semi-wormhole solutions~\cite{Bachas:1993kq}). Such models provide an important intermediate ``stepping stone", before tackling critical string theories.

We close with a few more important comments:

\paragraph{Higher dimensional messenger theories} - A higher-codimesion messenger theory can be used for our purposes\footnote{We thank the referee for raising this issue.}.
We would expect that the higher the dimension of the messenger theory, the weaker the interactions between the two boundary theories. On the other hand, we think that a messenger theory that is of the same dimension or lower than the other two, will generate interactions
that destroy the homogeneity of such boundary theories,  as the induced interactions must happen via lower-dimensional boundaries in this case.
When  the messenger theory is topological, the generalization seems straightforward.
For the case of a dynamical albeit weakly-coupled messenger theory,
discussed in section 4, one can also trivially generalize to a higher-dimensional
messenger theory, without important qualitative differences.

\paragraph{Multiple boundaries} - It is straightforward to generalise our class of models if we wish to describe manifolds with multiple asymptotic regions. One simply defines a multi-partite system composed out of a (quasi)-topological theory (such as BF theory or Chern-Simons theory) on a $d+1$ manifold $\mathcal{M}$ with n-boundaries. On each such boundary one places a holographic $d$-dimensional gauge theory coupled to the asymptotic value of the $d+1$ gauge field.

This naturally leads to a generalisation for the partition function that in the simplest examples is expected to take the form
\be\label{multipleboundpf}
Z \, = \, \sum_{S} \, e^{w(S)} \, Z^1_{S} \, Z^2_{S} \, ... \, Z^n_{S} \, ,
\ee
where $S$ labels the ``sectors" of each boundary theory\footnote{This is also reminiscent to the form of the partition function in terms of averaged theories studied in the literature, but here the average arises in a pefectly unitary single system setup (the label $S$ corresponds to representations and not a random Hamiltonian or disorder average). It is an interesting problem to compare and contrast the various approaches in more detail.}. In the case of the 2d YM theory on a sphere with $n$-holes coupled to n-copies of MQM, in analogy with the model of
section~\ref{D0D1}, the sectors correspond to representations and one simply finds a string of characters instead of
the simple case with the insertion of only two characters~\eqref{2dYMpf}. The partition function then does take the simple form \eqref{multipleboundpf}, with $S \rightarrow R$ a representation index and $w(S) \rightarrow C^{(2)}(R)$ the quadratic Casimir\footnote{One also needs to insert a factor $D_R^{2-2g-n}$ related to the dimension of the representation and the genus $g$ of the surface.}. Again the cross correlators do not exhibit short distance singularities being averages of lower point correlators of each boundary theory. For more complicated $(d+1)$-dimensional topologies we expect also the form \eqref{multipleboundpf} to become more complicated, a similar discussion in the context of Chern-Simons theory is~\cite{Balasubramanian:2016sro,Balasubramanian:2018por}.

\paragraph{Analytic continuation and Cosmologies} - As first proposed in~\cite{Maldacena:2004rf}, a radial analytic continuation of Euclidean wormhole geometries generically results in Cosmological space-times of the Big-Bang/Big-Crunch type. An appealing feature of the models studied in this paper is that it is possible that the relevant analytic continuation, does not involve the two Euclidean boundary QFT's, but merely the messenger theory connecting them.

The proposal then is that the $d+1$ dimensional bulk Cosmology is described by a tripartite system that involves two $d$-dimensional Euclidean theories connected via (the real time) transition amplitude of a $d+1$-Lorentzian, non-holographic messenger theory, between two boundary states~\cite{VanRaamsdonk:2020tlr}. Such systems therefore provide a generalisation of the dS/CFT proposal~\cite{Hull:1998vg,Strominger:2001pn,Witten:2001kn,Maldacena:2002vr,Anninos:2011ui}. We should also mention that a similar microscopic model of a Bang/Crunch universe in two dimensions was proposed earlier in~\cite{Betzios:2016lne}, but its relevance to the present setup became clear to us only recently (see appendix~\ref{orbifold}).

\paragraph{Wormholes and confinement} - Based on symmetry arguments and computations of expectation values of Wilson loop operators and cross-boundary loop correlators, it was argued in~\cite{worm} that two theories with a $U(N) \times U(N)$ factorised gauge symmetry in the UV, that cross interact in a way that only the diagonal $U_{diag.}(N)$ gauge symmetry remains intact in the IR could be good dual models of Euclidean wormholes (and Cosmologies as was exemplified in~\cite{VanRaamsdonk:2020tlr}). This fusion of the two gauge groups into a single one would be an avatar of a ``cross confining" type of physical behaviour. Confining theories are typically gapped in the IR and share this property with
(quasi) topological theories (even though they can generically differ in their ground state degeneracy properties). This was partly a motivation behind using a higher-dimensional (quasi) topological gauge theory as a messenger theory connecting the two boundary QFT's and ``tying" their individual gauge groups. For example in the MQM model of section~\ref{D0D1}, one does have to sum over the various $U_{diag.}(N)$ representations, but prior to the introduction of the messenger gauge field one has a factorised $U(N) \times U(N)$ gauge group structure (or a $U(N_1) \times U(n) \times U(N_2)$ structure for the more general bifundamental model of section~\ref{bifundmodel} ). Of course one might prefer to have a model where a ``cross-confining" behaviour would arise in a more dynamical fashion and not merely from selection and fusion rules of representation theory. Perhaps the Gaiotto-Witten type of models proposed in~\cite{VanRaamsdonk:2020tlr}, that we mentioned above, can exhibit such a behaviour.

\paragraph{On the factorization of the Hilbert space} - The fact that in Euclidean wormholes there is a bulk Gauss' law that correlates boundary global symmetries has implications for the structure of the total Hilbert space in the dual QFT. More strongly, from a boundary field theory perspective, we believe that there is a non-factorizability of the Hilbert space of the pair of interacting QFTs, similar to that appearing in non-abelian gauge theories. In non-abelian
gauge theories the Hilbert space does not spatially factorise into subregions (even though one can define commutant subalgebras of local operators), see~\cite{Casini:2013rba,Donnelly:2016auv} and references within. Nevertheless, it is true that one can embed this non-factorising Hilbert space into a larger factorising Hilbert space if additional edge degrees of freedom are introduced (that transform non-trivially under the symmetry group).

The precise statement in our construction is that while as in any other gauge theory the total state of the combined system is a ``global singlet", the subsystems on the two boundaries transform non trivially under $U(N)$. So one can consider the enlarged factorised Hilbert space $\mathcal{H} = \sum_{R_1} \mathcal{H}_{R_1} \otimes \sum_{R_2} \mathcal{H}_{R_2} $, with $|R_1 \rangle \otimes |R_2 \rangle$ describing a factorised state in representations $R_{1,2}$ respectively. After imposing the gauge constraint mediated by the topological messenger, there is a projection of the enlarged Hilbert space to the subspace containing only the states for which $R_1 = R_2$. This is not a tensor product anymore but takes the form of a direct sum of tensor products $\mathcal{H}_s = \sum_R \mathcal{H}^1_{R} \otimes  \mathcal{H}^2_{R} $\footnote{This could also be described via the use of intertwiners.}. We believe that this is what distinguishes the duals of Euclidean wormholes with those of the well understood Lorentzian black holes (ER bridge), which can be described by a simple tensor product Hilbert space of the two boundary CFT's (and whose Euclidean continuation factorises into the product of two cigar geometries).

\paragraph{Connection with black holes} - MQM models with non-singlet states are believed to be relevant for the description of long string excitations and black hole geometries in $c=1$ Liouville theory~\cite{Kazakov:2000pm,Maldacena:2005hi,Betzios:2017yms}. In this work we proposed that non-singlets are also relevant for the description of target space wormholes in Liouville theory\footnote{Again we emphasize the difference with the minimal and $c=1$ models studied in the recent literature, where the various topologies can also be interpreted as topologies of the worldsheet of a string. In this work we try to understand \emph{target space} wormholes (MQM automatically includes worldsheet wormholes and topologies~\cite{Betzios:2020nry}).}.

While we find some similarities with the model of non singlets in~\cite{Betzios:2017yms}, the partition function eqn. \eqref{bifundpartition} to the best of our knowledge does not correspond to that of an integrable system (it does not seem to be a $\tau$-function of a known hierarchy). This raises the interesting possibility that a variant of the model proposed here could also be relevant for the black hole - string transition. The picture we have in mind is that, for some parameter regime,  the saddles of such models correspond to the trivial linear dilaton background (or products of them in the case of multiple MQM copies), possibly with a long string condensate (star or ``fuzzball") on the background,  if the number of long string excitations is sufficiently large. This could correspond to an ``integrable phase" of the model, where the relevant low-energy excitations are described by an integrable model such as the Spin-Calogero type of model proposed in~\cite{Betzios:2017yms}. As we change parameters, the long-string condensate/star can undergo gravitational collapse and we enter the ``non-integrable phase" of the model,  whose leading saddles have a geometric interpretation in terms of black holes or wormholes\footnote{This is currently under investigation.}. Some simple models with similar integrable to chaotic behaviour can be found in~\cite{Zotos,Garcia-Garcia:2017bkg}.

\section*{Acknowledgements}\label{ACKNOWL}
\addcontentsline{toc}{section}{Acknowledgements}

We wish to thank Dio Anninos, Alex Belin, Adam Bzowski, Lorenz Eberhardt, Nava Gaddam, Victor Gorbenko, Diego Hofman, Kristan Jensen, Arjun Kar, Volodya Kazakov, Juan Maldacena, Francesco Nitti, Joao Penedones, Massimo Porrati, Mark van Raamsdonk, Theodore Tomaras and especially Costas Bachas and Ioannis Lavdas for useful discussions at various stages of this project.

We would also like to thank the organisers and participants of the online Workshop on Quantum Gravity, Holography and Quantum Information for stimulating discussions and comments on related topics and the anonymous referee for important questions and suggestions. E.K. would like to thank the organisers and participants of the PSI workshop for valuable comments. We acknowledge the hospitality of APC Paris, ICTP Trieste and the University of Crete, where a large part of this work was undertaken.

\noindent The research of P.B. is supported in part by the Natural Sciences and Engineering Research Council of Canada. P.B. and O.P. acknowledge support by the Simons foundation. Research at Perimeter Institute is supported in part by the Government of Canada through the Department of Innovation, Science and Economic Development and by the Province of Ontario through the Ministry of Colleges and Universities. This work is also supported in part by the Advanced ERC grant SM-GRAV, No 669288.

\newpage
\appendix
\renewcommand{\theequation}{\thesection.\arabic{equation}}
\addcontentsline{toc}{section}{Appendix\label{app}}
\section*{Appendices}

\section{Systems of interacting matrix models}\label{miscellaneousmatrix}

In this appendix we study the simplest tripartite systems in terms of two matrix models coupled via a matrix quantum mechanics (MQM) defined on a line segment, with the motivation to study the properties of the induced cross interaction between the two matrix models after we integrate out the messenger MQM. Appendix~\ref{orbifold} is of particular importance, since it describes a possible microscopic model for a two dimensional Bang/Crunch Cosmology~\cite{Betzios:2016lne} and its connection to the present work.

\subsection{Simple linear coupling}\label{linearcoupling}

The simplest version of the sandwich setup, is in terms of two zero-dimensional matrix models (minimal models) coupled through a messenger one-dimensional MQM system. The system is also a particular scaling limit of a matrix chain on an $A_r$ graph, the end-point nodes containing extra degrees of freedom. A analysis of multi-interacting matrix models from a similar perspective of (multi)-verses can be found in~\cite{KN1}

Our variables are $M(\tau)$ and $\Phi_{1,2}$, which are all $N \times N$ Hermitean matrices and $\tau$ is a Euclidean time variable with values on an interval $I$. We shall also gauge the $SU(N)$ symmetry (in order to project to the singlet sector and make the model solvable) and use a quadratic potential for MQM. The partition function then is
\be\label{F1}
Z = \int \, \mathcal{D} \Phi_1 \, \mathcal{D} M(\tau) \, \mathcal{D} A(\tau) \,\mathcal{D} \Phi_2 \, e^{-V(\Phi_1) -V(\Phi_2) + S_{int}} e^{- N \int \Tr (D_t M)^2 - V(M)}
\ee
with
\be
S_{int} = c \Tr \left( \Phi_1 M(\tau = \tau_1)  \right) + c \Tr \left( \Phi_2 M(\tau = \tau_2)  \right) \, .
\ee
If we adopt the natural $N$-scaling of matrix theories, all of the coefficients appearing should be $O(N)$.
For a quadratic potential, we find the MQM heat kernel/transition amplitude for the inverted harmonic oscillator (H.O.) $\tau_2 - \tau_1 = T$ between the states $| M_1 , \tau_1 \rangle$ and $| M_2 , \tau_2 \rangle$
\be
\langle M_2  |M_1 \rangle = \left(\frac{\omega}{2 \pi \sinh \omega T } \right)^{N^2/2} \exp \left(- \frac{N \omega \left[ (\Tr M_1^2 + \Tr M_2^2 ) \cosh \omega T \, - \, 2 \Tr M_1 M_2 \right]}{2 \sinh(\omega T)}  \right)
\ee
This means that the model can be written as an effective four matrix model as follows
\be
Z = \int \, \mathcal{D} \Phi_1 \, \mathcal{D} M_1 \, \mathcal{D} M_2 \,\mathcal{D} \Phi_2 \, e^{-V(\Phi_1) -V(\Phi_2) + c \Tr \left( \Phi_1 M_1  \right) + c \Tr \left( \Phi_2 M_2  \right)} \langle M_2  |M_1 \rangle \, ,
\ee
where $\langle M_2  |M_1 \rangle$ is the Heat/Mehler kernel for the matrix H.O.

It is possible to diagonalise the matrices and integrate over the angular variables. We need four-unitary matrices
$M_{1,2} = U_{1,2}^\dagger \mu_{1,2} U_{1,2}$ and $\Phi_{1,2} = \Omega_{1,2}^\dagger \lambda_{1,2} \Omega_{1,2}$. We then perform the three independent Itzykson-Zuber integrals for the relative rotations in the couplings appearing in the exponent. In the end we obtain a product of determinants in the integrand
\be
Z = \mathcal{N} \int \, \prod_{i=1}^N d \lambda_i^{1} \Delta(\lambda^1) \, d \lambda_i^{2} \Delta(\lambda^2)  \, \prod_{i=1}^N d \mu_i^{1}  \, d \mu_i^{2}  \, e^{-V(\lambda^1) -V(\lambda^2)} \det_{k l} \left( e^{  c  \lambda^1_k \mu_i^1} \langle \mu^1_i  | \mu^2_j \rangle  e^{  c \lambda^2_l \mu_j^2 } \right) \, ,
\ee
where now the propagator is that of $N$ particles (fermions) in an oscillator potential. This is similar to a continuous limit of a matrix chain. A difference with the matrix chain is that now we have a solvable propagator in the middle and two arbitrary potentials at the end-points.
We can then  use the identity
\be
\frac{1}{n!} \int \prod_i^n d x_i \det_{j k} \psi_j (x_k)  \det_{l m} \phi_{l}(x_m) = \det_{i j} \int d x  \, \psi_i(x) \phi_j (x) \, ,
\ee
sequentially to obtain
\be
Z = \mathcal{N} \int \, \prod_{i=1}^N d \lambda_i^{1} \Delta(\lambda^1) \, d \lambda_i^{2} \Delta(\lambda^2)    \, e^{-V(\lambda^1) -V(\lambda^2)} \det_{k l} K(\lambda_k^{1}, \lambda_l^{2} )  \, ,
\ee
with the reduction of the $2 \times N$ intermediate integrals to just two integrals
\be
K(\lambda_k^{1}, \lambda_l^{2} ) = \int d \mu^1 d \mu^2  e^{  c  \lambda^1_k \mu^1} \langle \mu^1  | \mu^2 \rangle  e^{  c \lambda^2_l \mu^2 }
\ee
Since they are Gaussian we can perform them explicitly to derive an effective action for the end-point eigenvalues
\bea
Z &=& \mathcal{N} \int \, \prod_{i=1}^N d \lambda_i^{1} \Delta(\lambda^1) \, d \lambda_i^{2} \Delta(\lambda^2)   \, e^{-V(\lambda^1) -V(\lambda^2)}  e^{ \frac{c^2}{2 N \omega \sinh(\omega T)} \left( \sum_i \left[ (\lambda^1_i)^2 + (\lambda^2_i  )^2 \right] \cosh \omega T  \right)}  \times \nn \\
&\times& e^{  \frac{c^2}{ \omega N \sinh(\omega T)} \sum_i  \lambda^1_i \lambda^2_i } \, .
\eea
In the last term there was a determinant that we replaced by the product of the diagonal terms due to the anti-symmetry of the vandermonde determinants. The end result is hence a simple two matrix model with a linear cross-coupling. The strength of the coupling now depends on the size of the extra distance $T$ and as expected diverges in the limit $T \rightarrow 0$ and vanishes as $T \rightarrow \infty$, where the two end-point matrix models decouple.

To completely solve such 2 matrix models (2MM) at finite-N one has to use the techniques of bi-orthogonal polynomials. In the double scaling limit they correspond to the $(p,q)$ minimal models coupled to gravity, depending on the properties of the potentials $V_1(\lambda^1), V_2(\lambda^2)$. After taking the double scaling limit,
the genus zero information is contained in the spectral curves expressed through two continuum resolvents ($\tilde{c} = c/N$ is the scaling variable)
\be\label{pqsaddles}
\omega_{1,2}(x) = \int d \lambda \frac{\rho_{{1,2}}(\lambda)}{x - \lambda}, , \quad \rho_{1,2}(\lambda) = \frac{1}{N} \sum_i \delta (\lambda - \lambda_i^{(1,2)}) \, ,
\ee
\bea
\frac{\tilde{c}^2}{ \omega \sinh(\omega T)} y + \frac{\tilde{c}^2 \cosh \omega T}{ \omega  \sinh(\omega T)} x = V_1'(x)  + \omega_1(x) \, , \nn \\
\frac{\tilde{c}^2}{ \omega  \sinh(\omega T)} x + \frac{\tilde{c}^2 \cosh \omega T}{ \omega  \sinh(\omega T)} y= V_2'(y) + \omega_2(y) \, .
\eea
These are spectral curves of the usual kind of $(p,q)$-models.
The novel-thing that happens here is that the model transitions to two decoupled $(2,p) \times (2,q)$ minimal models as $T \rightarrow \infty$. The relevant Riemann surface/specral curve is a $\mathcal{M}_{p,q}$ manifold with $(p-1)(q-1)/2$ singularities. The shape of the manifold depends on $T$ and splits into two decoupled manifolds in the limit $T \rightarrow \infty$. In some sense one could argue that there is a ``wormhole throat" on the Riemann-surface that the saddle point equations describe, that elongates and pinches off in the limit $T \rightarrow \infty$. Since it is known such such matrix models are also able to describe topological strings on Calabi-Yau manifolds it might be interesting to revisit this particular model from this perspective~\cite{Dijkgraaf:2002fc}.

\subsection{Matrix model with bifundamentals}\label{bifundamentalmatrix}

A shortcoming of the model analysed in the previous section~\ref{linearcoupling}, is that the messenger MQM and boundary gauge groups have to be the same. Another matrix model in which we can tune the relative gauge group sizes is the following model involving complex bifundamentals
(i.e. $M(\tau)$ is $N_3 \times N_3$ matrix and $B_{1,2}$ are $N_{MQM} \times N_{1,2}$ complex matrices )
\be\label{bifundmatrix}
Z = \int \, \mathcal{D} \Phi_1 \,\mathcal{D} \B_1  \, \mathcal{D} M(\tau) \, \mathcal{D} A(\tau) \,\mathcal{D} \Phi_2  \,\mathcal{D} \B_2 \, e^{-V(\Phi_1) -V(\Phi_2) - V(B_1^\dagger B_1) -  V(B_2^\dagger B_2) + S_{int} - n S_{MQM} }
\ee
with
\bea
S_{MQM} = \half \int d \tau \Tr (D_t M)^2 - V(M) \, , \nn \\
S_{int} = c \Tr_3 \left( B_1 \Phi_1 B_1^\dagger   \right) + c \Tr_3 \left( B_2 \Phi_2 B_2^\dagger  \right) \, + \nn \\
+ d \left( \Tr_1 (B_1^\dagger M(\tau = \tau_1) B_1) + \Tr_2 (B_2^\dagger M(\tau = \tau_2) B_2)  \right) \, .
\eea
Without great loss of generality we can set $N_1 = N_2 = N$ and $N_{MQM} = n$. Now there are two options $n \leq N$ or $N \leq n$, which need to be treated separately. We shall begin with the first option that is most relevant in the limit of soft cross interaction between the two boundary theories.

{\bf Analysis I: \\}
The rectangular complex matrices $B_{1,2}$ can be diagonalised with bi-unitary transformations as $B_{1,2} = \Omega^\dagger_{1,2} b_{1,2} V_{1,2}$ in terms of positive definite diagonal matrices $b_{1,2}$ (containing $n$ eigenvalues and a $N-n \times N-n$ sized block of zeros). We find the part of the action that interacts with MQM to take the form
\bea
S^d_{int} = d \left( \Tr_n ( b_1^2 \Omega_1 U_1^\dagger \mu_1 U_1 \Omega_1^\dagger ) +  \Tr_n b_2^2 \Omega_2 U_2^\dagger \mu_2 U_2 \Omega_2^\dagger  \right) \, .
\eea
After performing three independent Itzykson-Zuber (IZ) integrals over the relative rotations
\be
 R_1 = \Omega_1 U_1^\dagger\sp R_2 = \Omega_2 U_2^\dagger
  \ee
  and  the one between the initial and final state of MQM
  \be
  R_{12} = U_1 U_2^
\dagger\;,
 \ee
 we find a similar integral as before in terms of $b_{1,2}$ and $\mu_{1,2}$ eigenvalues. Since the MQM eigenvalues appear quadratically, we can also integrate them as before (using all the determinant tricks) to find  the effective action in the $b_{1,2}$ eigenvalue basis
\bea
S^d_{eff} &=& \frac{d^2}{2 \omega N \sinh(\omega T)} \left( \sum_i^n \left[ (b^{(1)}_i)^4 + (b^{(2)}_i  )^4 \right] \cosh \omega T    \right) \, + \nn \\
 &+& \,  \log \det_{i j} \exp \left[  \frac{d^2}{2 \omega N \sinh(\omega T)} (b^{(1)}_i)^2 (b^{(2)}_j)^2  \right] \, , \quad
\eea
where we denote the (positive definite) eigenvalues of the complex matrices $B_2$ and $B_1$ by $(b^{(2)}_i)$ and $(b^{(1)}_i)$. As discussed before, the last term can be reduced to the diagonal terms due to the bifundamental Vandermondes multiplying it.

The other two terms in the interaction action couple only to the first $n$-eigenvalues of the matrices $\Phi_{1,2} = W_{1,2}^\dagger \lambda_{1,2} W_{1,2}$ (due to the presence of the block of zeroes in $b_{1,2}$) at the endpoints
\be
S^c_{int} = c \Tr_N \left( b_1^2 V_1 W^\dagger_1 \lambda_1 W_1 V_1^\dagger    \right) + c \Tr_N \left(  b_2^2 V_2 W^\dagger_2 \lambda_2 W_2 V_2^\dagger    \right)
\ee
After performing the last two relative IZ integrals of matrices with unequal rank, we obtain either
\be
S^c_{eff} =   \left( \log \det_{n \times n} (c (b^1_i)^2 \lambda^1_j ) + \log \det_{n \times n} (c (b^2_i)^2 \lambda^2_j )  ) \right)  \, ,
\ee
or in the most general case of complex $\Phi_{1,2}$ matrices
\be
S^c_{eff} =   \left( \log \det_{n \times n} I_{0}(c (b^1_i)^2 \lambda^1_j ) + \log \det_{n \times n} I_{0}(c (b^2_i)^2 \lambda^2_j )  ) \right)  \, ,
\ee
where $I_{0}(x)$ is the Bessel-I function. This expression has the advantage of accommodating an arbitrary bi-fundamental potential. In full generality, it is impossible to integrate out the bifundamentals after this step, if they have an arbitrary potential. Nevertheless, we already see the inequivalent rank effect: Some of the end-point eigenvalues are coupled, some are not. This has the possibility of creating a ``quantum or small wormhole throat" in the scaling limit $1 < n \ll N$, certainly of smaller size than the one described in the saddle point equations eqns.~\eqref{pqsaddles}.

{\bf Analysis II: \\}
For a Gaussian potential for the bi-fundamentals, we can instead choose to first integrate them out completely using
\bea\label{determinantintegral1}
\int \mathcal{D} \B_1 e^{-b \Tr_n \left(B_1 B_1^\dagger \right) + c \Tr_n \left( B_1 \Phi_1 B_1^\dagger   \right)
+ d  \Tr_N \left( B_1^\dagger M_1 B_1 \right)}  =  \nn \\
= \det \frac{1}{b I_{n \times n} \otimes I_{N \times N} - c I_{n \times n} \otimes \Phi_1 - d M_1 \otimes I_{N \times N}  }
\eea
where the determinant is in the tensor product space. It is invariant under unitary rotations in both spaces. We can use this invariance to diagonalise the matrices appearing in the determinant.

After diagonalising $\Phi_{1,2}, M_{1,2}$, we need to perform only a single leftover MQM IZ integral to obtain
\bea
Z &=& \mathcal{N} \int \, \prod_{a,b=1}^N d \lambda_a^{1}  d \lambda_b^{2} \, \prod_{i=1}^n d \mu_i^{1}  d \mu_i^{2} \,  \Delta^2(\lambda^1) \Delta^2(\lambda^2)   \, e^{-V(\lambda^1) -V(\lambda^2)}  \nn \\
&\times&   \frac{\Delta(\mu^1)}{ \prod_{i = 1}^n \prod_{a=1}^N (b - d \mu^1_i - c \lambda_a^{1})} \det_{k l}  \langle \mu^1_k | \mu^2_l \rangle   \frac{\Delta(\mu^2)}{ \prod_{j = 1}^n \prod_{a=1}^N (b - d \mu^2_j - c \lambda_a^{2})}
\eea
Notice that the determinant in the MQM propagator after the IZ can be dropped out if we wish, due to the Vandermondes multiplying it,  that result into only the diagonal terms contributing. We notice that the coupling between the end-point matrices is  ``softer" than the direct linear cross coupling of the previous section and involves only the analogue of one loop determinants (simple determinants here) and not terms in the exponent. These factors can indeed be written in a determinental form, using a generalisation of the Cauchy identity given in~\cite{Honda:2013pea}.

We find
\be
 \frac{\Delta(\lambda^1)\Delta(\mu^1)}{ \prod_{i = 1}^n \prod_{a=1}^N (b - d \mu^1_i - c \lambda_a^{1})} =\det \begin{pmatrix}
\biggl(\frac{1}{(b - d \mu^1_i - c \lambda_a^{1})}\biggr)
_{\begin{subarray}{c} 1\leq i\leq n\\1\leq a\leq N\end{subarray}} \\
\bigl((c \lambda_a^{1})^{N-n-p}\bigr)
_{\begin{subarray}{c} 1\leq p\leq N-n\\1\leq a\leq N\end{subarray}}
\end{pmatrix}
\ee
The next step is to integrate out the eigenvalues $\mu_{1,2}$. Using determinant formulae of unequal ranks we obtain
\bea
\int \, \prod_{i=1}^n d \mu_i^{1}  d \mu_i^{2} \,  \det \begin{pmatrix}
\biggl(\frac{1}{(b - d \mu^1_i - c \lambda_a^{1})}\biggr)
_{\begin{subarray}{c} 1\leq i\leq n\\1\leq a\leq N\end{subarray}} \\
\bigl((c \lambda_a^{1})^{N-n-p}\bigr)
_{\begin{subarray}{c} 1\leq p\leq N-n\\1\leq a\leq N\end{subarray}}
\end{pmatrix}   \langle \mu^1_i | \mu^2_i \rangle   \,  \det \begin{pmatrix}
\biggl(\frac{1}{(b - d \mu^2_i - c \lambda_b^{2})}\biggr)
_{\begin{subarray}{c} 1\leq i\leq n\\1\leq b\leq N\end{subarray}} \\
\bigl((c \lambda_b^{2})^{N-n-p}\bigr)
_{\begin{subarray}{c} 1\leq p\leq N-n\\1\leq b\leq N\end{subarray}}
\end{pmatrix}  \nn \\
=  \det
\begin{pmatrix}
(m_{a b})
_{\begin{subarray}{c} 1 \leq a \leq N \\ 1 \leq b \leq N \end{subarray}}
& \bigl((c \lambda_a^{1})^{N-n-p}\bigr)
_{\begin{subarray}{c} 1\leq a\leq N\\1\leq p\leq N-n \end{subarray}}  \\
\bigl((c \lambda_b^{2})^{N-n-p}\bigr)
_{\begin{subarray}{c} 1\leq p\leq N-n\\1\leq b\leq N\end{subarray}}
& (0)
_{\begin{subarray}{c} N-n \times   N-n \end{subarray}}  \end{pmatrix}  \, , \nn
\eea
with
\be
m_{a b}(\lambda^{1}, \lambda^{2}) = \int_{-\infty}^\infty d \mu^1 d \mu^2  \frac{1}{(b - d \mu^1 - c \lambda_a^{1})}  \langle \mu^1 | \mu^2 \rangle \frac{1}{(b - d \mu^2 - c \lambda_b^{2})} \, .
\ee
Due to the zeros in the last block, some terms in the determinant do not contribute.

One could try to compute this integral exactly, but there is a contour issue, due to the singularities in the denominator. To understand this issue, in the paper~\cite{Betzios:2016lne} such integrals were defined as giving the Hilbert transform of the propagator eigenfunctions: Hermite polynomials or parabolic cylinder functions for the inverted oscillator. There are two options then: Either to take the large-$N,n$ limit and exploit saddle point techniques, similar to the previous section, or to pass over to the grand-canonical ensemble and find an operator whose spectrum dictates the physics of the model, similar to what happens in $c=1$ matrix models and ABJM~\cite{Betzios:2016lne,Marino:2011eh}. Since the first option is analogous to our analysis in the previous section, and the second route can in principle provide with non-perturbative information for the spectral curve of the model, we briefly sketch this second option.

To pass to the grand canonical ensemble, we first multiply with the two Vandermonde determinants remaining ($\Delta(\lambda^1) = \det V^1_{a b}\, , V^1_{a b} = (\lambda^1_a)^{b-1}$ etc.) to obtain
\be
Z_{N,n} = \mathcal{N} \int \, \prod_{a,b=1}^N d \lambda_a^{1}  d \lambda_b^{2} \,   \det
\begin{pmatrix}
((V^1 m V^2)_{a b})
_{\begin{subarray}{c} 1 \leq a \leq N \\ 1 \leq b \leq N \end{subarray}}
& \bigl((c V_{a c} \lambda_c^{1})^{N-n-p}\bigr)
_{\begin{subarray}{c} 1\leq a\leq N\\1\leq p\leq N-n \end{subarray}}  \\
\bigl((c V^2_{b c}\lambda_c^{2})^{N-n-p} \bigr)
_{\begin{subarray}{c} 1\leq p\leq N-n\\1\leq b\leq N\end{subarray}}
& (0)
_{\begin{subarray}{c} N-n \times   N-n \end{subarray}}  \end{pmatrix}
\ee
The grand canonical partition function is a Fredholm determinant with respect to $N$
\be
\sum_{N=0}^\infty e^{\beta \mu N} Z_{N,n} = \det \begin{pmatrix}
I + e^{\beta \mu} V^1 M V^2 & V^1 C \\
C V^2 & 0
\end{pmatrix}
\ee
Studying the spectrum of the operator appearing in the Fredholm determinant, could be the starting point for a further analysis of this class of matrix models.

\subsection{MQM on $S^1/Z_2$}\label{orbifold}

Matrix quantum mechanics on the orbifold $S_1/Z_2$~\cite{Betzios:2016lne} (once $Z_2$ is embedded into the gauge group) conjectured to describe a big-bang big-crunch universe, has also a similar structure to the model of the previous section~\ref{bifundamentalmatrix}. This model could provide an interesting arena for clarifying the expected relation between Euclidean wormholes and bang-crunch universes upon analytic continuation~\cite{Maldacena:2004rf}.

In particular, the partition function computed in~\cite{Betzios:2016lne}, is found to be that of two MQM models on line segments coupled via Vandermondes
\be
Z = \int \, \prod_{i=1}^N \prod_{a = 1}^n  \,d x_i \, d x_a' \, d y_i \, d y_a' \, \frac{\Delta(x) \Delta(x')}{\Delta(x,x')} \langle x_i | y_i  \rangle \langle x'_a | y'_a  \rangle \frac{\Delta(y) \Delta(y')}{\Delta(y,y')} \, ,
\ee
where we symbolise with $\Delta$'s  the Vandermonde determinants ($\Delta(x,x') = \prod_{i, a} (x_i - x_a')$ is the cross coupling Vandermonde determinant), and with $\langle x_i | y_i \rangle$ the oscillator transition amplitude for the corresponding matrix eigenvalues. In this expression, we have also simplified the determinants appearing in the matrix oscillator propagators, using once more the antisymmetry of the Vandermondes multiplying them.

One important aspect of this expression is that it can also be thought of as arising from two MQM models (with gauge group ranks $N,n$ respectively) on two Euclidean time intervals coupled by bifundamental fields at the endpoints of Euclidean time (``bifundamental instantons") that once integrated out bring the two determinants in the denominator as in eqn. \eqref{determinantintegral1}\footnote{This aspect of the model was briefly discussed in~\cite{Betzios:2016lne}.}. The model can therefore be thought to arise from
an action
\bea
S_{tot} &=& S_{MQM_1} + S_{MQM_2} + S_{int} \nn \\
S_{int} &=&  \Tr_n \left( B M_1(\tau = 0) B^\dagger   \right)
+   \Tr_N \left( B^\dagger M_2(\tau = 0) B \right) + \nn \\
&+& \Tr_n \left( C M_1(\tau = L) C^\dagger   \right)
+   \Tr_N \left( C^\dagger M_2(\tau = L) C \right) \, ,
\eea
where $B,C$ are $n \times N$ bifundamental matrices and $M_{1,2}(\tau)$ are the two MQM fields of size $n \times n$ and $N \times N$ respectively. We therefore find that it is remarkably a slightly modified instance of the class of models proposed by~\cite{VanRaamsdonk:2020tlr} and studied in this paper as duals to Euclidean wormholes. In this case, one has actually two complementary pictures, either the MQM's as messengers and the bifundamentals as boundary theories or the bifundamentals as lower-dimensional messengers coupling the two MQM's.

\section{On reflection positivity and spectral representation}\label{reflcpos}

The formal definition of reflection positivity of an operator $\hat{G}$ is that
\be
\langle f | \hat{\Theta} \hat{G} | f \rangle \geq 0 \, ,
\ee
where $f \in \mathcal{S}_+(\mathbb{R}^d_+)$ is a test function belonging in the (Schwartz) space with support $x_0 \geq 0$ and $\hat{\Theta}$ inducing reflections along the transverse hyperplane i.e. $x_0 \rightarrow - x_0$.

For the two-point function this means that
\be
\langle f | \hat{\Theta} \hat{G} | f \rangle \, = \, \int d^d x \int d^d y f^\dagger(- x_0 , \vec{x}) G(x-y) f(y_0, \vec{y}) \, \geq 0 \,  \, .
\ee
It is a standard exercise to show that the propagator of a free massive scalar with real mass $\mu$, results in a reflection positive two-point function $\langle f | \hat{\Theta} \hat{G}_{\mu^2} | f \rangle \geq 0$. Assuming then a K\"allen-Lehmann spectral representation for the general two-point function we find
\be\label{KLspectral}
G_{K.L.}(x-y) =  \int_0^\infty d \mu^2  \int \frac{d^d p}{(2 \pi)^d}   \frac{\rho(\mu^2)}{p^2 + \mu^2} e^{i p(x-y)} \, ,
\ee
so that
\bea
\langle f | \hat{\Theta} \hat{G}_{K.L.} | f \rangle \, &=& \, \int_0^\infty d \mu^2 \int \frac{d^d p}{(2 \pi)^d} \int d^d x e^{i p x}  f^\dagger(- x_0 , \vec{x}) \int d^d y e^{- i p y} f(y_0, \vec{y})    \frac{\rho(\mu^2)}{p^2 + \mu^2}  =  \nn \\
&=& \int_0^\infty d \mu^2 \rho(\mu^2) \, \langle f | \hat{\Theta} \hat{G}_{\mu^2} | f \rangle \, .
\eea
This means that a theory admitting a K\"allen-Lehmann spectral representation with positive weight $\rho(\mu^2) \geq 0$, satisfies reflection positivity.

One can reconstruct the spectral weight from the knowledge of the imaginary part of the correlator in momentum space (that is always well-defined)
\bea\label{dos}
\rho(\mu^2) = \frac{1}{\pi} \Im \, G(p) \, \vert_{p^2 = - \mu^2} \, &=& \frac{1}{2 \pi i} \, \left( G(p^2 = - \mu^2 - i \epsilon) - G(p^2 = - \mu^2 + i \epsilon) \right) \, \, , \nn \\
G(p) &=& \int_0^\infty d \mu^2    \frac{\rho(\mu^2)}{p^2 + \mu^2} \, .
\eea
Equations  \eqref{dos} form the dispersion relations for the two-point function (valid up to a possible finite number of subtractions). As an example for a CFT $G(p) \sim p^{2 \Delta - d}$ one finds the scaling behaviour
\be
\rho(\mu^2) \sim \m^{2 \Delta - d -2}
\ee
On the other hand, for an exponentially decaying propagator $p^{2 a} e^{-L p}$ (as $p \rightarrow \infty$ - UV in momentum space) we obtain that
\be
\rho(\mu^2) \sim \m^{2 a - 2} \cos(L \mu) \, .
\ee
This is clearly not a positive definite spectral weight and this explains why the cross correlator $G_{12}(p)$ in \eqref{a15} does not admit a positive definite spectral representation. The failure of the dispersion relation in this case stems from the fact that there are regions in the complex $p$-plane (negative half plane) for which the correlator is exponentially growing. For the dispersion relations to hold, one actually demands that the correlator is vanishing faster than $1/p$ in the upper or lower half complex plane in order to close the contour. A finite number of subtractions is also not able to remedy the exponential divergence in the negative half plane.

\section{A note on ghost decoupling}\label{Decouplingghosts}

When using axial gauge, one finds that the ghosts decouple from the action. In order to show this, we reintroduce gauge indices and denote the $U(N)$ gauge field by $A_\m^a$. The gauge fixing condition\footnote{There do exist subtleties when $n^2 = 0$, but are of no importance for the Euclidean examples we are interested in.}
\be
F^a = n_\mu A^a_\mu = 0
\ee
under a gauge transformation with parameter $\omega^a$
\be
\delta_\omega A_\m^a = \partial_\m \omega^a + g_{YM} f^{a }_{b c} A^b_\m \omega^c   \, ,
\ee
transforms as
\be
\delta_\omega F^a = n^\m \partial_\m \omega^a + g_{YM}  n^\m f^{a}_{b c} A^b_\m \omega^c  \, , \quad \Rightarrow \quad \frac{\delta F^a}{\delta \omega_b} = \delta^{a b} n^\m \partial_\m \, .
\ee
The last matrix is the one appearing in the Faddeev-Popov determinant and hence the ghosts decouple from the gauge field and only contribute in the overall normalisation of the path integral. The gauge field propagator nevertheless does exhibit singularities for momenta $n_\mu k^\mu = 0$, but these are remnants of the incomplete gauge fixing. In our two-dimensional example in section~\ref{noncompacttau} we shall perform a further gauge transformation into a complete gauge and no such subtleties will arise (no remaining dynamical kinetic term for the gauge field).

\section{$A_\tau = 0$ gauge}\label{alternategauge}

For the model of section~\ref{D0D1}, another gauge that can be used in the case where $\tau$ is non-compact is the gauge $A_\tau = 0$. Euclidean ``time" evolution in this gauge is defined along the $\tau$ direction which coincides with the evolution for the boundary MQM. In this case $F_{\t z} = \partial_\t A_z$ and the relevant equations of motion are
\bea\label{P10a}
\partial_\tau F^{\tau z}  = 0  \, , \quad \Rightarrow \quad A_z = f(z) + \t g(z) \, , \nn \\
\partial_z F^{z \tau} + i [A_z, F^{z \tau}] = \delta(z-L) g_{YM}^2 J_{MQM_{1}}^\tau(\tau) + \delta(z+L) g_{YM}^2 J_{MQM_{2}}^\tau(\tau)
\eea
The second equation of \eqref{P10a} is the constraint equation. It can also be written as
\be\label{P10}
- \partial_z \partial_\tau A_z - i [A_z, \, \partial_\tau A_z] = \delta(z-L) g_{YM}^2 J_{MQM_{1}}^\tau(\tau) + \delta(z+L) g_{YM}^2 J_{MQM_{2}}^\tau(\tau) \,  .
\ee
An analogous equation has appeared in a similar problem in~\cite{Minahan:1993mv}. To solve this equation
we introduce the path ordered exponential (Wilson line) and the dressing operator as
\be
W_a^b = \mathcal{P} e^{i \int_a^b d z A_z } \, , \quad D(z; a, b) = W_a^z \partial_\tau A_z W_z^b \, , \quad \partial_\tau W_{a}^b = i \int_a^b d z D(z; a, b)   \, .
\ee
We can then split the interval into three regions as
\be
z \in [-L , -L + \epsilon) \cup [-L + \epsilon, L - \epsilon] \cup (L - \epsilon, L] \, .
\ee
Using this split eqn. \eqref{P10} is regulated and can be written as
\be
- \frac{\partial_z D(z; -L, L)}{g_{YM}^2}  = \delta(z-L+ \epsilon)  W_{-L}^{L- \epsilon}  J_{MQM_{1}}^\tau(\tau) W_{L - \epsilon}^{L}  + \delta(z+L - \epsilon)   W_{-L}^{-L + \epsilon} J_{MQM_{2}}^\tau(\tau)  W_{-L + \epsilon}^{L} \, .
\ee
The reason for the split and small regulator $\epsilon$, is that $D(z)$ is a piecewise constant, jumping by the appropriate source factor once it reaches $z = \pm L$
\be
D(L- \epsilon) = D(L) + g_{YM}^2 W J_{MQM_{1}}^\tau \, , \qquad D(-L + \epsilon) = D(-L) + g_{YM}^2  J_{MQM_{2}}^\tau W \, ,
\ee
with $W = W_{-L}^L$. We further assume that the space terminates at $z = \pm L$.

This also means that picking any reference point $z \in [-L, L]$
\be
[W, \, \partial_\t A_z] = g_{YM}^2 W J_{MQM_{1}}^\tau  -   g_{YM}^2 J_{MQM_{2}}^\tau  W \, ,
\ee
and since
\be
\partial_\tau W = i \int_{-L}^L d z W_{-L}^z \partial_\tau A_z W_z^L = i \int_{-L}^L  d z D(z; -L, L) \, ,
\ee
we also find
\be\label{currentsmatch}
i J_{YM}^\tau = [W^{-1} , \, \partial^\t W] = i g_{YM}^2  2 L  J_{MQM_{1}}^\tau -  i g_{YM}^2  2 L  J_{MQM_{2}}^\tau  \, .
\ee
This equation has the interpretation of a constraint on the total current/charge of the system, that is also a constraint on the admissible representations. This constraint holds non-perturbatively and without using the dynamical equation of motion (first equation of~\eqref{P10a}). In physical terms $W \equiv W_{-L}^L$ corresponds to an open Wilson line that extends across the two boundaries and couples to the boundary charges. Evolution in this gauge is along the $\tau$ direction which coincides with the evolution of the boundary MQM.

In this gauge one can also find the $2d$ YM Hamiltonian (evolving along $\tau$) that corresponds to that for a particle on the group manifold~\cite{Minahan:1993mv}
\be
\hat{H}^\t_{A} \, = \, \frac{1}{g_{YM}^2} \int_{-L}^L d z \Tr (\partial_\t A_z)^2 \, = \, - \frac{1}{4 g_{YM}^2 L} \Tr \left(W^{-1} \partial_\t W \right)^2 \, .
\ee
Now this Hamiltonian has two separate left/right chiral symmetries under $SU(N)_L \times SU(N)_R$ with matrix generators~\cite{Polychronakos:2006nz}
\be
J_R = - i W^{-1} \, \partial_\t W \, , \quad  J_L = i  \partial_\t W \, W^{-1} \, , \quad J_{YM} = J_R + J_L \, .
\ee
It is then natural to interpret the constraint of \eqref{currentsmatch} as relating the left/right chiral currents and the two MQM currents on the two boundaries, so that $J_R = J_{MQM_1}$ and $J_L = J_{MQM_2}$. This is also physically motivated since the matter fields on the boundaries have a charge that sources the endpoints of the Wilson line stretching between them.
The Hamiltonian itself is written as
\be
\hat{H}_{A} = - \frac{1}{4 g_{YM}^2 L} \Tr J_L^2 =  - \frac{1}{4 g_{YM}^2 L} \Tr J_R^2 \, .
\ee
The eigenstates of the Hamiltonian are representation matrices $R_{a b}(W) = \langle W | R, ab\rangle$ ($a,b = 1 ..., d_R$), that transform under the right action in the rep $R$ of $U(N)$ and under the left action in the conjugate rep $\bar{R}$ of $U(N)$. For each rep $R$ there are $d_R^2$ energetically degenerate states with eigenvalue equal to the quadratic Casimir $C_R^{(2)}$. Since we are using functions on an interval and not a circle, the Hilbert space is built out of all the $| R, a b \rangle$ states (the states are not required to be class functions and carry uncontracted group indices). An equivalent description of the Hilbert space is in terms of open strings with Chan-Paton factors at their endpoints, the representation basis reorganises the $n$ open string Hilbert space in terms of irreducible representations of the permutation group $S_n$~\cite{Donnelly:2016jet,Blommaert:2018oro}, exchanging the strings.

\subsection{Wilson lines and compact $\tau$}\label{compacttauwilson}

In the case that $\tau$ is an $S^1$ there exist non-trivial winding modes and hence one can only work in the gauge $A_z = 0$.

In analogy with section~\ref{alternategauge}, we define the Wilson lines and dressing operator along $\tau$ as
\be
W_a^b = \mathcal{P} e^{i \int_a^b d \tau A_\tau } \, , \quad D(\tau , a, b ; z) = W_a^\tau \partial_z A_\tau W_\tau^b \, , \quad \partial_z W_a^b =  i \int_a^b d \tau  D(\tau , a, b ; z) \, .
\ee
Using the equation \eqref{PB2} one finds
\be
\partial_\tau D(\tau , a, b ; z) = 0 \, \, , \quad D(\tau , a, b ; z) = c(z) \, , \quad \Rightarrow \quad [W_a^b, \, \partial_z A_\tau] = 0 \, ,
\ee
the last equation arising from the independence on $\tau$. We then notice that due to \eqref{cases}, $D(\tau , a, b ; z)$  suffers a discontinuity at $z = \pm L$, since $A_\tau$ is defined piecewise. In particular the discontinuity causes
$\partial_z A_\t$ to jump by $J_{MQM_{1,2}}$ in all the formulae.

In order to obtain the Hamiltonian evolving along the $z$ direction of the cylinder, we can define $W_0^\beta = W$ as the Polyakov loop around the thermal circle. Since
\be
\partial_z W = i \int_{0}^\beta d \tau W_{0}^\tau \partial_z A_\tau W_\tau^\beta = i \int_{0}^\beta  d \tau D(\tau , 0, \beta ; z) \, ,
\ee
we also observe that due to constancy of $D(\tau)$ and the periodicity of $A_\tau$ in $\tau$
\be
\partial_z W = i \beta W \partial_z A_\tau  = i \beta \partial_z A_\tau  W \, , \quad \Rightarrow \quad [\partial_z W, \, W^{-1}] = 0 \, .
\ee
The 2d YM Hamiltonian evolving along $z$ can then be written as (it corresponds to a particle on the group manifold~\cite{Minahan:1993mv})
\be
\hat{H}_A^z \, = \, \frac{1}{g_{YM}^2} \int_0^\beta d \t (\partial_z A_\t)^2 \, = \, - \frac{1}{4 g_{YM}^2 \beta} \Tr (W^{-1} \partial_z W)^2
\ee
All the representation matrices $R_{a b}(W) , \,  W \in U(N)$ are eigenstates of the Hamiltonian. The $d_R^2$ states are energetically degenerate and have eigenvalue equal to the quadratic Casimir $C_R^{(2)}$. Since the base space is a circle, the gauge invariant eigenstates forming a basis of the physical Hilbert space are square integrable class functions, the characters $\chi_R(W) = \langle W | R \rangle$. For the generalised YM theories the Hamiltonian involves higher Casimirs and the $\theta$ term can be expressed in terms of the first Casimir~\cite{Frishman:2010tc}.


\section{Partitions}\label{partitions}

In this appendix we provide some terminology and simple examples of partitions.

\begin{itemize}
\item
A partition $\l$ is a sequence of non-increasing integers such that
\be
\l_1 \geq \l_2 ....\l_{\ell(\l)+1} = 0
\ee
The number of non-zero elements $\ell(\l)$ is called the length of the partition. The sum of all the elements $|\l|= \sum_{i \geq 1} \l_i$ is called the weight of the partition.
\item
The multiplicity $m_j(\l)$ of the positive integer j is how many times the number j appears in the partition $\l$ (such that $\l_i = j$).
\item
The partitions are labelled graphically using Young diagrams. They are an array of boxes where the $i$'th row contains $\l_i$ boxes. This means that the number of rows is the length of the partition $\ell(\lambda)$ and the number of columns  is just $\l_1$. The total number of boxes is then equal to the total weight $|\l|$.
\item
The conjugate or transpose of a partition $\l'$ or $\l^T$ is obtained by either reflecting the Young diagram along the diagonal exchanging rows and columns. As an example one obtains $\l_1' = \ell(\l)$.
\end{itemize}
As a simple example to have in mind the partition $(7,5,3^2,1^2)$ corresponds to the following Young tableaux
\be\label{partitionexample}
\yng(7,5,3,3,1,1)
\ee
The irreducible representations of the symmetric group $S_n$ are in one-to-one correspondence with the Young-diagrams $\lambda$. The rows are symmetrizers of the elements while the columns anti-symmetrizers. Furthermore the Young diagrams also parametrise the irreps of $GL(N)$, where the $\lambda_i$ are related to the highest weights $\Lambda_i$ and the diagram has at most $N$ non-empty rows (this just means that the length of the partition is $\ell(\lambda) = N$). For the group $SL(N)$ the diagram has instead at most $N-1$ non-empty rows. For the relation between highest weights $\Lambda_i$ and $\lambda_i$, see equation \eqref{mappingpartrep} and appendix~\ref{modulesbranching}.

Once the irrep/Young-diagram is specified one can introduce further information to it by filling its boxes with letters from an alphabet belonging to a totally ordered set. We then call the diagram a Young-Tableaux. These Tableaux can be used to enumerate the various states belonging to the $GL(N)$ or $SL(N)$ irrep modules, see appendix~\ref{modulesbranching}.  There exist two basic options. One can either have a standard Tableaux, when both the rows and columns have increasing elements from top to bottom and left to right. The other option is that of a semistandard Tableaux, or column strict, when the elements are weakly increasing along each row and strictly increasing along the columns. The Tableaux is then specified by two collections of elements $\lambda, \mu$. The first one dictates its shape and the second its content weight. The content weight $\mu$ counts the occurences of each element of the alphabet in the Tableaux (usually a collection of integers $1, ... N$). For example if $\lambda=(3,2)$ and $\mu = (1,1,2,1)$, then we get a Tableaux of the shape
\be \yng(3,2)\nn\ee
and we fill it with the numbers $1,2,3,4$. The number $1$,$2$,$4$ need to occur a single time, while the number $3$ has to occur exactly two times. It is easy to see graphically that there exist three possible semi-standard Tableaux of shape $\lambda$ and weight $\mu$. These are
\be
\young(123,34) \, , \qquad \young(133,24) \, , \qquad \young(124,33)
\ee
These possibilities/multiplicities are counted by the Kostka numbers $K_{\lambda, \mu}$ and in this specific example $K_{\lambda, \mu}=3$. The Kostka polynomials described in appendix~\ref{HallLittlewoodappendix} and in the main text~\ref{HallLittlewood}, are $q$ analogues of these multiplicities, see also appendix~\ref{modulesbranching} for the description of related branching functions.

Finally we shall present an example of zero weight states that are relevant for the MQM model of section~\ref{D0D1} (and the case when $k_1 = k_2 = 0$ of section~\ref{HallLittlewood}). Let us consider $SU(3)$. In this case we have two roots and the
fundamental representation contains three states with associated weights
\be
\young(1) \leftrightarrow (0,1) \, , \qquad \young(2) \leftrightarrow (-1,1) \, , \qquad \young(3) \leftrightarrow (0, -1) \, .
\ee
The fundamental does not contain any zero weight state and is therefore projected out. Following the rules above one finds $8$ semistandard Tableaux for the adjoint out of which
\be\label{examplezero}
\young(13,2) \, , \qquad \young(12,3) \, ,
\ee
have a weight $(0,0)$ (the weight is computed by summing the individual weights of each box). So the adjoint is a representation/highest weight module that contains a zero weight submodule. All the $SU(N)$ representations that do admit the presence of zero weights are described in~\cite{Klebanov:1991qa}.

\section{Representations and their continuum limit}\label{characterexpansion}

In this appendix we analyse directly the limit of large representations for the original form of the partition function
\eqref{pf1}. The most useful formulae are \eqref{conti} - \eqref{contf}, showing how the limit of continuous representations can be performed.

Labelling the representations in terms of a collection of integers (a partition) $R \equiv \lambda : (\lambda_1, ... \lambda_n)$, parametrising the number of boxes in each row of a Young diagram having $n$ rows in total (see also section~\ref{HallLittlewood} and appendix~\ref{partitions}), one finds the partition function for the first MQM~\cite{Boulatov:1991xz}
\be\label{MQMrep}
 Z_{R}^1(\beta) = \sum_{R_1} C_{R \, R_1}^{R_1} \, q^{\sum_{k=1}^{N_1} \left(\ell_k^{(1)} + \half \right)} \, , \quad \ell_k^{(1)} = \lambda_k^{(1)} + N_1 - k \, ,
\ee
where $\lambda_k^{(1)}$ are related to the highest weights of $R_1$ (that is a $U(N_1)$ representation) and the specific Littlewood-Richardson coefficients $C_{R \, R_1}^{R_1}$ capture the multiplicity that $R_1$ appears in the irrep decomposition of the tensor product $R \otimes R_1$. One can also describe them through an expression involving the integral of three characters
\be
C_{R_1 R_2}^{R_3} = \int D U \, \chi_{R_1}(U)  \, \chi_{R_2}(U) \, \overline{\chi}_{R_3}(U) \, .
\ee
In the specialised case where two of the reps appearing are the same, there is a bound on them $C_{R \, R_1}^{R_1} \leq D_R^{(0)}$ (with $D_R^{(0)}$ the dimension of the zero weight submodule of the highest weight module $R$, see~\ref{modulesbranching}). These $R_1$ representations give the leading contribution to the partition function.
In addition for high temperatures $\beta \rightarrow 0$, one finds
\be\label{special}
Z_R^1(\beta \rightarrow 0) \sim D_R^{(0)} (\omega \beta)^{-N_1} \, ,
\ee
so that the total partition function saddle is governed by the saddle point equations of 2d YM, as in~\cite{Douglas:1993iia}.

For more general temperatures, in order to study the limit of continuous representations, we define continuous variables for all the highest weights appearing
\be\label{conti}
\lambda(x) = \frac{\lambda_i}{N} \, , \quad x = \frac{i}{N}\, , \qquad \lambda(x) \geq \lambda(y) \, , \quad \text{if} \, \, x \leq y
\ee
and then we go to strictly decreasing adapted coordinates
\be
h(x) = \lambda(x) - x + \half \, , \qquad \frac{d h}{d x} \leq  - 1
\ee
and define a density of boxes $\mathcal{\rho}(h)$ via
\be
\mathcal{\rho}(h) \, = \, - \frac{d x(h)}{d h} \, \leq 1 \, .
\ee
The complete partition function takes the form
\be
Z(\beta) = \int D h D h_1 D h_2 \, e^{-N^2 S_{eff}(h, h_1, h_2)}  \, .
\ee
To find the effective action, one needs to exponentiate all the terms and replace them with their continuous version.
In particular the Casimirs become in the large-N limit
\be
\frac{1}{N} C^{(2)} \rightarrow N^2 \int_0^1 d x \, h^2(x)  \, ,  \quad C^{(1)} \rightarrow N^2 \int_0^1 d x \,  h(x)
\ee
The dimension is bounded by
\be\label{contf}
\log D_R^{(0)} \leq \log D_R \rightarrow \frac{N^2}{2} \int_0^1 d x \int_0^1 d y \log |h(x) - h(y)|
\ee
Unfortunately the large-N asymptotics of $C_{R R_1}^{R_1}$
\be
\log C_{R R_1}^{R_1} \rightarrow \, \, ??? \, \, \, .
\ee
do not seem to be known in the literature, and we can only bound them from below using the formula for the dimension.

The support of the eigenvalues should be then determined dynamically, solving the resulting saddle point equations. A complementary analysis of the model in terms of Hall-Littlewood polynomials is presented in section~\ref{HallLittlewood}. In particular it allows for a more precise derivation and analysis of the relevant saddle point equations, that are found to be described by eqn.~\ref{saddlefinal}.

\section{Hall Littlewood and symmetric functions}\label{HallLittlewoodappendix}

Here we shall list some properties of symmetric functions and Hall-Littlewood polynomials relevant for our analysis in section~\ref{HallLittlewood}.
Many additional formulae can be found in the references~\cite{Dorey:2016hoj,Betzios:2017yms,Barns-Graham:2017zpv} and most importantly in the book~\cite{Macdonald}.

Using the relation between representations $R$ of $U(n)$ and partitions $\lambda : (\lambda_1 , ... \lambda_n)$ with length $\ell(\lambda) = n$ (number of rows)\footnote{See appendix~\ref{partitions} for more details on partitions}, we can express the characters in terms of Schur polynomials $\chi_R (Z) \rightarrow s_\lambda (Z)$. The Schur polynomials have the concrete expression ($z_i$ are the eigenvalues of $Z$)
\be\label{Schur}
s_\lambda(Z) = \frac{\det_{1 \leq i, j \leq n} \left(z_i^{\lambda_j + n - j} \right)}{\det_{1 \leq i, j \leq n} \left(z_i^{ n - j} \right)} \, .
\ee
We then define the q-Hall inner product\footnote{To avoid confusion, notice that in the mathematical literature $t$ is the most common symbol associated to Hall-Littlewood polynomials and $q$ is usually an additional parameter of the more general $(q,t)$ Macdonald polynomials. Here $q = e^{- \omega \beta}$.} for two symmetric functions $f(Z) , g(Z)$
\be\label{inner}
\langle f \, , g \rangle_q = \frac{1}{N!} \left(\prod_{i=1}^N \frac{1}{2 \pi i} \oint \frac{d z_i}{z_i} \right) \frac{\prod_{i \neq j} (z_i - z_j)}{\prod_{i \neq j} (z_i - q z_j) } f(Z) g(Z^{-1})\, .
\ee
This inner product is useful because its kernel is essentially equivalent to the twisted MQM partition function \eqref{twistedMQM}, where $z_i = e^{i \theta_i}$ are the eigenvalues of the unitary matrix parametrising the holonomy of the gauge field around the thermal circle.

The orthogonal polynomials with respect to this measure are the Hall-Littlewood polynomials
\be
P_\lambda (Z ; q) = \frac{1}{\mathcal{N}_\lambda} \sum_{\sigma \in S_N} \sigma \left[ z_1^{\lambda_1}...z_N^{\lambda_N} \prod_{i<j} \frac{z_i-q z_j}{z_i - z_j} \right] \, ,
\ee
where $\lambda \in \mathcal{P}$ denotes the partition and the normalization is
\be
\mathcal{N}_\lambda = \frac{\phi_{N - \ell(\lambda)} \prod_{j \geq 1} \phi_{m_j(\lambda)}}{(1- q)^N} \, , \quad \quad \phi_m = \prod_{j = 1}^m (1- q^j) \, ,
\ee
with $m_j(\lambda)$ the multiplicity of the positive integer j in the partition $\lambda$ and $m_0 = N - \ell(\l) \geq 0$. One also defines the Q-Hall polynomials as $Q_\l (X; q) =b_\l(q) P_\l (X; q)$ with $b_\l (q) = \langle P_\l , P_\l \rangle_q^{-1} = \prod_{j \geq 1} \phi_{m_j(\lambda)} $. The orthogonality relation can then be written as
\be\label{orth2}
\langle P_\mu \, , P_\l \rangle_q = \frac{1}{\mathcal{N}_\mu} \delta_{\mu, \l}\, .
\ee
The Schur polynomials defined through eqn.~\ref{Schur}, are also a limit of the Hall-Littlewood polynomials for $q=0$ and orthonormal under the inner product~\ref{inner} upon setting $q=0$. There is also a relation
\be
s_\l(Z) = \sum_\mu K_{\l , \mu}(q) P_\mu (Z ; q)
\ee
with $K_{\l , \mu}(q)$ the Kostka-Foulkes polynomials. The inverse relation defines the Modified Hall-Littlewood (or Milne) polynomials
\be
Q'_\mu (Z ;q) = \sum_\l K_{\l , \mu} (q) s_\l(Z) \, , \quad \quad  \langle P_\l \, , Q'_\mu \rangle_{q=0} = \delta_{\l , \mu}
\ee
There is also a relation between the Modified Hall and the Q-Hall polynomials that reads
\be
Q'_\l (Z ; q) = Q_\l \left(\frac{Z}{1-q} ; q \right)
\ee
We note some useful properties of the Kostka polynomials
\begin{itemize}
\item $K_{\l , \mu}(q)=0$ unless $\l \geq \mu $. All the non-zero coefficients of the polynomial are positive.
\item They reduce to Kostka numbers for $q=1$, $\forall \l, \mu$.
\item $K_{\l, \mu}(0) = \delta_{\l, \mu}$.
\end{itemize}
We can also use the following identities ($(k^N)$ is the partition with $ N$-non zero parts equal to $k$.)
\bea\label{idenntitiespolynomials}
\prod_{j=1}^N \frac{1}{z_j^k} = P_{(k^N)}(Z^{-1}; q) \nn \\
\sum_\lambda s_\l (\Omega) s_\l (Z) &=& \sum_{\l , \rho} s_\l(\Omega) K_{\l , \rho}(q) P_\rho(Z ; q) = \sum_\rho   Q'_\rho(\Omega ; q) P_\rho(Z ; q)\, \nn \\
 &=& \prod_{\a=1}^{n} \prod_{j=1}^N (1- \omega_\a z_j)^{-1} \, , \nn \\
\sum_\lambda s_{\l^T} (\Omega) s_{\l} (Z) &=&  \sum_\lambda Q_{\l^T} (\Omega) Q_{\l} (Z) = \sum_\lambda P_{\l^T} (\Omega) P_{\l} (Z) = \prod_{\a=1}^{n} \prod_{j=1}^N (1+ \omega_\a z_j)  \, , \nn \\
\sum_\l P_\l(X ; q) Q_\l(Y ; q) &=& \prod_{\a, \b \geq 1}^{n} \frac{1 - q x_\a y_\b}{1 - x_\a y_\b}\, ,
\eea
with $\l^T$ the transpose partition to $\l$. These should be thought of as completeness relations with respect to the inner product~\ref{inner}, the first two for $q=0$ and the last for non-zero $q$. The summands vanish unless $\ell(\l) \leq min \lbrace N, n \rbrace$.

\section{Algebras and Branching functions}\label{modulesbranching}

In this appendix we collect various useful definitions and formulae regarding (affine) Lie algebras and branching functions, relevant for the discussion in section~\ref{HallLittlewood} of the main text.

\paragraph{Algebras and weights} - For the affine algebra $\hat{A}_{n-1}$ we have the following set of commutation relations in the Chevalley generator basis $\lbrace h^i, e^i, f^i ; i = 0, ... n-1 \rbrace$
\be
[h^i, h^j] = 0 \, , \quad [h^i , e^j] = \hat{A}_{j i} e^j \, , \quad [h^i, f^j] = - \hat{A}_{j i } f^j \, , \quad [e^i, f^j] = \delta_{i j} h^i \, ,
\ee
with $\hat{A}_{i j}$ being the Cartan matrix.  The basis elements with $i>0$ generate the $A_{n-1}$ subalgebra. We shall denote the roots of ${A}_{n-1}$ by $\a$ (the co-roots are denoted by $\a^\vee$), those of $\hat{A}_{n-1}$ are simply given by $\hat{\a} = \a + m \delta$, with $\delta$ the imaginary root of the affine algebra and $m \in \mathbb{Z}$. More explicitly they are given by $\a^{i j} = e^i - e^j \, , i \neq j >0 $ in the non-affine case. We can also define positive roots by the stricter restriction $i< j$ and simple roots that are the basic building blocks of the root system by $\a^i = e^i - e^{i + 1}$. The Weyl group $W$ describes the reflection symmetries of the root system. Using positive roots, we also define the Weyl vector as a sum
\be
\rho = \half \sum_\a \a \, , \quad \a >0 \, .
\ee
In our semi-simple setting we can also take $\rho$ to be the sum of fundamental (dominant) positive weights that are defined by ${\Omega}_i = \sum_{j=1}^i e_j$ for $A_{n-1}$. The fundamental weights of $\hat{A}_{n-1}$ can be expressed in terms of fundamental weights of the $A_{n-1}$ sub-algebra as
\be
\hat{\Omega}_i = \hat{\Omega}_0 + {\Omega}_i \, , \quad i>0 \, ,
\ee
where $\hat{\Omega}_0$ is called the basic fundamental weight of the affine Lie algebra.

The general weights of an integrable representation are expanded in terms of fundamental weights as
\be
\hat{\Psi} = \sum_{i = 0}^{n-1} \hat{\psi}_i \hat{\Omega}_i + m \delta \, , \qquad {\Psi} = \sum_{i > 0}^{n-1} {\psi}_i \Omega_i  \, ,
\ee
with $\hat{\psi}_i, \, \psi_i, \, m$ integers ($m$ is related to the grading) and $\delta$ the additional imaginary root in the affine case. The coefficients ${\psi}_i $ are called Dynkin labels\footnote{There is also the notion of conjugate representations for which the weights are the negatives of the original one.}. For the affine algebras $\hat{A}_{n-1}$ the marks and comarks\footnote{These are the coefficients in the expansion of the highest root $\theta$ in roots and co-roots.} are equal to one, so the sum of all the Dynkin labels $k = \sum_i \hat{\psi}_i$ is an integer denoted as the level of the affine Lie algebra. It is also clear that different weights can have different levels.

The Dynkin labels are also the eigenvalues of the Chevalley generators of the Cartan subalgebra
\be
h^i | \Psi \rangle = \psi_i(h) | \Psi \rangle \, , \quad i>0 \, ,
\ee
the affine algebra version of which is similar, with the addition of an extra equation
\be
- L_0 |\Psi \rangle = m | \Psi \rangle \, .
\ee
This last operator is also called the grading operator.

Every finite-dimensional representation has a unique highest weight state $| \Lambda \rangle$ on which
\be
h^i | \Lambda \rangle = \Lambda_i(h) | \Lambda \rangle \, .
\ee
The eigenvalues $\Lambda_i(h)$ are highest weight Dynkin labels and are positive (they belong to the positive root Weyl chamber $P_+$). An $A_{n-1}$ highest weight with Dynkin labels  $\Lambda(h) = \Lambda_1(h), ... \Lambda_{n-1}(h)$ can also be mapped  to a partition
\be\label{mappingpartrep}
\lambda = \lbrace \l_1, ... \l_{n-1} \rbrace \, , \quad \l_i = \Lambda_i(h) + \Lambda_{i+1}(h) + ... \Lambda_{n-1}(h) \, .
\ee
where $\lambda_i$ label the boxes in each row of the Young diagram ($i$ corresponds to the row). While the highest weights belong to the positive chamber $\Lambda \in P_+ $ the weights in general belong to the space $P(\Lambda)$ that can be obtained by repeated action with elements of the Weyl group $W$. This has as a consequence that partitions can only be assigned to highest weights. If we wish to describe all the states in a representation, we take the Tableaux of shape $\lambda$ and fill it with integers according to rules consistent with the Freudenthal multiplicity formula \eqref{freudenthalmult1}. We then construct a semistandard Tableaux that describes a state in the module $M(\Lambda)$.
A description of partitions and semistandard Tableaux can be found in appendix~\ref{partitions}. In some of these modules there exists a submodule that contains only states having zero weights, in the sense that the associated Dynkin labels are zero $\psi_i = 0$. This submodule is relevant for the discussion in appendix~\ref{characterexpansion} and in the main text. A description of this submodule in terms of Gel'fand-Zetlin patterns is presented in appendix A. of~\cite{Boulatov:1991xz}. An example in terms of the more canonical labelling of semistandard Young-Tableaux is presented in appendix~\ref{partitions}, see eqn. \eqref{examplezero}.

\paragraph{Characters} - We shall also need the definition of characters. In the Chevalley basis the affine character is defined as
\be
\chi_{\hat{\Lambda}}(\zeta ; \tau) = \Tr_{M(\hat{\Lambda})} e^{2 \pi i \tau L_0} e^{- 2 \pi i \sum_{j=1} \zeta_j \hat{h}^j} \, ,
\ee
with $M(\hat{\Lambda})$ the affine module, $h^j$ the Chevalley (Cartan) generators and $L_0$ the grading operator. Using the notation of the main text we shall also denote this as $\chi_{\hat{\Lambda}}(\zeta ; \tau) \equiv \chi_{\hat{\Lambda}}(Z ; q) $. The non-affine character involves a similar expression with only the presence of the Cartans $h^i$.

\paragraph{Branching Rules} - Consider the highest weights belonging to the positive root Weyl chamber ${\Lambda} \in P_+$ of the Lie algebra ${\mathfrak{g}}$. We can then decompose the Lie algebra $\mathfrak{g}$ modules into subalgebra $\mathfrak{h}$ modules as $M^{\mathfrak{g}}({\Lambda}) = \oplus_{\Lambda'} \dim M({\Lambda})_{\Lambda'} \otimes M^\mathfrak{h}(\Lambda')$ with the relevant restricted subspaces being defined by
\be
M({\Lambda})_{\Lambda'} = \lbrace |v\rangle \in M({\Lambda}) \, : \, \, h | v \rangle = \Lambda'(h) | v \rangle  \, , \, \forall h \in \mathfrak{h} \, \rbrace \, .
\ee
The coefficients of the decomposition $ \dim M({\Lambda})_{\Lambda'}$ are called branching coefficients and correspond to the dimension of the restricted subspace $ M({\Lambda})_{\Lambda'}  $. We can equivalently write the decomposition in terms of partitions as $\lambda \mapsto \oplus_{\mu \in P_+} b^\l_\m  \, \m$. When applied to characters, this decomposition is generalised to an expansion in terms of branching functions that depend on the character parameter $q$. In the case where the Lie algebra $\hat{\mathfrak{g}}$  is affine, there is an additional subtlety. The existence of the imaginary root $\delta$ means that there exist weights that are indistinguishable from the point of view of the subalgebra $(\Lambda - m \delta)( h) = \Lambda(h) $ with $m$ an integer. This will be important for the examples we shall consider.

The specific branching functions we are interested in the main text, that describe the $\hat{A}_{n-1}/A_{n-1}$ coset submodule involve only the grading operator and are expressed as~\cite{Dorey:2016hoj} (we denote $\Lambda \equiv \Lambda(\lambda)$ since it belongs to $P_+$ and is in $1-1$ correspondence with the partition $\lambda$)
\be
b_{\Lambda(\lambda)}^{\hat{\Lambda}} (q) = \Tr_{M^{\hat{\Lambda}}_{\Lambda(\lambda)}} q^{- L_0} \, .
\ee
They also admit an explicit expansion in terms of branching coefficients~\cite{Hatayama:1998eeq}
\be\label{branchingfunct}
b_{\Lambda(\lambda)}^{\hat{\Lambda}} (q) \equiv \sum_{m=0}^\infty \dim M^{\hat{\Lambda}}_{\Lambda(\lambda) - m \delta} \, q^m \, ,
\ee
where $\dim M^{\hat{\Lambda}}_{\Lambda(\lambda) - m \delta}$ is the dimension of the branching coefficients/multiplicity of the particular embedding\footnote{Notice that this restricted module is reminiscent to the one appearing in the string functions, the difference being that we consider only $\Lambda \in P_+$ in our case.}
\be
M^{\hat{\Lambda}}_{\Lambda(\lambda) - m \delta} = \lbrace  |v\rangle \in M(\hat{\Lambda}) \, : \, \, h^i |v\rangle = \Lambda_i | v \rangle \, , \, e^i |v\rangle = 0  \, (i \neq 0) \, , \, |v\rangle = | \Lambda(\lambda) - m \delta \rangle \rbrace
\ee
and where $\hat{\Lambda} = k \Lambda_{(C)}$ are the specific highest weights as described in the main text.

A formula giving the multiplicities (Freudenthal) is the following
\bea\label{freudenthalmult1}
 \dim M^{\Lambda_1}_{\Lambda_2} = \frac{2}{\mathcal{N}} \sum_{\a > 0} \sum_{k=1}^\infty M^{\Lambda_1}_{\Lambda_2+ k \a} (\Lambda_2+ k \a, \a) \, , \nn \\
\mathcal{N} = \left((\Lambda_1 + \rho, \Lambda_1 + \rho) - (\Lambda_2 + \rho, \Lambda_2 + \rho) \right) \, ,
\eea
where with $(x,y)$ we symbolise the Killing Cartan bilinear form and $\rho$ the Weyl root. This gives a recursion relation to determine the multiplicities starting from $M^{\Lambda_1}_{\Lambda_1} = 1$.

There is also another formula due to Konstant that gives the multiplicities/branching coefficients using a summation over the Weyl group elements~\cite{Klimyk}
\be\label{Konstantmultipl}
\dim M^{\Lambda_1}_{\Lambda_2} = \sum_{w \in W} \epsilon(w) K\left((\Lambda_2 + \rho) - w(\Lambda_1 + \rho) \right)
\ee
In this formula $\rho$ the Weyl vector and $\e(w) = (-1)^{\ell(w)}$ the determinant (i.e. $\pm 1$) of the Weyl group element $w$, that can also be expressed in terms of $\ell(w)$: the length of the Weyl group element. $K$ is Konstant's partition function. Using this expression, one can derive an interesting asymptotic formula of the restricted module that shows exponential growth (Cardy growth). In particular for $A_{n-1}$~\cite{Klimyk}
\be\label{asymptoticrestricted2}
\dim M^{\hat{\Lambda}}_{\Lambda(\lambda) - m \delta}  \rightarrow_{ m \rightarrow \infty}  (c/6)^{n/4} b m^{-(n+2)/4} e^{\pi \sqrt{2 c m /3  }} \, ,
\ee
with $c$ the central charge and $b$ a coefficient that can be computed explicitly depending on the specific algebra and the level $k$.
In this limit we have kept $\Lambda$ fixed.

For completeness we also present the description of the string functions
\bea
c^{\Lambda_1}_{\Lambda_2}(q) = q^{\mathfrak{m}_{\Lambda_1}(\Lambda_2)} \sum_{m=0}^\infty \dim M^{\Lambda_1}_{\Lambda_2 - m \delta} q^m
\eea
The difference with the previous branching coefficients is that $\Lambda_2$ is more generally in $P$ and not only in the space of (dominant) positive weights $P_+$. $\mathfrak{m}_{\Lambda_1}(\Lambda_2)$ is called the modular anomaly given by
\be
\mathfrak{m}_{\Lambda_1}(\Lambda_2) = \mathfrak{m}_{\Lambda_1} - \frac{|\Lambda_2|^2}{2 k} \, , \qquad
\mathfrak{m}_{\Lambda} = \frac{|\Lambda + \rho|^2}{2 (k+g)} - \frac{| \rho|^2}{2 g} \, .
\ee

\paragraph{Some results for $A_1$} - We  now specialise to the case of $A_1$ ($SU(2)$) and its affine version. In Chevalley basis we have
\be
[e, f] = h \, , \quad [h , e] = 2 e \, , \quad [h, f] = - 2 f \, .
\ee
We have a single root $\a_1$ and fundamental weight $\Omega_1$. We therefore have a single highest weight $\Lambda_1$ to consider (usually denoted as the spin $j$ of the representation).

The simple roots of $\hat{A}_1$ are
\be
\a_0 = \delta - \a_1 \, , \quad \a_1 \, ,
\ee
The complete set of roots is spanned by
\be\label{A1roots}
\hat{\Delta} =  \lbrace n_0 \a_0 + n_1 \a_1 \, : \, |n_0-n_1| \leq 1 \, , \, n_0, n_1 \in \mathbb{Z}  \rbrace \, .
\ee
One can also restrict these numbers so that one describes only $P_+$ (positive roots) as follows
\be\label{A1proots}
\hat{\Delta}_+ =  \lbrace n_0 \a_0 + n_1 \a_1 \, : \, |n_0-n_1| \leq 1 \, , \, n_0, n_1 \in \mathbb{Z}_+ \cup{0} \, , (n_0,n_1) \neq (0,0)  \rbrace \, .
\ee
We also note that the restriction to $P_+$ becomes less important as we increase the level $k \rightarrow \infty$ (the ``cone" of $P_+$ inside $P$ opens up).

Konstant's partition function \eqref{Konstantmultipl} for the Lie algebra $\hat{A}_1$ simplifies into~\cite{Klimyk}
\be
K(n_0 \a_0 + n_1 \a_1 ) = \sum_{r=0}^\infty (-1)^r p^{(3)} \left((r+1)n_0 - r n_1  - \half r (r+1) \right) \, ,
\ee
where $n_0$ and $n_1$ are as in \eqref{A1proots}. The function $p^{(3)}(n)$ is defined via
\be
\chi_{\hat{U}(1)^r} = \phi^{-r}(q) = \sum_{n \in \mathbb{Z}} p^{(r)} (n) q^{n} \, , \quad |q|<1 \, .
\ee
with $\phi(q)$ the Euler function. This character is also the character of an $r$-fold module of a Heisenberg algebra.
In addition one also finds the identity
\bea
 \sum_{j=0}^\infty K(n_0 \a_0 + n_1 \a_1 + j \delta ) q^j \, = \,  \nn \\
= \sum_{j=0}^\infty  \sum_{r=0}^\infty (-1)^r p^{(3)} \left((r+1)n_0 - r n_1  - \half r (r+1) + j \right) q^j \, = \, \nn \\
= \phi^{-3}(q) \sum_{r=0}^\infty (-1)^r q^{-(r+1)n_0 + r n_1 + r(r+1)/2 }  \, .
\eea

Using this form of the Konstant partition function, one finds that eqn. \eqref{asymptoticrestricted2} takes the form~\cite{Klimyk}
\be
\dim M^{\hat\Lambda}_{\tilde{\Lambda} = \Lambda - m \delta}  \longrightarrow_{m \rightarrow \infty}  \frac{\sin \pi \frac{N+1}{k+2}}{2 m (k+2)} e^{\pi \sqrt{\frac{2 k m}{k+2}}} \, .
\ee
As a simpler example for $k=1$ one has $\hat{\Lambda} = \Lambda_0$ so that
\be
\dim M^{\Lambda_0}_{\tilde{\Lambda} = \Lambda_0 - m \delta} = p(j) \sim \frac{1}{4 \sqrt{3} m} e^{\pi \sqrt{2 m/3}} \, .
\ee

Another procedure to derive the branching functions is the following:
For $A_1$ the characters are
\be\label{SU2char}
 \chi_j(z) = \frac{z^{j+1} - z^{-j - 1}}{z - z^{-1}} = z^{j} + ... z^{-j} \, , \quad z = e^{2 \pi i \zeta} \, .
\ee
For the affine $\hat{A}_1$ we find the characters expressed in terms of theta functions
\be
 \chi_{\lambda_1}^{(k)} = \frac{\Theta^{(k+2)}_{\lambda_1 + 1} - \Theta^{(k+2)}_{-\lambda_1 + 1}}{\Theta^{(2)}_{1} - \Theta^{(2)}_{- 1}} \, ,
\ee
with the theta function ($z = e^{2 \pi i \zeta}, q = e^{2 \pi i \tau}$)
\bea
\Theta^{(k)}_{\lambda_1}(z ; q ; t) \equiv \Theta^{(k)}_{\lambda_1}(\zeta ; \tau ; t) \, = \,  e^{- 2 \pi i k t} \, \sum_{n \in \mathbb{Z} + \lambda_1/ 2 k} \, e^{2 \pi i  k \tau n^2 }\, e^{- 2 \pi i k n \zeta} \, , \nn \\
=  e^{- 2 \pi i k t} \, \sum_{n \in \mathbb{Z} } \, e^{2 \pi i (  k \tau n^2 + \lambda_1^2 \tau/4 k ) }\, e^{- 2 \pi i (k n \zeta - \lambda_1 n \tau + \half \lambda_1 \zeta)} \,  .
\eea
A useful expansion (in terms of $\hat{U}(1)_k$ characters) involves the use of string functions $ c_m^j (q) = c^j_{-m}(q)$
\be\label{U1expansion}
\chi^{(k)}_j (z; q) = \sum_{m = - k + 1}^k  c_m^j (q) \Theta^{(k)}_m(z;q) = \sum_{m = - k + 1}^k \sum_{n \in \mathbb{Z} }  c_m^j (q) z^{k n + m/2} q^{k (n + m/2k)^2}  \, ,
\ee
that describe the branching functions of the coset $\hat{SU}(2)_k/\hat{U}(1)_k$.

The normalised $\hat{U}(1)_k$ characters are
\be
\chi^{(k)}_j (q) = \frac{1}{\eta(q)} \Theta^{(k)}_j(z=0 ; q)  =  \frac{1}{\eta(q)}  \sum_{n \in \mathbb{Z}} q^{k (n+ j/2k)^2}
\ee
We would like to similarly express the affine in terms of the non-affine $SU(2)$ characters. In order to do so, we need to invert \eqref{SU2char} and feed the result into \eqref{U1expansion}. The best we can do is to solve
\be
z^j + z^{-j} = \chi_j(z) - \chi_{j-1}(z) \, .
\ee
We then find
\bea
\chi^{(k)}_j (z; q) &=& \sum_{m=0}^{k-1} \sum_{r = \frac{m}{2} \, mod \, k} c_m^j(q) \left(\chi_{r}(z) - \chi_{r-1}(z) \right) q^{r^2/k} \, = \, \nn \\
&=& \sum_{m=0}^{k-1} \sum_{r = \frac{m}{2} \, mod \, k} c_m^j(q) \left(q^{r^2/k} - q^{(r+1)^2/k} \right)  \chi_{r}(z) \, .
\eea

\section{The free field transition amplitude coupled to sources}\label{freetrans}

In this appendix we study the real time transition amplitude of a (messenger) free field theory coupled to external sources. We also consider the case where the sources correspond to fields of two Euclidean theories at the endpoints in time. When these field theories are integrated out in the path integral, they define appropriate boundary states for the
messenger free field.

The free field transition amplitude \eqref{free1} can be written as
\be\label{sourcetrans}
\left\langle\Phi(t_b)\vert\Phi(t_a)\right\rangle^{J} = \int_{\Phi(t_a)}^{\Phi(t_b)} \mathcal{D}{\Phi} \, \exp\left[i (S_{cl} + S_{cl}^J + S_{fl}^J)\right] \, ,
\ee
where it is convenient to split the field into a classical piece and a fluctuating piece $\Phi = \Phi_{cl} +\delta \Phi$, that is to be integrated over in the path integral (it is actually easy to obtain the Euclidean transition amplitude results by a simple continuation $(t_b - t_a) = - i L$).

The classical solution can be found solving
\be
\partial^{2}_{t} \Phi + k^2 \Phi+M^2\Phi =0 \, , \quad
\omega_k^2 = k^2 + M^2 \, ,
\ee
that yields
\be
\Phi_{cl}(t,k )=\frac{\Phi_{b}(k) \sin\omega_k (t-t_a)+\Phi_{a}(k) \sin\omega_{k}(t_b - t)}{\sin\omega_k (t_b - t_a)} \, ,
\ee
so that
\be
S_{cl}= \int \frac{d^{d}k}{(2 \pi)^d} \frac{\omega_k  \left[\left(\Phi_b(k)\Phi_b(-k)+\Phi_{a}(k)\Phi_{a}(-k)\right) \cos\omega_{k}(t_b -t_a)-2\Phi_{b}(k)\Phi_{a}(-k)\right]}{2 \sin \omega_k (t_b-t_a)} \, ,
\ee
\be\label{classcoupling}
S_{J, cl}=\int_{t_a}^{t_b} dt\int \frac{d^{d}k  \left[\Phi_a(k)\sin\omega_k (t_b-t)+\Phi_b (k)\sin\omega_k (t-t_a)\right]  }{(2 \pi)^d \, \sin\omega_k (t_b-t_a)}J(t,-k) \, ,
\ee
the two classical pieces. These are to be supplemented with the fluctuating piece depending on the Green's function
\be
G_{\omega_k}(t,t^{\prime}) =D_{\omega_k}(t,t^{\prime})^{-1}= (-\partial^{2}_{t}-\omega_k^2)^{-1}\delta (t-t^{\prime})\,,\quad t,t^{\prime}\in (t_a ,t_b) \, .
\ee
All the ambiguity of the result is hidden in the boundary conditions one imposes for the Green's function.

In particular the fluctuating piece can be written as
\be
S_{fl}^J = \int_{t_a}^{t_b}dt\,dt^{\prime}\int \frac{d^d k}{(2 \pi)^d } \left[\frac{1}{2}\delta\tilde{\Phi}(t,k)D_{\omega_k}(t,t^{\prime})\delta\tilde{\Phi}(t^{\prime},-k)-\frac{1}{2 }J(t,k)G_{\omega_k}(t,t^{\prime}) J(t^{\prime},-k)\right] \, ,
\ee
where we defined
\be
\delta\tilde{\Phi}(t,k)=\delta\Phi(t,k)+ \int_{t_a}^{t_b}dt^{\prime}\int \frac{d^{d}k}{(2 \pi)^d}\, G_{\omega_k}(t,t^{\prime}) J(t^{\prime},k) \, .
\ee
We can readily perform the path integral over the shifted fluctuations to obtain
\be\label{transitsource}
\left\langle\Phi(t_b)\vert\Phi(t_a)\right\rangle^{J} =  \prod_k \sqrt{\frac{\omega_k}{\sin\omega_k(t_b -t_a)}}e^{i S_{cl} + i S_{cl}^J + i A_{fl}^J }
\ee
where
\be\label{baregreen}
A_{fl}^J = \int_{t_a}^{t_b}dt\,dt^{\prime}\int \frac{d^dk}{(2 \pi)^d} \left[-\frac{1}{2 } J(t,k)G_{\omega_k}(t,t^{\prime}) J(t^{\prime},-k)\right]
\ee
One choice for the Green's function is that of Dirichlet boundary conditions (we use $< , >$ symbols to keep track of the ordering between $t, t'$)
\bea\label{DirichletGreen}
G^D(t,t^{\prime}) =  \frac{1}{\omega_k  \sin \omega_k (t_b - t_a)} \sin \omega_k(t_b - t_{>}) \sin \omega_k (t_{<} - t_a) \, , \qquad \nn \\
 G^D(t,t^{\prime})\vert_{t = t_b} =   G^D(t,t^{\prime})\vert_{t^{\prime} = t_a} = 0 \, , \quad
\eea
another choice being the Neumann one
\bea
G^N(t,t^{\prime}) = - \frac{1}{\omega_k  \sin \omega_k (t_b - t_a)} \cos \omega_k(t_b - t_{>}) \cos \omega_k (t_{<} - t_a) \, , \qquad \nn \\
 \partial_t G^N(t,t^{\prime})\vert_{t = t_b} =  \partial_{t^{\prime}} G^N(t,t^{\prime})\vert_{t^{\prime} = t_a} = 0 \, . \quad
\eea

A simplification in our case comes from the fact that the sources are only defined at the interval endpoints, that is
\be\label{boundarysource}
J(t, k) = g \delta(t - t_a)  \phi_1(k) + g \delta(t - t_b) \phi_2(k) \, .
\ee
One then finds that the classical piece eqn. \eqref{classcoupling}, simply reduces to the classical coupling action of the main text eqn. \eqref{intpart} as expected. The fluctuating piece in the case of Dirichlet boundary conditions simply vanishes. Upon integrating over the boundary values $\Phi_a(k), \Phi_b(k)$ one finds the induced effective action for $\phi_{1,2}(k)$ (with this method we can also obtain the field independent prefactor/vacuum energy)
\be\label{Dirichleteffective}
S_{eff} = g^2 \int \frac{d^d k}{(2 \pi)^d} \frac{\left[ \phi_1(k) \phi_1(-k) + \phi_2(k) \phi_2(-k) \right] \cos \omega_k(t_b - t_a) + 2 \phi_1(k) \phi_2(-k)  }{2 \omega_k \sin \omega_k (t_b - t_a)}   \, .
\ee
that matches the computation performed in the main text eqn.\eqref{intervalself} and \eqref{intervalself12}.

We would like now to perform a computation of a messenger correlation function in the presence of the two boundary theories, which we should therefore integrate out in the path integral. These boundary theories define then a certain pair of in/out boundary states $| B(t_a) \rangle$, $\langle B(t_b) |$, and the transition amplitude is between these two boundary states. This computation is relevant if we wish to understand if there is any relation between the notion of messenger boundary time and that of bulk time (in the case where the dual geometry would be that of a Bang/Crunch type of universe we should similarly observe some peculiarities in the boundary theory correlation functions as we reach the end of time). We therefore need to add to the term \eqref{boundarysource} an additional arbitrary source term, so that we can measure correlators at arbitrary times $t$ by taking functional derivatives. Integrating out the boundary fields $\phi_{1,2}$ we find the boundary states (from now on we do not keep track of the overall normalisation, since it is irrelevant for the correlators)\footnote{In case the boundary fields $\phi_{1,2}$ are interacting, they would also introduce non-linearities in these boundary states.}
\bea
\langle \Phi_a | B(t_a) \rangle = \mathcal{N} \exp \left(- {g^2} \int \frac{d^dk}{(2 \pi)^d} \frac{\Phi_a(k) \Phi_a(-k)}{k^2 + m^2}   \right) \, , \nn \\
\langle B(t_b)  | \Phi_b \rangle = \mathcal{N} \exp \left(- {g^2} \int \frac{d^dk}{(2 \pi)^d} \frac{\Phi_b(k) \Phi_b(-k)}{k^2 + m^2}   \right)  \, .
\eea
Any correlator should then be computed as an expectation value between these two boundary states. In particular it can be written in terms of the source transition amplitude of the free field $\Phi$ \eqref{sourcetrans} and \eqref{transitsource}, as
\bea
\langle B(t_b) | e^{i S_J} | B(t_a) \rangle = \int \mathcal{D} \Phi_a \mathcal{D} \Phi_b  \, \langle B(t_b)  | \Phi_b \rangle \, \langle \Phi_b | \Phi_a \rangle^J \,  \langle \Phi_a | B(t_a) \rangle  \, = \, \nn \\
=  \, \mathcal{N}'    \int \mathcal{D} \Phi_a \mathcal{D} \Phi_b  \, e^{\left(- {g^2} \int \frac{d^dk}{(2 \pi)^d} \frac{\Phi_b(k) \Phi_b(-k)}{k^2 + m^2}   \right)} \, e^{i S_{cl} + i S_{cl}^J + i A_{fl}^J } \,  e^{ \left(- {g^2} \int \frac{d^dk}{(2 \pi)^d} \frac{\Phi_a(k) \Phi_a(-k)}{k^2 + m^2}   \right)} \, = \,  \nn \\
= \, \mathcal{N}'    \int \prod_{i = a, b} \mathcal{D} \Phi_i   \, \exp \left(i A_{fl}^J +  i \int \frac{d^dk}{(2 \pi)^d} \frac{\half \Phi_i(k)  M_{i j} \Phi_j(-k) +  \Phi_i(k) J_i(-k)}{\sin\omega_k (t_b-t_a)}  \right) \, , \nn \\
\eea
with $A_{fl}^J$ given by \eqref{baregreen} and where we defined
\bea
J_{i = b} (-k) = - \int_{t_a}^{t_b} d t  \sin \omega_k (t - t_b) J(t, -k) \, , \quad  J_{i = a} (-k) =  \int_{t_a}^{t_b} d t  \sin \omega_k (t - t_a) J(t, -k) \, , \nn \\
M_{i j} =  \begin{pmatrix}
\omega_k \cos\omega_k (t_b-t_a) +  \frac{g^2 {\sin\omega_k (t_b-t_a)} }{k^2+ m^2}   & - \omega_k  \\
- \omega_k  & \omega_k \cos\omega_k (t_b-t_a) +  \frac{g^2 {\sin\omega_k (t_b-t_a)} }{k^2+ m^2}
\end{pmatrix} \, . \qquad
\eea
Performing the integral over $\Phi_a, \Phi_b$ we find a correction to the Green's functions due to boundary effects captured by the inverse matrix $M^{-1}$.

In particular we get for the two point function
\bea
\langle B(t_b) | \Phi(t, k) \Phi(t', - k) | B(t_a) \rangle =  G^D(t,t^{\prime} ; k) \, - \, \nn \\
 - \frac{2 G^D(t,t^{\prime} ; k)  \omega^2_k  }{\left( \omega_k^2  -\frac{g^4}{(k^2+ m^2)^2} \right)\sin^2 \omega_k (t_b-t_a) - 2 \frac{g^2 \omega_k }{k^2+ m^2} \sin \omega_k (t_b-t_a)  \cos \omega_k (t_b-t_a) } \, - \, \nn \\
 - \half \frac{ (\sin^2 \omega_k (t' - t_b) + \sin^2 \omega_k (t - t_a) \, + t \leftrightarrow t') \left( \frac{g^2}{k^2 + m^2} +  \omega_k \cot \omega_k (t_b - t_a) \right) }{\left( \omega_k^2  -\frac{g^4}{(k^2+ m^2)^2} \right)\sin^2 \omega_k (t_b-t_a) - 2 \frac{g^2 \omega_k }{k^2+ m^2} \sin \omega_k (t_b-t_a)  \cos \omega_k (t_b-t_a) } \nn \\
\eea
The first thing to notice is that this correlator does not depend only on $t-t'$ (due to finite size boundary effects).
The second is that as we approach one operator on the endpoint, we find  that while $G^D$ vanishes, the two point function does not vanish, but approaches
\bea
\langle B(t_b) | \Phi(t_b, k) \Phi(t', - k) | B(t_a) \rangle =  \qquad \nn \\
= - \half \frac{ (\sin^2 \omega_k (t' - t_b) + \sin^2 \omega_k (t_b - t_a)  +  \sin^2 \omega_k (t' - t_a)) \left( \frac{g^2}{k^2 + m^2} +  \omega_k \cot \omega_k (t_b - t_a) \right) }{\left( \omega_k^2  -\frac{g^4}{(k^2+ m^2)^2} \right)\sin^2 \omega_k (t_b-t_a) - 2 \frac{g^2 \omega_k }{k^2+ m^2} \sin \omega_k (t_b-t_a)  \cos \omega_k (t_b-t_a) } \nn \\
\eea
Finally we observe the presence of a remnant piece even if we set $t' = t_b$, which is equivalent to that of setting $t' = t_a$. Upon fourier transforming $k \rightarrow x$ this will give a position space correlation when the two points are at the end of time as well as a correlation between the past and future endpoints. In some sense one can argue that the first correlation arises because the boundary state has a built in correlation.

\newpage
\addcontentsline{toc}{section}{References}
 
\end{document}